\documentclass[%
reprint,
superscriptaddress,
amsmath,
amssymb,
longbibliography,
floatfix,
]{revtex4-1}

\usepackage{graphicx}% Include figure files
\usepackage{dcolumn}% Align table columns on decimal point
\usepackage{bm}% bold math
\usepackage{color}

\begin{document}

\newcommand\newver[1]{{\color{blue}#1}}
\newcommand\comment[1]{{\color{black}#1}}

%\preprint{APS/123-QED}

\title{Attosecond Streaking in the Water Window: A New Regime of Attosecond Pulse Characterization 
}% Force line breaks with \\

\author{Seth L. Cousin}
\affiliation{ICFO--Institut de Ciencies Fotoniques, The Barcelona Institute of Science and Technology, 08860 Castelldefels (Barcelona), Spain}

\author{Nicola Di Palo}
\affiliation{ICFO--Institut de Ciencies Fotoniques, The Barcelona Institute of Science and Technology, 08860 Castelldefels (Barcelona), Spain}

\author{B\'arbara Buades}
\affiliation{ICFO--Institut de Ciencies Fotoniques, The Barcelona Institute of Science and Technology, 08860 Castelldefels (Barcelona), Spain}

\author{Stephan M. Teichmann}
\affiliation{ICFO--Institut de Ciencies Fotoniques, The Barcelona Institute of Science and Technology, 08860 Castelldefels (Barcelona), Spain}

\author{M. Reduzzi}
\affiliation{Dipartimento di Fisica Politecnico Milano Piazza Leonardo da Vinci 32, 20133 Milano Italy}
\affiliation{Instituto di Fotonica e Nanotecnologie (CNR-IFN) Politecnico Milano Piazza Leonardo da Vinci 32, 20133 Milano Italy}

\author{M. Devetta}
\affiliation{Instituto di Fotonica e Nanotecnologie (CNR-IFN) Politecnico Milano Piazza Leonardo da Vinci 32, 20133 Milano Italy}

\author{A. Kheifets}
\affiliation{Research School of Physics and Engineering, The Australian University, Canberra, ACT 2601, Australia}

\author{G. Sansone}
\affiliation{Dipartimento di Fisica Politecnico Milano Piazza Leonardo da Vinci 32, 20133 Milano Italy}
\affiliation{Instituto di Fotonica e Nanotecnologie (CNR-IFN) Politecnico Milano Piazza Leonardo da Vinci 32, 20133 Milano Italy}
\affiliation{Institute of Physics, University of Freiburg, Hermann-Herder-Strasse 3, 79104 Freiburg, Germany}

\author{Jens Biegert}
 \email{Jens.Biegert@icfo.eu}
\affiliation{ICFO--Institut de Ciencies Fotoniques, The Barcelona Institute of Science and Technology, 08860 Castelldefels (Barcelona), Spain}
\affiliation{ICREA, Pg. Llu\'{\i}�s Companys 23, 08010 Barcelona, Spain}

\date{\today}% It is always \today, today,

\begin{abstract}
We report on the first streaking measurement of water-window attosecond pulses generated via high harmonic generation, driven by sub-2-cycle, CEP-stable, 1850\,nm laser pulses. Both the central photon energy and the energy bandwidth far exceed what has been demonstrated thus far, warranting the investigation of the attosecond streaking technique for the soft X-ray regime and the limits of the FROGCRAB retrieval algorithm under such conditions. We also discuss the problem of attochirp compensation and issues regarding much lower photo-ionization cross sections compared with the XUV in addition to the fact that several shells of target gases are accessed simultaneously. Based on our investigation, we caution that the vastly different conditions in the soft X-ray regime warrant a diligent examination of the fidelity of the measurement and the retrieval procedure. 
%\begin{description}
%\end{description}
\end{abstract}

%Use showkeys class option if keyword
\maketitle

%\tableofcontents

\section{\label{sec:Intro}Introduction}
Excitation, scattering and relaxation of electrons are core phenomena that occur during the interaction between light and matter. It is essential to study the temporal dynamics of electrons \cite{Corkum2007} since their behavior determines how a chemical bond forms or breaks, their confinement and binding determines how energy flows, and these phenomena govern the efficiency of modern organic solar cells or the speed of electronics, all alike. Furthering our understanding therefore requires the capability to localize an excitation in a molecule, or a material, and to follow the flow of energy with attosecond temporal resolution. Spectacular progress has been made in attoscience research since the first real time observation of the femtosecond lifetime of the M-shell vacancy in krypton in 2002 \cite{Drescher2002}. Electron tunneling \cite{Uiberacker2007}, time delays in photoemission \cite{Schultze2010,Klunder2011}, an atom's response during photoabsorption \cite{Ott2013}, electron localization during molecular dissociation \cite{Sansone2010}, ultrafast charge transfer in biomolecules \cite{Calegari2014}, phase transitions \cite{Schiffrin2013} and electron dynamics in condensed matter \cite{Schultze2014} were investigated. However, despite these achievements, isolated attosecond pulses \cite{Hentschel2001,Sansone2006,Goulielmakis2008,Ferrari2010,Zhao2012} were only generated in the extreme ultraviolet range (EUV or XUV), i.e. at photon energies lower than 124\,eV \cite{Fieß2010}, thereby confining investigations largely to valence electron dynamics.\\ 
Isolated attosecond pulses at higher photon energies than the XUV (10 to 124\,eV), in the soft X-ray range (SXR), permit localizing the initial excitation step through ionization of a specific core level of a distinct target atom. This “fingerprinting” capability is essential for localization of the flow of energy and excitation since it permits interrogating the entire electronic structure of an atom, molecule or solid with ultrafast temporal resolution and element selectivity. Such soft X-ray attosecond pulses will give access to a plethora of fundamental processes like intra-atomic energy transfer \cite{Svensson1988} and its range \cite{Sisourat2010}, charge induced structural rearrangement \cite{Tseng2010}, change in chemical reactivity \cite{Remacle2006}, photo damage of organic materials \cite{Griffini2013}, emergence of reflectivity \cite{Ross1985}, and carrier scattering \cite{Stahler2010}, recombination \cite{Shanmugam2013} and exciton dynamics \cite{Benjamin2013}, just to name a few. Attosecond pulses in the SXR spectral range will also give access to fundamental ultrafast electron transfer between adsorbates and surfaces which have been accessible so far only by alternative approaches such as core hole spectroscopy \cite{Fohlisch2005}. Another enticing possibility with attosecond soft X-ray pulses is the realization of ultrafast electron diffraction from within the molecule itself \cite{Boll2013}. This concept requires the measurement of the electron interference pattern arising from an initially well-defined and confined electronic wavepacket inside the molecule, thus calling for an initially localized ionization process. Soft X-ray pulses with duration at, or below, the atomic unit of time (24\,as) would provide a universal tool to access the timescales of exchange and correlation which characterize the universal response of electronic systems after the sudden removal of an electron and which occur on the characteristic timescale of less than 50\,as \cite{Breidbach2005}. Such pulses will also enable, for the first time, to time-resolve the few attosecond dynamics inside tailored materials, whose sub-femtosecond dynamics are largely unexplored \cite{Reed2010}.\\
Here, we present the generation of attosecond-duration pulses with photon energy in the soft X-ray range up to 350\,eV. This photon energy is sufficiently high to fully cover the K-shell absorption edge of carbon at 284\,eV for spectroscopic applications \cite{Cousin2014a,Cousin2015, Biegert2016}. We demonstrate the first attosecond streaking measurement in the soft X-ray water-window regime and confirm an upper limit of the duration of the isolated pulse of 322\,as. These results herald a new era of attosecond science in which sub-femtosecond temporal resolution is paired with element selectivity by reaching the fundamental absorption edges of important constituents of matter such as carbon.\\
\section{\label{sec:PulseCharacterization}Femtosecond and attosecond pulse generation and characterization}
With the initial generation of ultrashort pulses (defined here as pulses which are too short to be measured directly using modern detectors and electronics), an entire field of optics has arisen to address the challenge of characterizing them. Frequency resolved optical gating (FROG) \cite{Kane1993} and spectral phase interferometry for direct electric-field reconstruction (SPIDER) \cite{Walmsley1996} solved this challenge, by using replicas of the pulse itself combined with non-linear interactions to either reduce the problem to a solvable phase retrieval issue in the case of FROG, or to directly extract the phase through Fourier analysis of the the non-linear signals recorded in the case of SPIDER. Due to the relationship between spectral bandwidth and pulse duration, further significant reduction of pulse durations to the sub-femtosecond regime is not possible directly from laser media, however  through novel pulse compression schemes such as filamentation and hollow-core fibre pulse compression \cite{Nisoli1996,Schenkel2003,Hauri2004,Couairon2005,Hauri2005a,Guandalini2006}, pulse durations continued to decrease to the single-cycle limit. Sub-cycle pulses have also been synthesized \cite{Wirth2011} in schemes, where discrete broad spectra can be combined coherently. Another approach to the generation of sub-femtosecond radiation comes fortuitously and intrinsically from the process of high harmonic generation (HHG) \cite{Chen2014}. HHG results from the interaction of an ultrashort, intense laser pulse with typically a gas-phase medium. The laser pulse first facilitates the tunnel ionization of an electron from the gas atom, where after it accelerates the electron initially away and then back to the parent ion \cite{Corkum1993}. In the event of recombination with the ion, the kinetic energy obtained during its excursion is released in the form of high harmonic radiation. The energy range and cut-off that is achieved during HHG is dependant predominantly on the wavelength of the driving laser. In the past few years, there has been a significant push to generate bright HHG radiation at higher photon energies / shorter wavelengths by driving the process with longer wavelength sources. The HHG process repeats every half cycle of the  driving laser, which at 800\,nm corresponds to 2.7\,fs per cycle. Not all electrons ionize or recombine at the same time, which in turn implies that the high harmonic photons are too born at different times. A pulse in which different wavelengths are temporally dispersed is by definition a chirped pulse and in this specific context the dispersion is called attochirp. Even in the case of maximum attochirp, the sub-cycle nature of the process guarantees sub-cycle high harmonic pulse durations. Another consequence of the half-cycle repetition frequency of the process is that every half cycle with the intensity to initiate the process, can result in a  burst of this attochirped radiation, so for a multi-cycle pulse i.e. 30\,fs at 800\,nm multiple attosecond bursts will be present, which results in a train of attosecond pulses. The obvious path to an isolated attosecond pulse is then to ensure that only one half-cycle dominates the HHG process which can be achieved through various gating techniques \cite{Goulielmakis2008,Ferrari2010,Sansone2006,Zhao2012,Kim2013}. A fortuitous temporal gating arises as a consequence of successful phase matching of the HHG in particular conditions \cite{Chen2014} especially relevant to this work. These gating techniques, in combination with control of the carrier-to-envelope phase (CEP) of the laser pulses can be used to select the pulse's electric field shape in which predominantly only one half-cycle dominates and only one repeatable isolated attosecond pulse is generated.\\
Extending existing pulse characterization techniques to attosecond pulses is not as straight forward as using the same non-linear interactions as used for few femtosecond pulses i.e. second harmonic generation in bulk is not possible at the short X-ray wavelengths arising from HHG. What can be exploited however is the perturbation of electrons (ionized from atoms irradiated by the X-rays) by a phase-locked copy of the pulse which generated the harmonic radiation. Photo-electron spectrograms generated both by attosecond pulse trains and isolated pulses can be evaluated using different techniques, namely attosecond streaking \cite{Itatani2002} for isolated attosecond pulses and reconstruction of attosecond beating by interference of two-photon transitions (RABBIT) \cite{Paul2001} to yield the pulse duration of the harmonic radiation.\\
The experimental basis of the attosecond streaking measurement is an interferometer in which one arm carries the X-ray radiation and the other arm carries a fraction of the fundamental radiation that was used to generate the X-ray radiation. After the arms of the interferometer are recombined, the combined components irradiate a gas target. The X-rays have sufficient energy to photoionize the gas target whereas the weaker fundamental radiation just has the energy to perturb the ejected photoelectrons. By measuring the time-of-flight of the photoelectrons as a function of delay between the two arms of the interferometer, the momentum shift of the photoelectrons is recorded in the form of a streaking spectrogram. Encoded in the  spectrogram is the phase of the electron wavepacket which is assumed to be identical to the X-ray photon wavepacket. An iterative phase retrieval algorithm such as frequency-resolved optical gating for complete reconstruction of attosecond bursts FROGCRAB~\cite{Mairesse2005} can be used to reconstruct the full electric field of the X-ray pulse and hence its pulse duration.\\
\section{\label{sec:AttoStreaking}Attosecond Streaking}
\begin{figure*}
\includegraphics[width=1\textwidth]{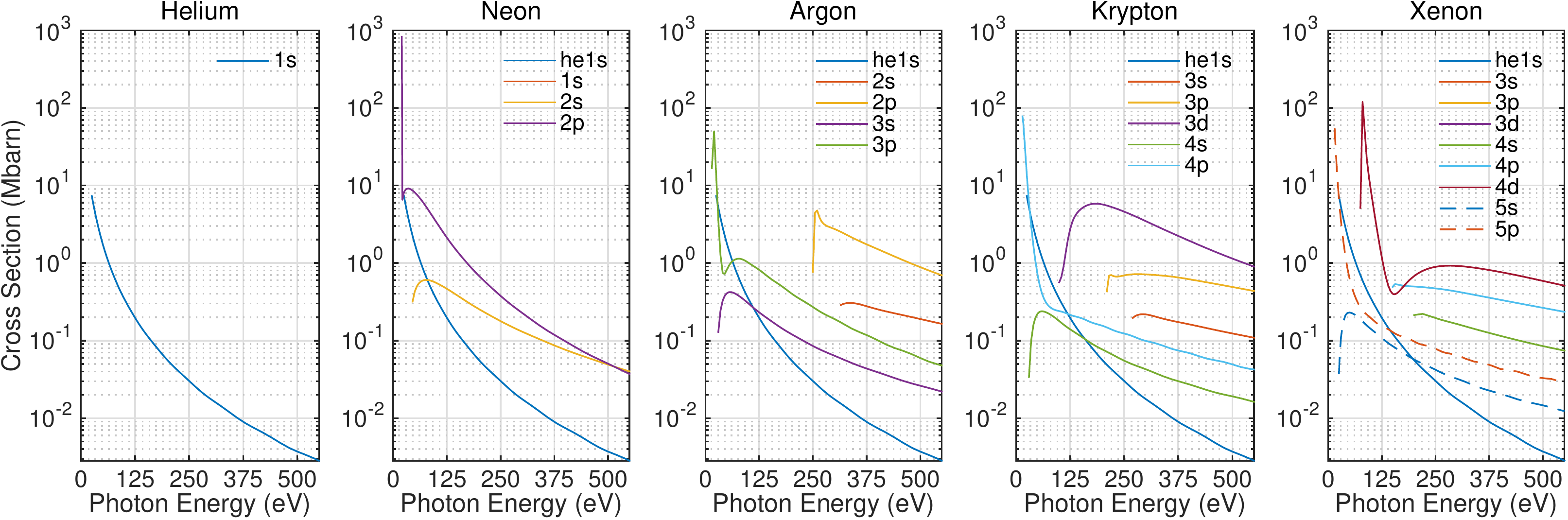}
\caption{\label{fig:fig1}Photoionization cross sections \cite{Yeh1985}, for the typically used target gases in attosecond streaking experiments. Helium 1s cross section is included in dark blue in each plot as a point of reference.}
\end{figure*}

To date, virtually all attosecond streaking measurements reconstructed by the FROGCRAB algorithm, have been performed on attosecond pulses that have been generated by HHG, driven with 800\,nm titanium sapphire (Ti:Sa) lasers \cite{Kienberger2004,Hentschel2001,Sansone2006,Goulielmakis2007,Cavalieri2007,Uiberacker2007,Goulielmakis2008,Schultze2010}. Recent research into using these algorithms suggests that care needs to be taken if very accurate reconstruction is essential and that other algorithms may perform better \cite{Zhao2017,Pabst2016,Pabst2017}. Due to the scaling of the HHG cut-off energy \cite{Krause1992} the spectra supporting these pulses has been restricted to XUV photon energies and bandwidths well below 50\,eV.
To access higher photon energies efficiently, the only way forward is to use longer wavelength radiation to drive HHG. Technically, this means either the addition of wavelength conversion schemes after the Ti:Sa laser system, or a complete new laser architecture such as optical parametric amplification (OPA) / optical parametric chirped pulse amplification (OPCPA). Both of these options pose significant technical challenges. There are only a handful of systems world-wide capable of supplying the stable and extreme radiation needed to generate repeatable isolated attosecond pulses in the water window range. i.e. central wavelengths larger than 800\,nm, few-cycle, CEP controlled laser pulses with sufficient energy to facilitate phase-matched HHG.\\
Due to the wavelength scaling of the single atom response ($\lambda^{-5}$ to $\lambda^{-9}$) \cite{Frolov2008,Austin2012}, phase-matched HHG using longer wavelength radiation relies on multi-atmosphere gas pressure to balance the phase mismatch between the fundamental driving laser and the generated harmonics \cite{Popmintchev2009}. These multi-atmosphere gas pressures are achieved either via capillary wave-guide targets or through the correct engineering of vacuum pumping systems to facilitate an easy-to-align free-space target. Again there are only a handful of systems reported capable of achieving these environments.\\
In the case of XUV attosecond pulses, photoionization occurs from the valence or inner valence shells, while for soft X-ray (for example in the water window 283-543\,eV) photoelectrons, core-shell electrons are preferentially released in the continuum. Even though gaining access to these levels is interesting for the purpose of electronic processes triggered by a core vacancy, it can be a hindrance for the implementation of streaking measurements, due to the photoionization cross sections for core levels, which are significantly lower than for valence shells (see Fig. \ref{fig:fig1}). This effect, combined with the lower flux of high photon energy in the soft X-ray, potentially makes the acquisition of a streaking spectrogram a lengthy process.\\
The broad spectra associated with isolated attosecond pulses in the water window spans over multiple core shells of the typical gases used. The overlap of streaking traces from two different shells heavily complicates interpretation of the streaking spectrogram. Using helium as the target gas can mitigate this issue, however the ionization cross section is extremely low. The 3d shell of krypton stands out as a viable option for water-window streaking as it has a relatively high cross section compared to krypton's other core shells and has a relatively flat response over a broad bandwidth.
\section{\label{sec:AttoChirp}Attochirp Compensation}
\begin{figure}
\includegraphics[width=0.48\textwidth]{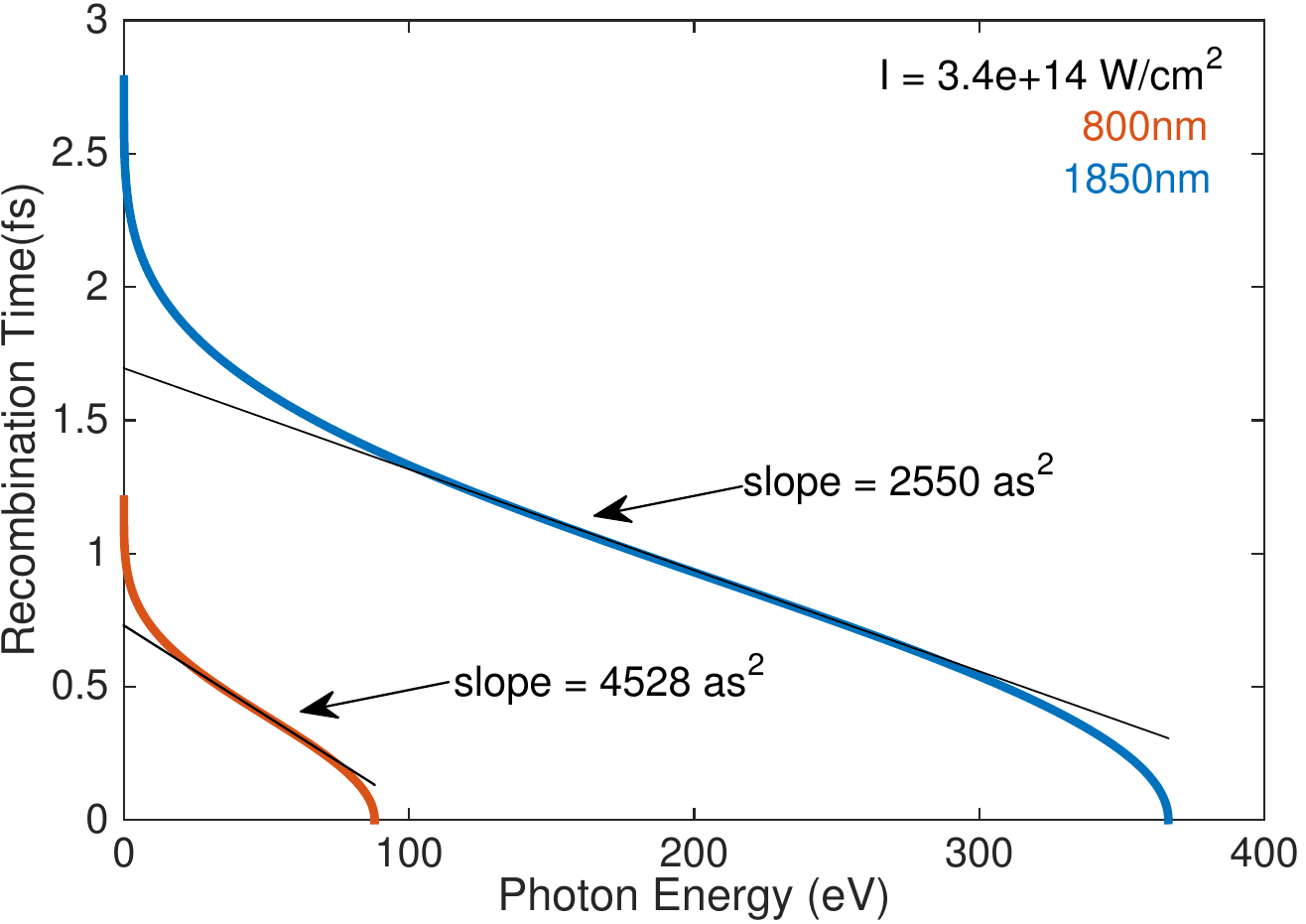}%
\caption{\label{fig:fig2}Attochirp as a function of driving wavelength, 800\,nm (red curve) and 1850\,nm (blue curve).}
\end{figure}

Attochirp is intrinsic to the HHG process and in order to compensate for the chirp, either a suitable post-compression technique is needed or a method \cite{Biegert2014} to mitigate attochirp. Short trajectory harmonics have traditionally been selected via placement of the target with respect to the focus. Attosecond pulses emanating from short trajectory harmonics are positively chirped (low energy electrons recombine before the high energy electrons). Negative chirp can be introduced by the thin metal foil filter \cite{Kim2004} used to reject the fundamental laser from the copropagating XUV pulse. One of the benefits of driving HHG with long wavelength drivers is that the magnitude of the intrinsic attochirp scales inversely with wavelength \cite{Doumy2009} and intensity \cite{Mairesse2003a}. Figure \ref{fig:fig2} highlights this, showing the calculated slope of recombination for both 800\,nm and 1850\,nm driven harmonics. It has also been proposed that atomic gas can be used as an attochirp compressor \cite{Ko2012}, however unlike in the XUV regime, beyond 200\,eV, chirp compensation is hard to come by. The group delay dispersion (GDD) of the materials used for compression is far less effective as shown in Tab. \ref{tab:tab1}, showing GDD and transmission for different materials at 50\,eV and at 250\,eV. 
\begin{table}
\centering
\caption{GDD and transmission of gases and solids for attochirp compensation.}
\label{tab:tab1}
\begin{ruledtabular}
\begin{tabular}{lcccc}
Material                  & \multicolumn{2}{c}{GDD (as$^2$)}  & \multicolumn{2}{c}{Transmission (\,\%)} \\ \cline{1-1}
\textbf{Solid (100\,nm)}    & \textbf{50\,eV} & \textbf{250\,eV} & \textbf{50\,eV}     & \textbf{250\,eV}    \\ \hline
Aluminium (Al)            & -673          & -7             & 83                & 34                \\
Zirconium (Zr)            & -1975         & -18            & 0.6               & 17                \\
Titanium (Ti)             & 11633         & -13            & -                 & 62                \\
Tin (Sn)                  & 1087          & -13            & 0.09              & 58                \\
Chromium (Cr)             & 3114          & -18            & 0.3               & 36                \\
\textbf{Gas (1\,mm, 100\,mbar)} &               &                &                   &                   \\ \hline
Neon (Ne)                 & 133           & -4             & 17                & 87                \\
Xenon (Xe)                & -1168         & -6             & 72                & 69                \\
Krypton (Kr)              & -978          & 5              & 68                & 30                \\
Argon (Ar)                & -1083         & 1              & 81                & 31               
\end{tabular}
\end{ruledtabular}
\end{table}
Chirped attosecond pulses with central energies below 100\,eV have already been compressed using metallic foils \cite{Sansone2006}. For a chirped attosecond pulse with a central energy of 250\,eV, to compensate for 2500\,as$^2$ of chirp using zirconium, we would need a $\sim$13\,$\mu$m thick foil, or in neon gas we would need $\sim$6\,bar in a 10\,mm cell. The transmission through either however would be negligible. Attochirp compensation of pulses beyond 200\,eV may not be possible using the traditional post compression schemes, but more novel schemes may be needed.\\
\section{\label{sec:ExpSetup}Experimental Setup}
\begin{figure*}
\includegraphics[width=1\textwidth]{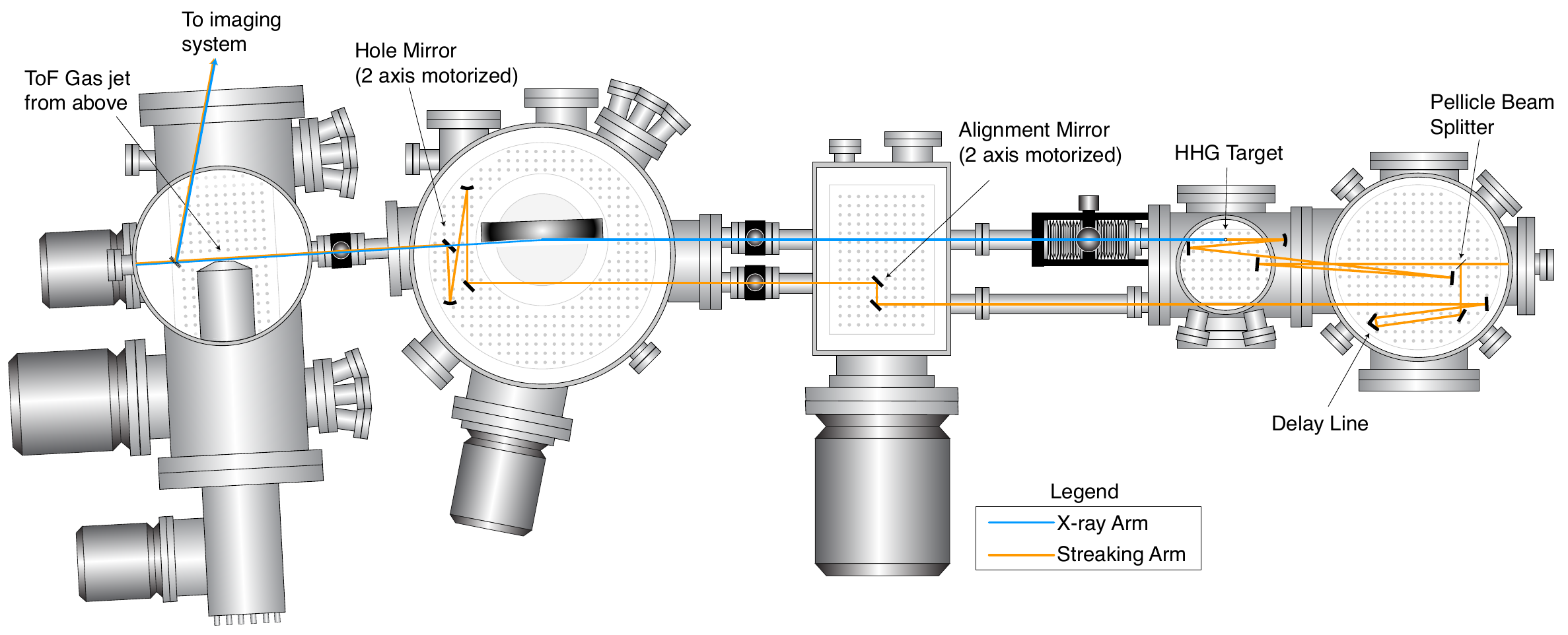}
\caption{\label{fig:fig3}Schematic layout of the attosecond streaking beamline. Five interconnected vacuum chambers make up the attosecond streaking beam line. Following the beam propagation direction, from right to left, the first two chambers are only rough-pumped to facilitate the high gas pressures needed for phase matched HHG, the last three chambers are turbo-pumped to minimize reabsorption of the X-rays and ensure the required pressure environments needed by the diagnostics in the last chamber (left).}
\end{figure*}

To fulfil the stringent requirements of the laser source needed to generate isolated attosecond pulses at high photon energies, we have dedicated extensive time and effort to our light source. The system is based upon a cryogenically-cooled, two stage, titanium sapphire amplifier delivering stable, robust,  40\,fs, 7\,mJ pulses at 1\,kHz with immaculate beam quality. These pulses are used to seed a TOPAS-HE OPA in which three stages of white-light seeded amplification results in passively CEP stabilized 45\,fs idler pulses at a center wavelength of 1850\,nm. A hollow-core fiber pulse compressor is then used to spectrally broaden these pulses to support sub 2-cycle pulse durations, which are compressed in bulk to give 0.4\,mJ, 12\,fs CEP stable pulses. An additional slow feedback loop ensures that slow drifts of CEP fluctuations can be mitigated over arbitrary time durations.\\
These extreme laser pulses enter an attosecond streaking beamline depicted in Fig. \ref{fig:fig3} where the majority of their energy is devoted to harmonic generation.  They are focused down to 54\,$\mu$m to achieve an intensity of 4.3$\times$10$^{14}$W/cm$^2$ in a free-space target with an interaction region $<$1\,mm long and a backing pressure of 3\,bar of neon. See ref. \cite{Cousin2014a} for more details regarding HHG, target geometry and water-window spectroscopy with this source). Soft X-ray radiation is generated well into the water-window range  resulting in flux of 5.6$\times$10$^{5}$ photons/s from 284\,eV to 350\,eV on target. The remaining energy is split to be used as the perturbing streaking field in the streaking measurement, ultimately achieving an intensity of 3.2$\times$10$^{11}$W/cm$^2$ in the focus.\\ 
The X-rays are refocused to a time-of-flight (ToF) spectrometer using an ion-beam polished, grazing incidence, ellipsoidal X-ray optic, possessing a measured surface roughness of less than 0.5\,nm over the whole surface (260\,mm$\times$50\,mm). A gold mirror with a 3\,mm through-hole, drilled at 45$^\circ$ facilitates the recombination and co-alignment of the 1850\,nm streaking pulses (reflecting from the front surface) and the X-rays which propagate unabated through the drilled aperture due to their low divergence. This co-alignment is made possible under vacuum due to the use of motorized mirrors for beam steering, as well as a in-vacuum actuator controlled beam sampling mirror extracting the combined beams to an imaging system. When the mirror is inserted a beam profiling camera can either image the plane of the hole-drilled mirror, or by translating the camera further away, a far-field image can be seen. Collinearity is verified when 1850\,nm beams from both arms are centered on the hole and remain spatially overlapped in the far field. The ToF spectrometer (Stefan Kaesdorf ETF10), can be operated in either an electron or ion collection mode and incorporates a 461\,mm field free drift tube. The spectrometer offers a ToF resolution of T/DT = 100, which translates to an energy resolution of E/DE = 50. We discuss the impact of spectral resolution in section \ref{sec:SXRFROGCRAB}. An Einzel lens is employed to focus charged particles to a micro channel plate without changing their energies. Optimal focusing of the electrons is achieved when a lens voltage of around 5 times the expected electron energy is applied. Energy acceptance for this lens voltage is approximately Gaussian with a central energy around 200\,eV and a full width half max (FWHM) bandwidth of around 150\,eV. In this configuration however, the normal acceptance angle of the ToF spectrometer (30$^{\circ}$ full cone) is replaced by a spherical acceptance volume with a diameter of 200\,$\mu$m. This required diligent alignment as it it was imperative to co-align the soft X-rays, the streaking IR and the gas jet all within the small volume.\\
\section{\label{sec:SXRStreaking}Attosecond Streaking in the SXR}
Figure \ref{fig:fig3} depicts the implementation of the attosecond streaking interferometer. Interferometric stability was qualitatively evaluated by monitoring the fringes generated by the interference of 1850\,nm radiation in both arms of the interferometer, measured at the location of the ToF gas jet location. Without any active stabilization the fringe stability was excellent, suggesting no need to stabilize the system. The beamline design specifically isolates the chambers from the optical breadboards inside, assisting with the stability. Note that no infrared attenuation filter is used and hence no attochirp compensation is expected. This was done to maximize the flux on target, however various tests were carried out to affirm that any remnant infrared would not influence the streaking trace. Firstly, the remnant infrared was well below the intensity needed to ionize the krypton ToF gas jet, estimated to be three orders of magnitude lower than that of the streaking infrared. Next, even when assuming the remnant 1850\,nm radiation is co-propagating with the soft X-rays, it would be phase locked with the attosecond burst. In the case that there is significant intensity in the infrared, electrons ionized in the ToF gas jet by the soft X-rays would experience a constant momentum shift, which would manifest in the photo electron spectra as a constant offset in electron energy. We confirmed this was not the case, by examining the photo electron spectra generated in identical conditions, however first with a thin filter combination of 200\,nm carbon / 200\,nm chromium blocking the infrared and then with no filter. No shift in electron energy was detected, only a significant attenuation of the signal.\\ 
\subsection{Recording the streaking trace --- SXR generation in Ne and detection in Kr}
\begin{figure}
\includegraphics[width=0.48\textwidth]{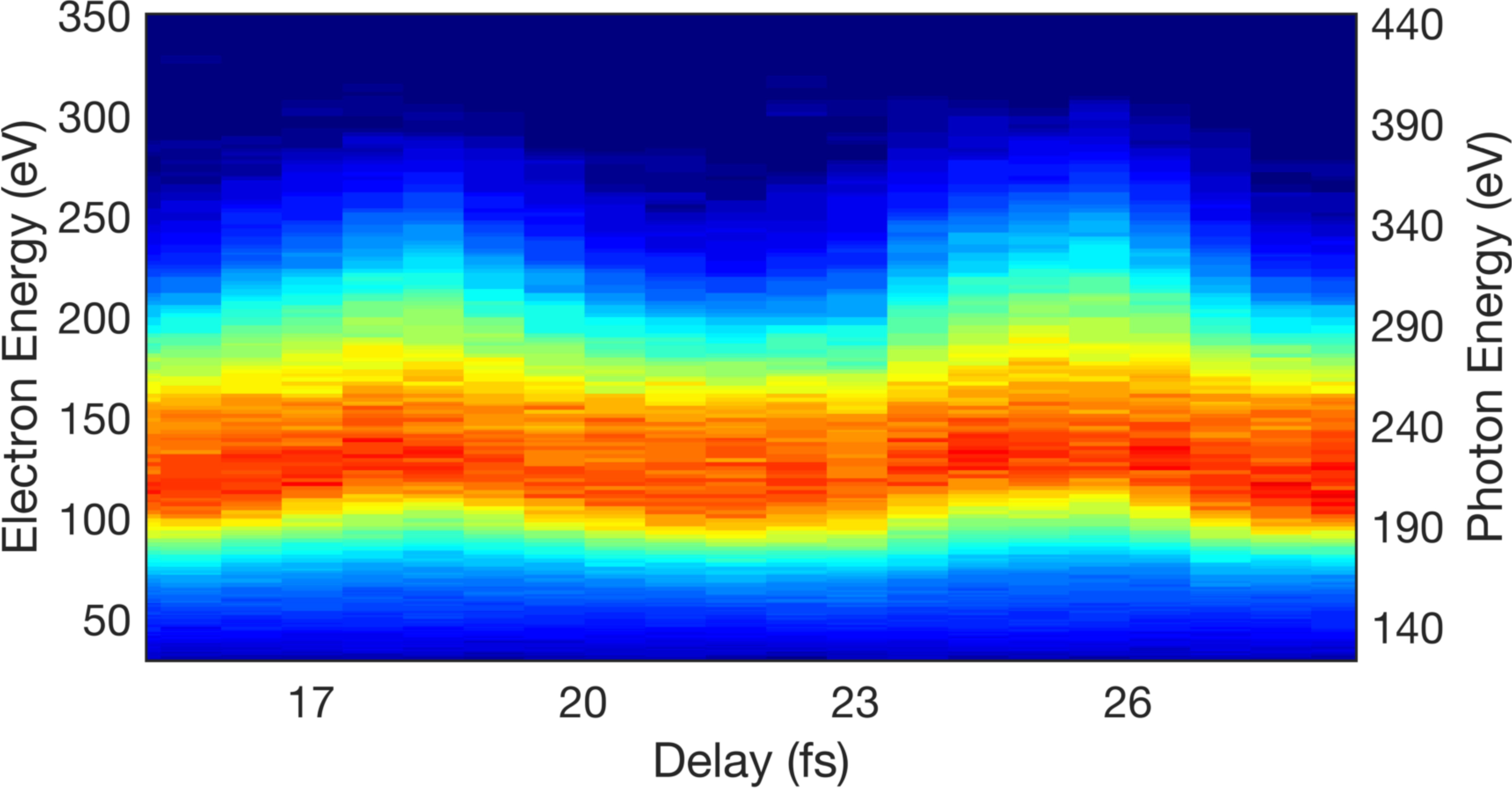}
\caption{\label{fig:fig4}Streaking trace (linear scale) from the water-window attosecond pulse recorded in Kr.}
\end{figure}

The ToF spectrometer records the streaking spectrogram as we vary the delay between the two arms of the interferometer.  The raw data recorded over a period of 10\,hours is shown in Fig. \ref{fig:fig4} in which ToF spectrometer calibration is applied to convert flight times to electron and photon energies. The electron count rate during the data acquisition was optimized to yield 300 counts/s. Specific attention can be drawn to some of the unprecedented features of this streaking spectrogram, namely i) the central electron-energy (150\,eV) translating to a central photon energy of $\sim$250\,eV (ionized from 3d shell with an ionization potential of 94\,eV), ii) the broad bandwidth ($>$100\,eV) which supports a pulse duration of 20\,as (well below the atomic unit of time).
Other features that can be inferred from this raw data before any algorithmic processing is performed, which will be discussed in detail through the rest of this article include iii) The streaking excursion albeit on the order of 50\,eV is relatively low compared to the bandwidth, iv) the spectrogram does not exhibit any clear asymmetry on the leading vs. the trailing edges of the modulations (asymmetry is visual manifestation of attochirp) implying that we have an unchirped pulse and v) there is no indication of multiple ``ghosted'' phase offset traces. 
\subsection{Multiple emissions and multiple shells --- XUV generation in Ar and detection in Kr}
\begin{figure}
\includegraphics[width=0.48\textwidth]{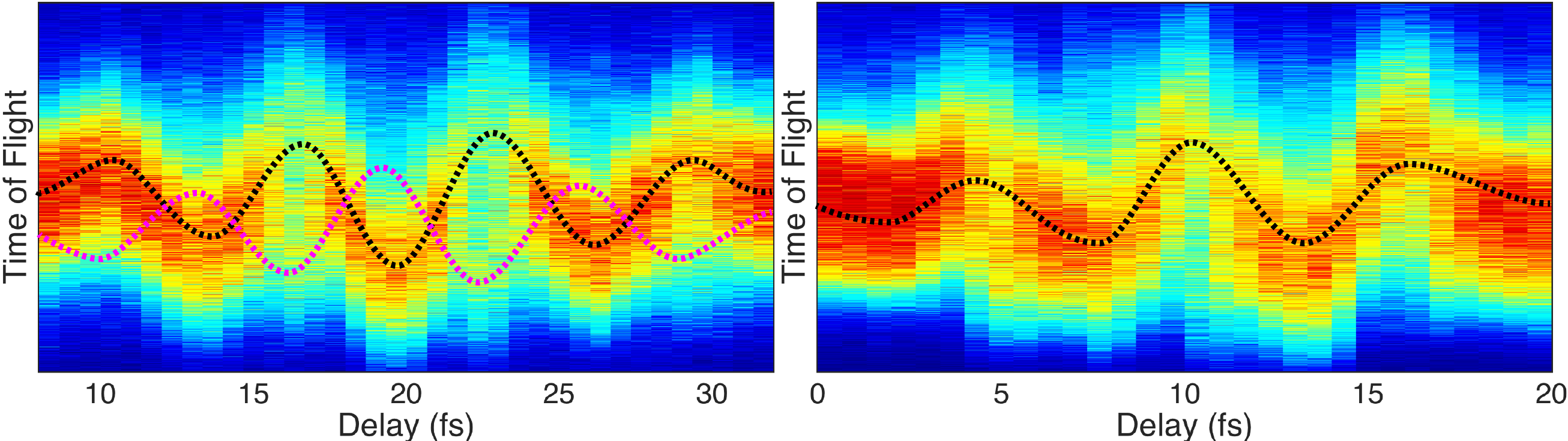}
\caption{\label{fig:fig5}Streaking traces performed using the same laser source, but a different HHG gas (argon) which offers higher flux at the expense of photon energy (centered  100\,eV). The trace on the left for a randomly selected value of CEP clearly shows more than one attosecond pulse (highlighted with the dotted black and magenta curves). By selecting a value of CEP that results in a spectral continuum and maximum cut-off, the trace on the right is recorded, where only one attosecond pulse is detected. Argon gas is used primary for alignment, as the photon yield is higher and traces are acquired quicker i.e. the full traces are acquired in 5 minutes and 20 minutes respectively}
\end{figure}

In order to investigate the possibility of emission from different shells, we switch from Ne to Ar since it will generate much lower photon energies that will access  multiple shells in the detection gas Kr. Figure~\ref{fig:fig5} shows streaking spectrograms taken with identical laser parameters but for HHG in argon (at 1\,bar), thus resulting in radiation with a 100\,eV central energy). The figure displays traces for two values of CEP, with clear evidence of multiple attosecond pulses vs. an isolated pulse. A ghosted trace that is a half cycle out of phase with the primary trace could predominantly come from two sources: 1) a 2nd attosecond burst generated by a preceding or proceeding, lower intensity, half cycle, or 2)  a $\pi$ CEP phase shift of the laser pulses during the streaking trace acquisition. Measurements of CEP fluctuations prove that it is not the latter. At this photon energy we expect multiple shells (3d (95\,eV), 4s (27.5\,eV) and 4p (14.1\,eV)) to contribute due to their comparable ionization cross-sections. If emission originated from multiple shells though, their corresponding emitted bandwidths would overlap, however we a) do not observe interference between them, b) expect the delay between emission from different shells to be on the order of a couple of tens of attoseconds \cite{Schultze2010}, so this could not explain what is seen in Fig. \ref{fig:fig5}.  Double pulse ghosting resulting from two attosecond bursts  generated by subsequent half cycles has been investigated theoretically (see Fig. 2 of \cite{Gagnon2009} and \cite{Gagnon2008}). This implies that if there are any satellite attosecond pulses, they are of significantly lower intensity and their contribution to the photoelectron spectra is negligible. In these conditions, a sufficiently isolated, water-window attosecond pulse has been generated and partially characterized.\\
\subsection{Auger processes in SXR streaking}
\begin{figure}
\includegraphics[width=0.48\textwidth]{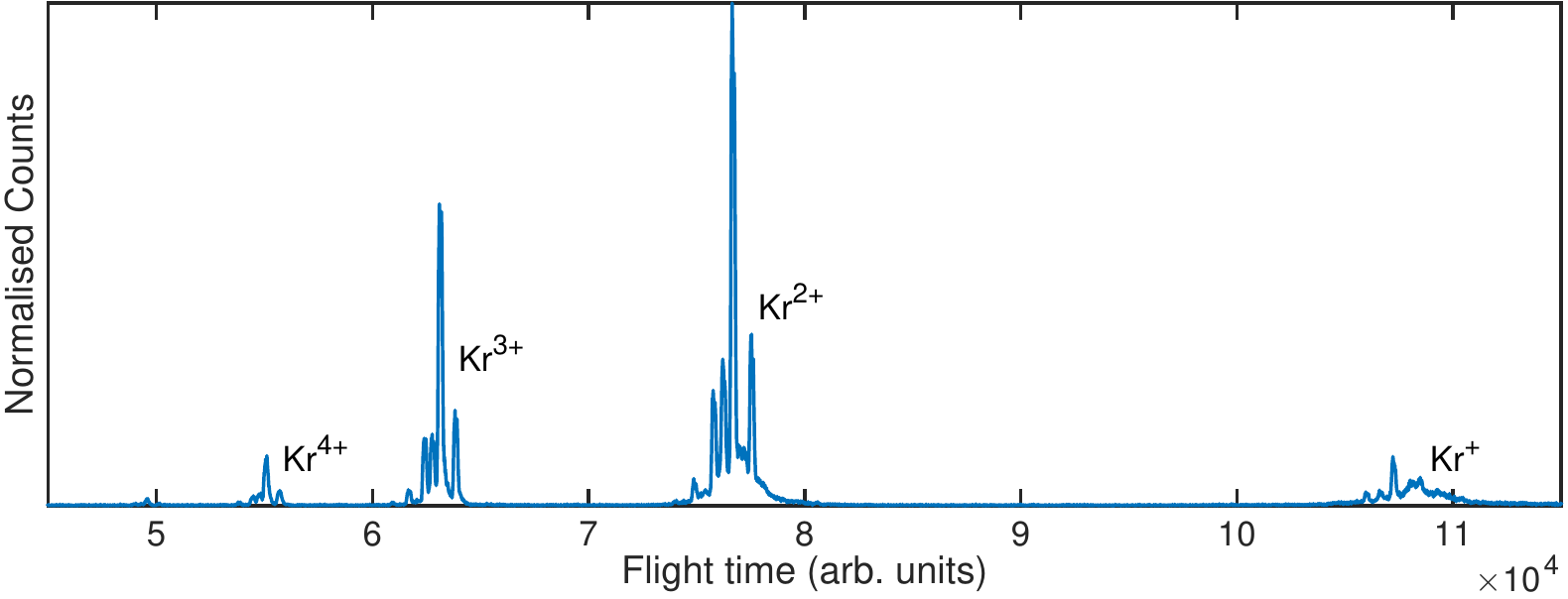}%
\caption{\label{fig:fig6}Time-of-flight spectrum measured in krypton ionized by the soft X-ray continuum. The presence of Kr$^{2+}$, Kr$^{3+}$ and Kr$^{4+}$ clearly indicates the occurrence of Auger relaxation mechanism after the initial photoionization. The isotope structure of Kr is also visible.}
\end{figure}

\begin{figure}
\includegraphics[width=0.48\textwidth]{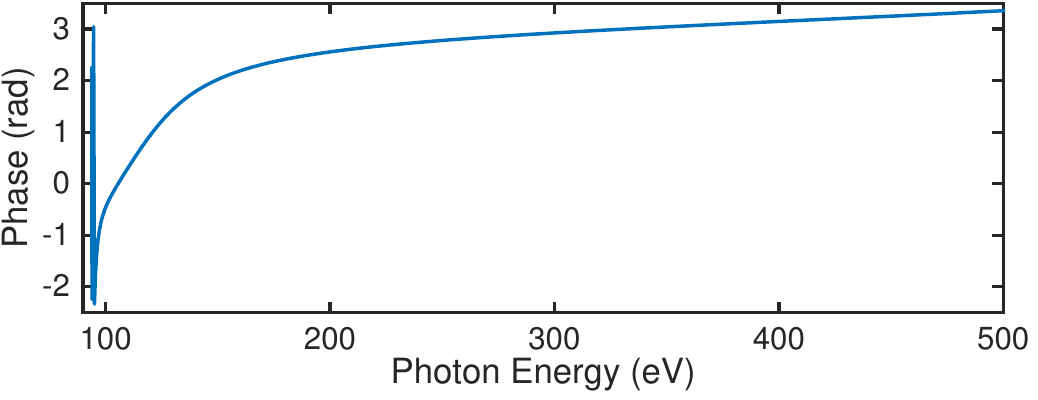}
\caption{\label{fig:fig7}Calculated dipole emission phase for Kr 3d as function of photon energy \cite{Kheifets2013}.}
\end{figure}

As described above, the energy range and bandwidth of our X-ray pulses can access core shells of krypton. From Fig.~\ref{fig:fig1}, we find that the cross section for photo ionization from the 3d shell at 300\,eV is roughly one order of magnitude larger than from the 3p shell and two orders of magnitude larger than from the 4p shell. The asymmetry parameter of the 3d and 3p shells cross and are close to 1 at 300\,eV photon energy. The asymmetry parameters diverge slightly for lower and higher photon energies within the bandwidth of the pulse. Due to this behaviour and because of the vastly different probability for ionization, we can safely assume that photoemission originates predominantly from the 3d shell and that the asymmetry of the emission presents no impediment to our measurement. Still it is critical to rule out any other sources of photoelectrons from ionization phenomena. Fig. \ref{fig:fig6} shows the ion ToF spectrum for krypton ionized by the soft X-ray continuum. The higher order ionization indicates the presence of Auger processes.  
To quantify the contributions to the electron spectra from the Auger processes that are expected from core level ionization, we operate the time of flight spectrometer in an ion detection mode and record the ion spectra.  Figure~\ref{fig:fig6} shows the time of flight spectra illustrating clearly the 2nd, 3rd and 4th ionization of the krypton gas, which is evidence of Auger relaxation processes. We perform a thorough analysis of the expected processes to ascertain whether our recorded streaking spectrogram contains contributions from them. Analysis of the partial ionization cross sections \cite{Becker1996} indicates that ejection of a 3d electron is the dominant ionization mechanism in krypton in the photon energy range between 94\,eV (threshold at 94.20\,eV) and 300\,eV. For very high photon energy, ionization from the 3p shell is energetically allowed (threshold at 214.80\,eV \cite{Lindle1986}. In the energy range between 214\,eV and 300\,eV the photoionization cross section from the 3p shell is about 5-10\,\% \cite{Ricz2003} of the one from the 3d shell, thus making this contribution negligible. The initial photoionization (from the 3d and 3p shell) step leads with high efficiency to the emission of one (or more) electron by Auger decay. This conclusion is confirmed by the ion spectra, which present multiple charged ion states Kr$^{2+}$, Kr$^{3+}$ and Kr$^{4+}$ (see Fig.~\ref{fig:fig6}). For the interpretation of the streaking data, it is fundamental to understand in which energy range the Auger lines will appear and the different branching ration between the different channels. For the sake of clarity, we will discuss separately the two cases: i) Auger decay from 3d-1 shell - Ionization from the 3d shell leads to single or cascaded Auger decay determining the formation of Kr$^{2+}$ and Kr$^{3+}$ ions, respectively. For both mechanisms, the energy of the Auger electrons is always below 60\,eV (see also Fig.3 in ref.~\cite{Verhoef2011} and it does not affect the streaking trace considering that only photoelectrons with energies larger that  90\,eV were efficiently collected. ii) Auger decay from 3p shell - Experimental data on different Auger relaxation pathways of the core hole 3p are reported in ref.~ \cite{Jauhiainen1995}. In particular, three groups of Auger lines were analyzed; their main characteristics are summarized in Tab.~\ref{tab:tab2}. These Auger decay fall out of the main energetic window of the TOF and are not efficiently collected by our electron spectrometer. The relative intensity of M$_{2,3}$-M$_{4,5}$N$_{2,3}$ this group is, according to ref.~\cite{Jauhiainen1995} about 75.4\,\%. According to ref.~\cite{Jonauskas2011}, the relative intensity of these decay channels is given by: M$_2$-M$_{4,5}$N$_{2,3}$ 50.6\,\% and M$_3$-M$_{4,5}$N$_{2,3}$ 61.4\,\%. Therefore, the Auger lines are only up to 6\,\% of the total 3d signal and can be safely neglected. Finally, the ratio of the M$_{2,3}$-NN Auger lines with respect to the direct photoionization from the 3d shell is given by: M$_{2,3}$-NN/3d = 0.66\,\% and can be neglected. The main conclusion of our analysis is that the Auger decay processes do not appreciably contribute to the photoelectron spectrum measured in our experimental conditions; the electrons measured are emitted by single photon ionization from the 3d shell.\\ 
\begin{table}
\centering
\caption{Analysis of the Auger decay following photoionization from the 3p core shell}
\label{tab:tab2}
\begin{ruledtabular}
\begin{tabular}{lcc}
 \textbf{Auger Process}            & \textbf{Threshold (eV)} & \textbf{Branching ratio} \\ \hline
\textbf{M$_{2,3}$-M$_{4,5}$N$_1$}   & 57-90                   & 21.3\,\,\%                   \\
\textbf{M$_{2,3}$-M$_{4,5}$N$_{2,3}$} & 85-105                  & 75.4\,\,\%                   \\
\textbf{M$_{2,3}$-NN}       & 150-185                 & 2.3\,\,\%                   
\end{tabular}
\end{ruledtabular}
\end{table}
It is important to note that any attosecond photo--electron--based streaking method retrieves the electron wavepacket and not the optical pulse directly and irrespective of the used retrieval method. The typical assumption is therefore the identical mapping of the optical pulse into the measured electron wavepacket. Due to the large bandwidth of the electron wavepacket, we investigated whether the typical assumption holds or if any significant time lag is expected for photoemission from the 3d shell of krypton. Figure \ref{fig:fig7} shows the calculated dipole emission phase as function of photon energy \cite{Kheifets2013}. The 3d phase is mostly featureless across the relevant photon energy range for which we extract a GDD of -10.7\,as$^2$ at 243\,eV. This negligible contribution of the dipole emission phase means that the retrieved electron wavepacket is indeed an accurate representation of the soft X-ray attosecond pulse.\\
\subsection{\label{sec:SXRFROGCRAB}Retrieval of attosecond SXR pulses}
In the following, we used the FROGCRAB~\cite{Mairesse2005} algorithm to retrieve information about the water-window pulse from our measured streaking spectrogram. We note that while different algorithms are nowadays available to improve on various aspects of FROGCRAB, we decided to use FROGCRAB as it is the most widely applied method in the attosecond field. An important approximation made in many algorithms, and also in FROGCRAB, is the central momentum approximation (CMA) which requires that the pulse's spectral bandwidth is not larger than the central photon energy of the spectrum~\cite{Calegari2013}. While there exists no hard boundary for the validity of the CMA, we note that the extreme bandwidth of our pulse approaches this limit but does not clearly violate it. We like to stress however, that similarly to FROG retrievals of optical pulses, an extreme pulse bandwidth and any measurement with low signal-to-noise, as currently only possible in the SXR regime, places stringent demands on the sampling conditions and convergence criteria.\\
It is important to note that interpolation and filtering of spectrogram traces is a common practice, but each case should be carefully considered before applying such measures. In our example, optimal spectral sampling $\delta$E, according to Tab. \ref{tab:tab3}, would require measuring with an order of magnitude better ToF resolution, but the corresponding retrieval grid would become unpractically large to handle numerically. Without such fine sampling, one may however neglect to reveal spectral interferences originating from multiple pulses or substructures of the attosecond pulse. To rule out such possibilities, we took additional photon spectra with a resolution of 0.5 eV. These spectra did not reveal any fringes or fine structures, hence supporting the measurement of an isolated water-window SXR pulse. In addition, we simulate a single-cycle streaking spectrogram with $\delta$E of 0.2\,eV, which is then interpolated to $\delta$E of 0.6\,eV. We choose 0.6\,eV instead the optimally required 0.3\,eV since reconstruction time is still tractable but already amounts to 48 hours. The interpolated trace is processed by the FROGCRAB algorithm and satisfactorily reconstructs the simulated pulse. The simulated trace is then downsized to have $\delta$E of 10\,eV to emulate a low resolution spectrometer and afterwards re-interpolated to a $\delta$E of 0.6\,eV. Moreover, we have verified numerically that the interpolation factor used still results in a reliable FROGCRAB reconstruction. This new heavily interpolated trace is then processed by the FROGCRAB algorithm and similarly reconstructs and supports a single pulse. In general, such procedure should be considered on a case by case basis since, it is a common mistake to use insufficient sampling points and strong filtering, thereby neglecting the fine details within a spectrogram. Quoting marginals and FROG errors are also only meaningful if the measurement grid is sufficiently populated with data above the noise level. Whatever the conditions may be, temporal structures should be well resolved and reflected also in the spectral domain, and visa versa.
\\
The principal component generalized projections algorithm (PCGPA) iterative loop used by the FROGCRAB algorithm relies on several numerical constraints for the dataset to be processed \cite{Gagnon2008}. One of these requires the data matrix to be squared, with the number of points on each axis $N$ being a power of 2 and satisfying the sampling criterion, which connects the resolution of the frequency and time delay axis with the relation:
\begin{equation}
\delta\tau\delta E = \frac{2\pi\hbar}{eN},
\end{equation}
where $E$ is the energy expressed in eV and $e$ is the unit electron charge. For our experimental trace, bandwidth exceeds an energy range of 200\,eV, setting a lower limit for the total energy (frequency) range where the data needs to be interpolated. For the delay axis, without knowing the pulse duration, we set the time delay resolution to be smaller than the Fourier transform limit (20\,as) so as to have sufficient points for the reconstruction. Finally, the accuracy of the reconstruction increases with the number of points $N$ but so obviously does the computational time needed by the computer to run the algorithm. Taking into account all these features, we finally chose the parameters listed in Tab. \ref{tab:tab3} to perform the FROGCRAB retrievals.
\begin{table}[]
\centering
\caption{FROGCRAB Parameters}
\label{tab:tab3}
\begin{ruledtabular}
\begin{tabular}{ll}
Parameter                             & Value  \\ \hline
$\delta$ E                            & 0.34\,eV \\
$\delta\tau$                          & 5.9\,as  \\
N                                     & 2048   \\
$\Delta\tau$ (total time delay range) & 12\,fs   \\
$\Delta$E (total energy range)       & 700\,ev 
\end{tabular}
\end{ruledtabular}
\end{table}

\begin{figure*}
\includegraphics[width=1\textwidth]{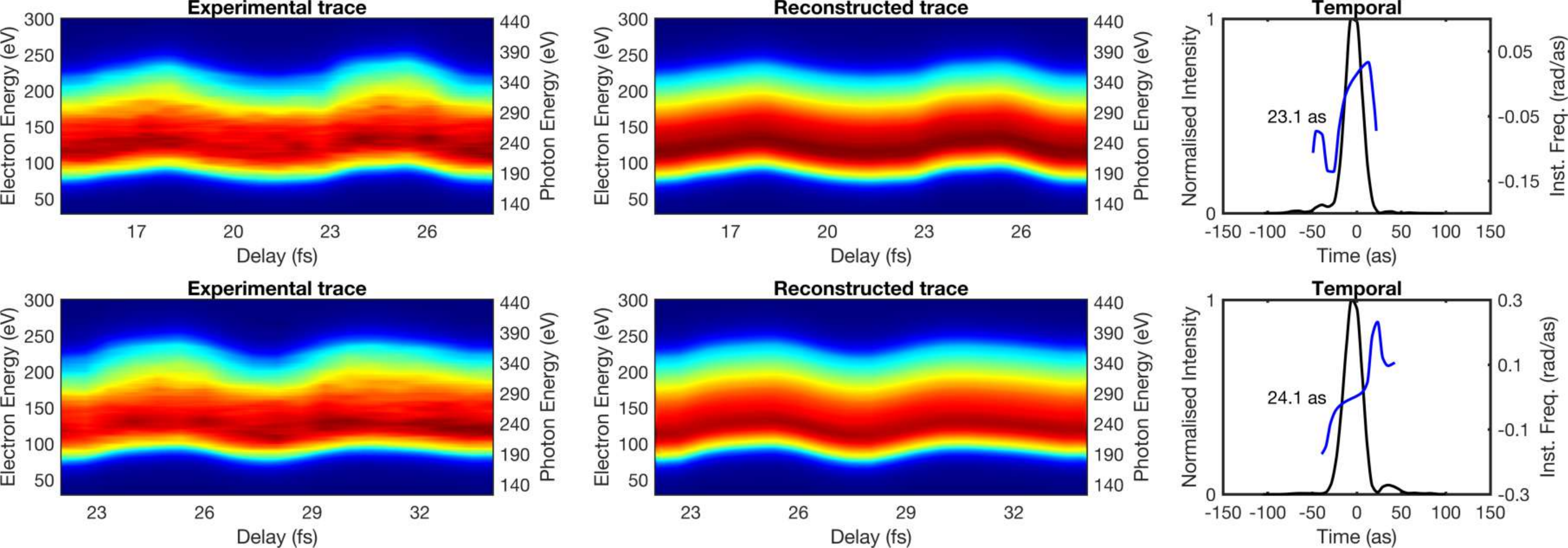}
\caption{\label{fig:fig8}FROGCRAB retrievals of the raw data shown in Fig.~\ref{fig:fig4}. Top: 2 cycles from 15fs to 27fs and bottom: 2 cycles from 22fs to 34fs. Left plots show the measured traces which have had some filtering applied and the right plots show the FROGCRAB retrieved traces.}
\end{figure*}

Based on these parameters, we apply the FROGCRAB algorithm separately to each of the two cycles shown in the streaking trace in Fig.~\ref{fig:fig4}. Within the error, both of the retrievals should give the same information about the SXR pulse. Figure~\ref{fig:fig8} shows the results in which we show the experimental trace for each cycle (top and bottom row) next to the reconstructed trace and the temporal profile with instantaneous frequency. We extract pulse durations of 23.1\,as and 24.1\,as, and lowest order phase of 152\,as$^2$ and 274\,as$^2$, respectively. 

\subsection{Contributions to chirp compensation}

As a next step we tried to investigate possible sources which would lead to chirp compensation and the generation of a SXR pulse below the atomic unit of time. In contrast to the XUV, this is a difficult undertaking in the SXR regime as the ubiquitous reference measurement with purposely added chirp by means of a metal filter is not applicable anymore in the high photon energy range above 250 eV. Moreover, relying on simulations is also very limited as we are not aware of any simulation that can describe HHG under such high pressures in combination with full 3D propagation. As a first crude measure, we hence calculated the classical attochirp as expected for a single emitter to 2550\,as$^2$. Investigating any possible contribution to phase compensation, we first turned to the remnant gas flowing from the high pressure gas target.  Simulations based on computational fluid dynamics of the gas pressure on axis, reveal a value of -165\,as$^2$. Moreover, we find that dispersion contributions from the plasma are negligible at these photon energies. Lastly, the dipole phase associated with the ionization of the krypton streaking target gas is both relatively flat as well as negligible over our bandwidth.

\subsection{Temporal gating at SXR generation conditions}

\begin{figure}
\includegraphics[width=0.49\textwidth]{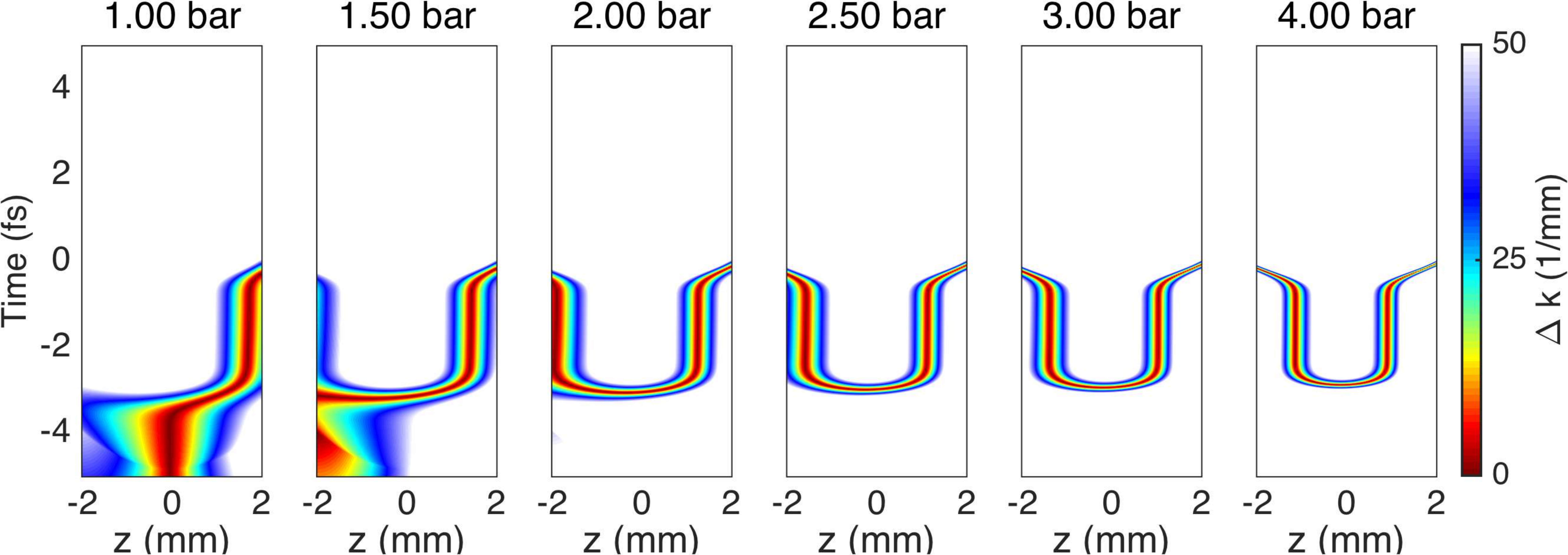}
\caption{\label{fig:fig_PM}Spatio-temporal phase matching maps as a function of HHG target pressure in Ne. Calculated on-axis phase mismatch as a function of propagation position and time within the pulse for 300\,eV radiation generated in neon for our experimental conditions and target pressures of 1 to 4 bar. The intensity in our target only has a field strength sufficient to generate 300 eV radiation between -1 to 1 mm in the propagation direction.}
\end{figure}

Despite the absence of any obvious post--generation chirp compensation, we like to mention that recent research has presented evidence that the high--pressure conditions for SXR pulse generation provide the possibility for near-instantaneous temporal gating which could lead to the emission of extremely short, isolated SXR pulses. First investigations~\cite{Chen2014} showed a transition from attosecond pulse train generation to isolated attosecond pulse generation despite using a multi--cycle pulse when increasing pressure. Our own investigation \cite{Teichmann2016}, for the much higher pressures (10 bar) used in our case in He, suggested that through the interplay of gas pressure, intensity and wavelength, a short temporal phase matching window exists. In the regime of long wavelength, high intensity laser pulses and a high gas pressure target, the temporal phase matching window can be much less than half-a-cycle of the laser pulse. Here, we perform a similar simulation of the exact experimental conditions. Our phase-matching calculations consider dispersion of both wavelengths from neutral gas and free electrons, the geometric phase (Gouy phase) of the fundamental as well as the dipole phase of the short trajectories. We do not consider long trajectories since they are effectively suppressed during propagation due to their larger divergence emission, filtered by the various pumping apertures of our beamline. Absorption of the propagating radiation is also considered. For the radiation generation, a semi-classical model of HHG is considered with the strong field approximation. A Ne atom is tunnel-ionized by a linearly polarized electric field with a peak intensity of  4.3$\times$10$^{14}$W/cm$^2$, a central wavelength of 1850\,nm and a pulse duration of 12\,fs FWHM with a Gaussian temporal profile and a Gaussian focus with a waist of 54\,$\mu$m. These parameters are used to find the time-dependent fraction of free electrons. High harmonics are generated from a single electron from the outer shell of each atom. The calculations then considers the ion in the ground state. The ionization rates are calculated using the ADK formula. We use a continuous wave approximation where the electric field amplitude varies very slowing during the tunnelling process and the dipole approximation where the driving laser field wavelength is significantly larger than the electron wavefunction. The tunnel-ionized electron is considered not to interact with the remaining ion and to have zero initial kinetic energy. It is assumed that the ion remaining after ionization does not interact with the laser field. The effect of the magnetic field of the driving laser is neglected. The results are show in Fig.~\ref{fig:fig_PM}, illustrating that for our experimental conditions, the phase mismatch ($\Delta$k) approaches zero, supporting the conditions for an extremely short isolated attosecond burst.

While these simulations, and the previous publications, are not conclusive to place a value on the expected duration of the attosecond pulse, they clearly point at the different generation conditions in the SXR regime and our measurement is the first attosecond streaking-based experimental verification of SXR gating via temporal phase matching, resulting in an isolated water--window attosecond pulse. Additional evidence for the emission of an isolated attosecond burst comes from the spectral continuum generated during HHG. The spectrometer has a resolution below 0.5\,eV, which is less than the spacing between discrete harmonics driven by 1850\,nm radiation (1.4\,eV). The lack of spectral modulations/discrete harmonics suggests an isolated attosecond pulse is generated. Without a direct possibility to infer the duration of the isolated attosecond pulse, we can however place an upper limit of 322 as on the pulse duration based on the classically estimated attochirp of 2385\,as$^2$ (intrinsic attochirp, minus the remnant gas dispersion) and the pulse bandwidth.

\subsection{\label{sec:Sims}Requirements for streaking at high photon energies}

Without limitation to the possibility of intrinsic temporal gating, we investigate additional reasons why no asymmetry in the streaking trace could be visible. Clearly, the ultra-broad bandwidth responsible for these attosecond pulses pushes the streaking technique and the FROGCRAB algorithm into a new and extreme regime. Here, we investigate theoretically the influence of two important and interlinked parameters: bandwidth and streaking field intensity. The streaking excursion is proportional to the field strength of the streaking pulse; we used a 1850\,nm pulse at a streaking field intensity of $10^{11}$\,W/cm.

We first examine the influence of the streaking field intensity on a streaking spectrogram having a fixed and broad bandwidth similar to our experimentally measured bandwidth. 
\begin{figure}
\includegraphics[width=0.48\textwidth]{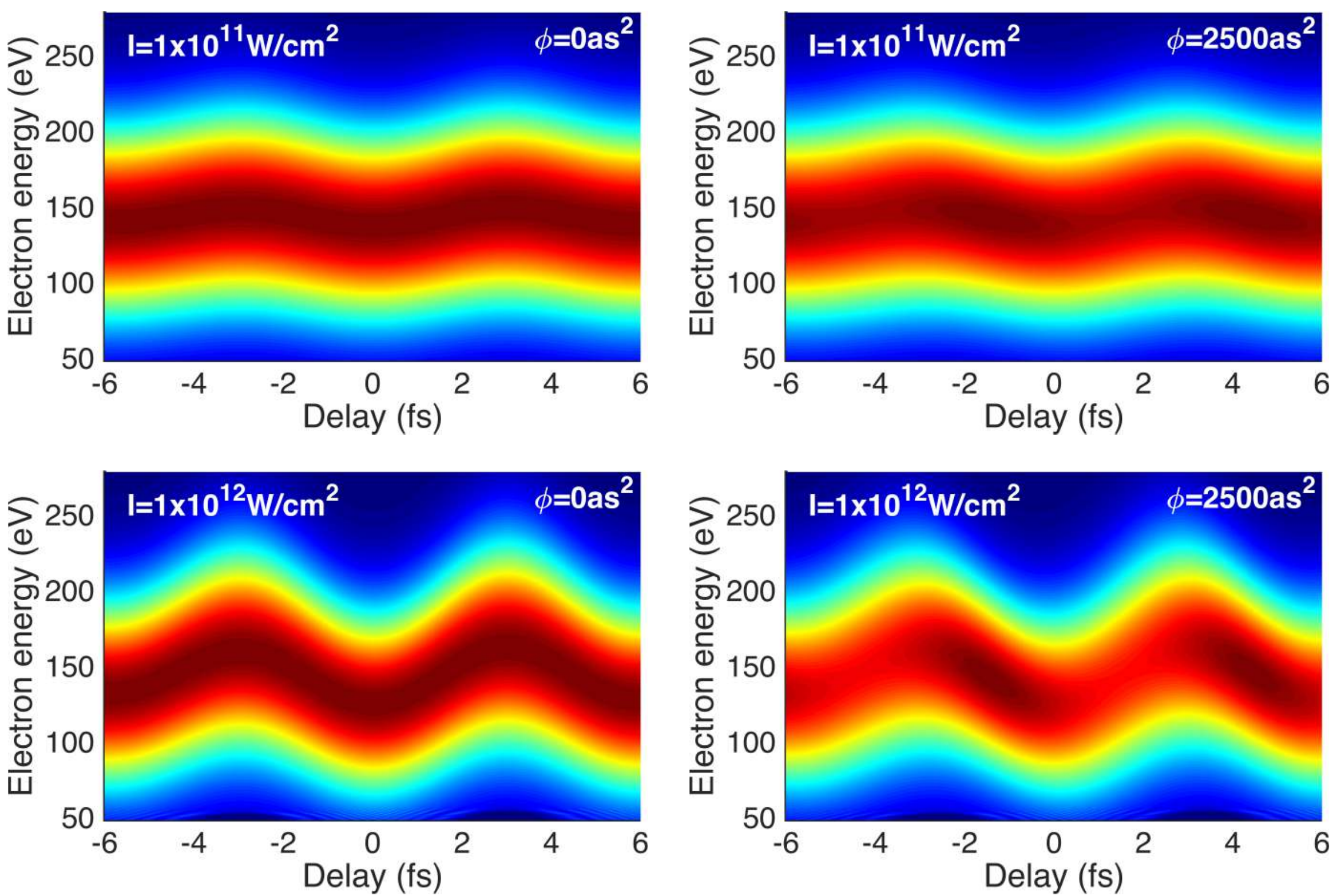}
\caption{\label{fig:fig9}Noise free streaking field intensity simulations. Top: low intensity and bottom: high intensity. Left: no attochirp simulated. Right: 2500\,as$^2$ simulated.}
\end{figure}
Figure \ref{fig:fig9} shows theoretically generated, noise-free streaking spectrograms comprised of two cycles for our bandwidth. The left plots have no chirp applied and the right plots have 2500\,as$^2$ applied.  The results show that for an intensity of 10$^{11}$\,W/cm$^2$ the streaking asymmetry (leading vs. trailing edge) is barely visible, whereas for 10$^{12}$\,W/cm$^2$ it is clearly visible. Without adding experimental noise, this may suggest that a lack of asymmetry arises from insufficient streaking intensity. However, a detrimental requirement is to avoid direct ionization from the streaking field which places an upper limit to the intensity. To investigate this dependency further, we simulate the influence of noise on generated streaking traces based on our experimental bandwidth and chirp parameters. Three simulation sets shown in Fig. \ref{fig:fig10} are generated having 1250\,as\,$^2$, 2500\,as\,$^2$ and 5000\,as\,$^2$ chirp respectively. A line-out of the photoelectron spectrum is taken for an un-streaked value. Error bars are assigned to the line-out based on Poissonian statistics. Line-outs taken at the maximum and zero crossings are then compared to the error bars. For a statistically relevant asymmetry at the zero-crossings, the yellow and blue curves need to equal/exceed the error bars. The results suggest that for a low signal-to-noise ration, even when asymmetry is fairly obvious, an attochirp of 2500\,as$^2$ is barely statistically discernible, whereas at 5000\,as$^2$ it is.\\ 
\begin{figure}
\includegraphics[width=0.48\textwidth]{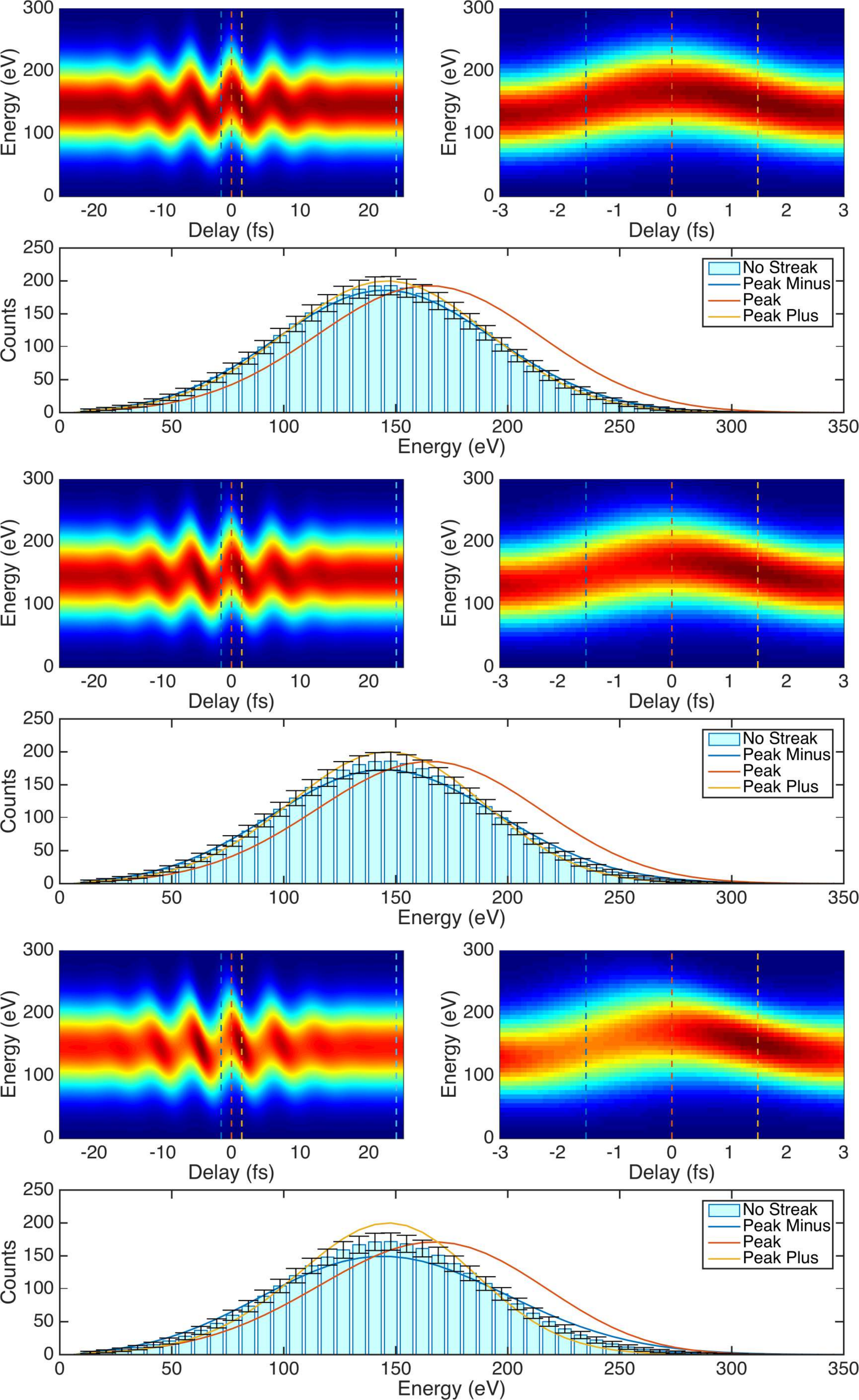}
\caption{\label{fig:fig10}Statistical noise analysis of an attosecond pulse with similar bandwidth to our experimental data. Top: 1250\,as$^2$ simulated, middle: 2500\,as$^2$ simulated and bottom: 5000\,as$^2$ simulated.}
\end{figure}
To gain further understanding of the limits due to the CMA, we investigate the capability of the FROGCRAB algorithm as well as the Least-Squares Generalized Projections Algorithm (LSGPA) \cite{Gagnon2008} to retrieve the pulse duration as a function of increasing bandwidth. The results are shown in Fig. \ref{fig:fig11} in which bandwidth is increased from 20\,eV to 100\,eV. For bandwidths up to 60\,eV, both algorithms satisfactorily reconstruct the theoretical pulses even with a moderate streaking field intensity of 10$^{11}$\,W/cm$^2$. For a 100\,eV bandwidth and streaking field intensity of 10$^{11}$\,W/cm$^2$, both reconstruction algorithms fail. This highlights the need for a sufficiently intense streaking pulse to facilitate successful reconstruction of the attosecond pulse. This is illustrated by the bottom set of plots in which a broad spectrum similar to our measured spectrum, combined with sufficient streaking field intensity (10$^{12}$\,W/cm$^2$) facilitates the successful reconstruction with both the FROGCRAB and LSGPA algorithm.\\

\begin{figure}
\includegraphics[width=0.48\textwidth]{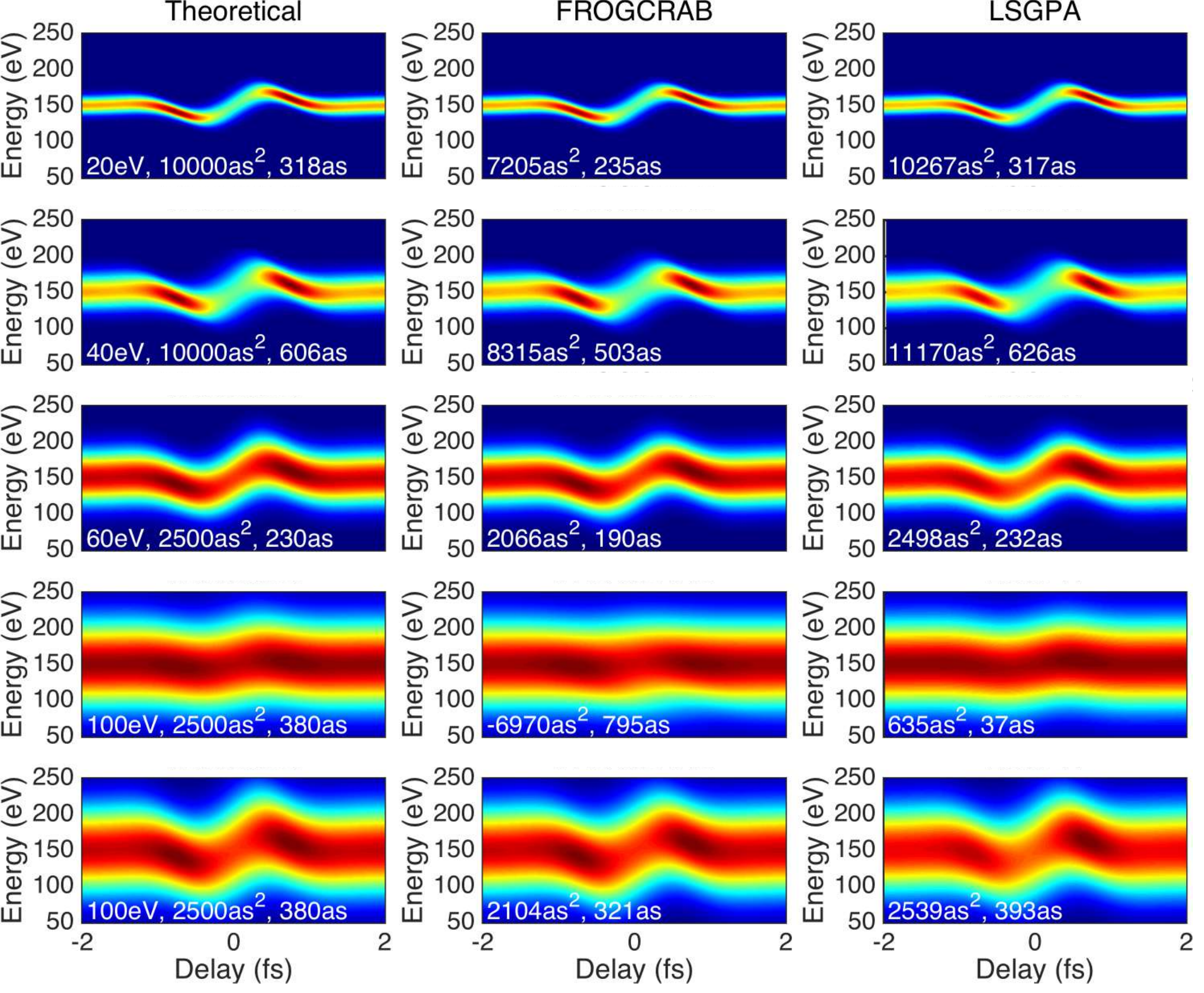}
\caption{\label{fig:fig11}Bandwidth retrievals. Left column (theoretically generated streaking spectrograms): plots are annotated at the bottom left, with the simulated bandwidth, simulated applied attochirp and corresponding simulated pulse duration. Middle (FROGCRAB algorithm) and left (LSGPA algorithm) columns are annotated with the retrieved attochirp and the retrieved pulse duration. Top four rows: 10$^{11}$\,W/cm$^2$, Bottom row: 10$^{12}$\,W/cm$^2$}
\end{figure}

We note that although we have focused on the most well known retrieval algorithm (FROGCRAB), it is important to stress that the conclusions drawn in this work pertain to the physics of the experimental data acquisition. Independent of the retrieval algorithm chosen to process the data, in this regime of extremely broad spectra and high photon energy, diligence and attention to detail is required in the choice of photo-electron source gas, streaking field intensity and signal-to-noise ratio. Ignoring any of these factors could lead to misleading pulse retrievals from all of the available algorithms.

\section{\label{sec:Summary}Summary}
We have performed the first streaking measurement of water-window photon-energy pulses. The pulses are generated via HHG in Ne and are driven by sub-2-cycle CEP stable laser pulses with a central wavelength of 1850\,nm. The streaking traces acquired do not exhibit any obvious sign of attochirp which we may attribute to either issues with the measurement process or a newly found generating condition at the unprecedentedly high pressures and ionization conditions~\cite{Teichmann2016}. Notwithstanding these limitations, we are able to confirm the generation of an isolated attosecond water--window SXR pulse and we can place an upper bound on the pulse duration of 322\,as.
Simulations indicate that in this ultra-broadband regime, the streaking excursion needs to be similar in magnitude to the bandwidth and that good signal-to-noise has to be experimentally achieved. Further simulations indicate that attosecond streaking combined with either the FROGCRAB or the LSGPA algorithms, reconstruction in an ultra-broadband regime is possible assuming the streaking excursion and signal-to-noise ratio requirements are met. Our next step will be to increase the streaking field intensity and improve the signal-to-noise ratio in order to fully characterize the water-window pulses that we are generating. 

\section{\label{sec:Acknowledgments}Acknowledgments}
Spanish Ministry of Economy and Competitiveness ``Severo Ochoa'' Programme for Centres of Excellence in R\&D (SEV-2015-0522), FIS2014-56774-R, Catalan Agencia de Gesti\'{o} d'Ajuts Universitaris i de Recerca (AGAUR) SGR 2014-2016, Fundaci\'{o} Cellex Barcelona, CERCA Programme / Generalitat de Catalunya, Laserlab-Europe (EU-H2020 654148).\\
We acknowledge fruitful discussions with Dr. Stefan Pabst, Dr. Jan Marcus Dahlstr\"{o}m, Prof. Chii-Dong Lin, Prof. Zenghu Chang and Prof. Michael Chini.\\ 
\bibliography{lib.bib}

\begin{thebibliography}{74}%
\makeatletter
\providecommand \@ifxundefined [1]{%
 \@ifx{#1\undefined}
}%
\providecommand \@ifnum [1]{%
 \ifnum #1\expandafter \@firstoftwo
 \else \expandafter \@secondoftwo
 \fi
}%
\providecommand \@ifx [1]{%
 \ifx #1\expandafter \@firstoftwo
 \else \expandafter \@secondoftwo
 \fi
}%
\providecommand \natexlab [1]{#1}%
\providecommand \enquote  [1]{``#1''}%
\providecommand \bibnamefont  [1]{#1}%
\providecommand \bibfnamefont [1]{#1}%
\providecommand \citenamefont [1]{#1}%
\providecommand \href@noop [0]{\@secondoftwo}%
\providecommand \href [0]{\begingroup \@sanitize@url \@href}%
\providecommand \@href[1]{\@@startlink{#1}\@@href}%
\providecommand \@@href[1]{\endgroup#1\@@endlink}%
\providecommand \@sanitize@url [0]{\catcode `\\12\catcode `\$12\catcode
  `\&12\catcode `\#12\catcode `\^12\catcode `\_12\catcode `\%12\relax}%
\providecommand \@@startlink[1]{}%
\providecommand \@@endlink[0]{}%
\providecommand \url  [0]{\begingroup\@sanitize@url \@url }%
\providecommand \@url [1]{\endgroup\@href {#1}{\urlprefix }}%
\providecommand \urlprefix  [0]{URL }%
\providecommand \Eprint [0]{\href }%
\providecommand \doibase [0]{http://dx.doi.org/}%
\providecommand \selectlanguage [0]{\@gobble}%
\providecommand \bibinfo  [0]{\@secondoftwo}%
\providecommand \bibfield  [0]{\@secondoftwo}%
\providecommand \translation [1]{[#1]}%
\providecommand \BibitemOpen [0]{}%
\providecommand \bibitemStop [0]{}%
\providecommand \bibitemNoStop [0]{.\EOS\space}%
\providecommand \EOS [0]{\spacefactor3000\relax}%
\providecommand \BibitemShut  [1]{\csname bibitem#1\endcsname}%
\let\auto@bib@innerbib\@empty
%</preamble>
\bibitem [{\citenamefont {Corkum}\ and\ \citenamefont
  {Krausz}(2007)}]{Corkum2007}%
  \BibitemOpen
  \bibfield  {author} {\bibinfo {author} {\bibfnamefont {P~B}\ \bibnamefont
  {Corkum}}\ and\ \bibinfo {author} {\bibfnamefont {Ferenc}\ \bibnamefont
  {Krausz}},\ }\bibfield  {title} {\enquote {\bibinfo {title} {{Attosecond
  science}},}\ }\href {http://dx.doi.org/10.1038/nphys620} {\bibfield
  {journal} {\bibinfo  {journal} {Nat Phys}\ }\textbf {\bibinfo {volume} {3}},\
  \bibinfo {pages} {381--387} (\bibinfo {year} {2007})}\BibitemShut {NoStop}%
\bibitem [{\citenamefont {Drescher}\ \emph {et~al.}(2002)\citenamefont
  {Drescher}, \citenamefont {Hentschel}, \citenamefont {Kienberger},
  \citenamefont {Uiberacker}, \citenamefont {Yakovlev}, \citenamefont
  {Scrinzi}, \citenamefont {Westerwalbesloh}, \citenamefont {Kleineberg},
  \citenamefont {Heinzmann},\ and\ \citenamefont {Krausz}}]{Drescher2002}%
  \BibitemOpen
  \bibfield  {author} {\bibinfo {author} {\bibfnamefont {M}~\bibnamefont
  {Drescher}}, \bibinfo {author} {\bibfnamefont {M}~\bibnamefont {Hentschel}},
  \bibinfo {author} {\bibfnamefont {R}~\bibnamefont {Kienberger}}, \bibinfo
  {author} {\bibfnamefont {M}~\bibnamefont {Uiberacker}}, \bibinfo {author}
  {\bibfnamefont {V}~\bibnamefont {Yakovlev}}, \bibinfo {author} {\bibfnamefont
  {a}~\bibnamefont {Scrinzi}}, \bibinfo {author} {\bibfnamefont
  {Th}~\bibnamefont {Westerwalbesloh}}, \bibinfo {author} {\bibfnamefont
  {U}~\bibnamefont {Kleineberg}}, \bibinfo {author} {\bibfnamefont
  {U}~\bibnamefont {Heinzmann}}, \ and\ \bibinfo {author} {\bibfnamefont
  {F}~\bibnamefont {Krausz}},\ }\bibfield  {title} {\enquote {\bibinfo {title}
  {{Time-resolved atomic inner-shell spectroscopy.}}}\ }\href {\doibase
  10.1038/nature01143} {\bibfield  {journal} {\bibinfo  {journal} {Nature}\
  }\textbf {\bibinfo {volume} {419}},\ \bibinfo {pages} {803--7} (\bibinfo
  {year} {2002})}\BibitemShut {NoStop}%
\bibitem [{\citenamefont {Uiberacker}\ \emph {et~al.}(2007)\citenamefont
  {Uiberacker}, \citenamefont {Uphues}, \citenamefont {Schultze}, \citenamefont
  {Verhoef}, \citenamefont {Yakovlev}, \citenamefont {Kling}, \citenamefont
  {Rauschenberger}, \citenamefont {Kabachnik}, \citenamefont {Schr{\"{o}}der},
  \citenamefont {Lezius}, \citenamefont {Kompa}, \citenamefont {Muller},
  \citenamefont {Vrakking}, \citenamefont {Hendel}, \citenamefont {Kleineberg},
  \citenamefont {Heinzmann}, \citenamefont {Drescher},\ and\ \citenamefont
  {Krausz}}]{Uiberacker2007}%
  \BibitemOpen
  \bibfield  {author} {\bibinfo {author} {\bibfnamefont {M}~\bibnamefont
  {Uiberacker}}, \bibinfo {author} {\bibfnamefont {Th}~\bibnamefont {Uphues}},
  \bibinfo {author} {\bibfnamefont {M}~\bibnamefont {Schultze}}, \bibinfo
  {author} {\bibfnamefont {a~J}\ \bibnamefont {Verhoef}}, \bibinfo {author}
  {\bibfnamefont {V}~\bibnamefont {Yakovlev}}, \bibinfo {author} {\bibfnamefont
  {M~F}\ \bibnamefont {Kling}}, \bibinfo {author} {\bibfnamefont
  {J}~\bibnamefont {Rauschenberger}}, \bibinfo {author} {\bibfnamefont {N~M}\
  \bibnamefont {Kabachnik}}, \bibinfo {author} {\bibfnamefont {H}~\bibnamefont
  {Schr{\"{o}}der}}, \bibinfo {author} {\bibfnamefont {M}~\bibnamefont
  {Lezius}}, \bibinfo {author} {\bibfnamefont {K~L}\ \bibnamefont {Kompa}},
  \bibinfo {author} {\bibfnamefont {H-G}\ \bibnamefont {Muller}}, \bibinfo
  {author} {\bibfnamefont {M~J~J}\ \bibnamefont {Vrakking}}, \bibinfo {author}
  {\bibfnamefont {S}~\bibnamefont {Hendel}}, \bibinfo {author} {\bibfnamefont
  {U}~\bibnamefont {Kleineberg}}, \bibinfo {author} {\bibfnamefont
  {U}~\bibnamefont {Heinzmann}}, \bibinfo {author} {\bibfnamefont
  {M}~\bibnamefont {Drescher}}, \ and\ \bibinfo {author} {\bibfnamefont
  {F}~\bibnamefont {Krausz}},\ }\bibfield  {title} {\enquote {\bibinfo {title}
  {{Attosecond real-time observation of electron tunnelling in atoms.}}}\
  }\href {\doibase 10.1038/nature05648} {\bibfield  {journal} {\bibinfo
  {journal} {Nature}\ }\textbf {\bibinfo {volume} {446}},\ \bibinfo {pages}
  {627--32} (\bibinfo {year} {2007})}\BibitemShut {NoStop}%
\bibitem [{\citenamefont {Schultze}\ \emph {et~al.}(2010)\citenamefont
  {Schultze}, \citenamefont {Fiess}, \citenamefont {Karpowicz}, \citenamefont
  {Gagnon}, \citenamefont {Korbman}, \citenamefont {Hofstetter}, \citenamefont
  {Neppl}, \citenamefont {Cavalieri}, \citenamefont {Komninos}, \citenamefont
  {Mercouris}, \citenamefont {Nicolaides}, \citenamefont {Pazourek},
  \citenamefont {Nagele}, \citenamefont {Feist}, \citenamefont
  {Burgd{\"{o}}rfer}, \citenamefont {Azzeer}, \citenamefont {Ernstorfer},
  \citenamefont {Kienberger}, \citenamefont {Kleineberg}, \citenamefont
  {Goulielmakis}, \citenamefont {Krausz},\ and\ \citenamefont
  {Yakovlev}}]{Schultze2010}%
  \BibitemOpen
  \bibfield  {author} {\bibinfo {author} {\bibfnamefont {M}~\bibnamefont
  {Schultze}}, \bibinfo {author} {\bibfnamefont {M}~\bibnamefont {Fiess}},
  \bibinfo {author} {\bibfnamefont {N}~\bibnamefont {Karpowicz}}, \bibinfo
  {author} {\bibfnamefont {J}~\bibnamefont {Gagnon}}, \bibinfo {author}
  {\bibfnamefont {M}~\bibnamefont {Korbman}}, \bibinfo {author} {\bibfnamefont
  {M}~\bibnamefont {Hofstetter}}, \bibinfo {author} {\bibfnamefont
  {S}~\bibnamefont {Neppl}}, \bibinfo {author} {\bibfnamefont {a~L}\
  \bibnamefont {Cavalieri}}, \bibinfo {author} {\bibfnamefont {Y}~\bibnamefont
  {Komninos}}, \bibinfo {author} {\bibfnamefont {Th}~\bibnamefont {Mercouris}},
  \bibinfo {author} {\bibfnamefont {C~a}\ \bibnamefont {Nicolaides}}, \bibinfo
  {author} {\bibfnamefont {R}~\bibnamefont {Pazourek}}, \bibinfo {author}
  {\bibfnamefont {S}~\bibnamefont {Nagele}}, \bibinfo {author} {\bibfnamefont
  {J}~\bibnamefont {Feist}}, \bibinfo {author} {\bibfnamefont {J}~\bibnamefont
  {Burgd{\"{o}}rfer}}, \bibinfo {author} {\bibfnamefont {a~M}\ \bibnamefont
  {Azzeer}}, \bibinfo {author} {\bibfnamefont {R}~\bibnamefont {Ernstorfer}},
  \bibinfo {author} {\bibfnamefont {R}~\bibnamefont {Kienberger}}, \bibinfo
  {author} {\bibfnamefont {U}~\bibnamefont {Kleineberg}}, \bibinfo {author}
  {\bibfnamefont {E}~\bibnamefont {Goulielmakis}}, \bibinfo {author}
  {\bibfnamefont {F}~\bibnamefont {Krausz}}, \ and\ \bibinfo {author}
  {\bibfnamefont {V~S}\ \bibnamefont {Yakovlev}},\ }\bibfield  {title}
  {\enquote {\bibinfo {title} {{Delay in photoemission.}}}\ }\href {\doibase
  10.1126/science.1189401} {\bibfield  {journal} {\bibinfo  {journal}
  {Science}\ }\textbf {\bibinfo {volume} {328}},\ \bibinfo {pages} {1658--1662}
  (\bibinfo {year} {2010})}\BibitemShut {NoStop}%
\bibitem [{\citenamefont {Kl{\"{u}}nder}\ \emph {et~al.}(2011)\citenamefont
  {Kl{\"{u}}nder}, \citenamefont {Dahlstr{\"{o}}m}, \citenamefont
  {Gisselbrecht}, \citenamefont {Fordell}, \citenamefont {Swoboda},
  \citenamefont {Gu{\'{e}}not}, \citenamefont {Johnsson}, \citenamefont
  {Caillat}, \citenamefont {Mauritsson}, \citenamefont {Maquet}, \citenamefont
  {Ta{\"{i}}eb},\ and\ \citenamefont {L'Huillier}}]{Klunder2011}%
  \BibitemOpen
  \bibfield  {author} {\bibinfo {author} {\bibfnamefont {K}~\bibnamefont
  {Kl{\"{u}}nder}}, \bibinfo {author} {\bibfnamefont {J~M}\ \bibnamefont
  {Dahlstr{\"{o}}m}}, \bibinfo {author} {\bibfnamefont {M}~\bibnamefont
  {Gisselbrecht}}, \bibinfo {author} {\bibfnamefont {T}~\bibnamefont
  {Fordell}}, \bibinfo {author} {\bibfnamefont {M}~\bibnamefont {Swoboda}},
  \bibinfo {author} {\bibfnamefont {D}~\bibnamefont {Gu{\'{e}}not}}, \bibinfo
  {author} {\bibfnamefont {P}~\bibnamefont {Johnsson}}, \bibinfo {author}
  {\bibfnamefont {J}~\bibnamefont {Caillat}}, \bibinfo {author} {\bibfnamefont
  {J}~\bibnamefont {Mauritsson}}, \bibinfo {author} {\bibfnamefont
  {A}~\bibnamefont {Maquet}}, \bibinfo {author} {\bibfnamefont {R}~\bibnamefont
  {Ta{\"{i}}eb}}, \ and\ \bibinfo {author} {\bibfnamefont {A}~\bibnamefont
  {L'Huillier}},\ }\bibfield  {title} {\enquote {\bibinfo {title} {{Probing
  Single-Photon Ionization on the Attosecond Time Scale}},}\ }\href
  {https://link.aps.org/doi/10.1103/PhysRevLett.106.143002} {\bibfield
  {journal} {\bibinfo  {journal} {Physical Review Letters}\ }\textbf {\bibinfo
  {volume} {106}},\ \bibinfo {pages} {143002} (\bibinfo {year}
  {2011})}\BibitemShut {NoStop}%
\bibitem [{\citenamefont {Ott}\ \emph {et~al.}(2013)\citenamefont {Ott},
  \citenamefont {Kaldun}, \citenamefont {Raith}, \citenamefont {Meyer},
  \citenamefont {Laux}, \citenamefont {Evers}, \citenamefont {Keitel},
  \citenamefont {Greene},\ and\ \citenamefont {Pfeifer}}]{Ott2013}%
  \BibitemOpen
  \bibfield  {author} {\bibinfo {author} {\bibfnamefont {Christian}\
  \bibnamefont {Ott}}, \bibinfo {author} {\bibfnamefont {Andreas}\ \bibnamefont
  {Kaldun}}, \bibinfo {author} {\bibfnamefont {Philipp}\ \bibnamefont {Raith}},
  \bibinfo {author} {\bibfnamefont {Kristina}\ \bibnamefont {Meyer}}, \bibinfo
  {author} {\bibfnamefont {Martin}\ \bibnamefont {Laux}}, \bibinfo {author}
  {\bibfnamefont {J{\"{o}}rg}\ \bibnamefont {Evers}}, \bibinfo {author}
  {\bibfnamefont {Christoph~H}\ \bibnamefont {Keitel}}, \bibinfo {author}
  {\bibfnamefont {Chris~H}\ \bibnamefont {Greene}}, \ and\ \bibinfo {author}
  {\bibfnamefont {Thomas}\ \bibnamefont {Pfeifer}},\ }\bibfield  {title}
  {\enquote {\bibinfo {title} {{Lorentz Meets Fano in Spectral Line Shapes: A
  Universal Phase and Its Laser Control}},}\ }\href
  {http://science.sciencemag.org/content/340/6133/716.abstract} {\bibfield
  {journal} {\bibinfo  {journal} {Science}\ }\textbf {\bibinfo {volume}
  {340}},\ \bibinfo {pages} {716 LP -- 720} (\bibinfo {year}
  {2013})}\BibitemShut {NoStop}%
\bibitem [{\citenamefont {Sansone}\ \emph {et~al.}(2010)\citenamefont
  {Sansone}, \citenamefont {Kelkensberg}, \citenamefont {Perez-Torres},
  \citenamefont {Morales}, \citenamefont {Kling}, \citenamefont {Siu},
  \citenamefont {Ghafur}, \citenamefont {Johnsson}, \citenamefont {Swoboda},
  \citenamefont {Benedetti}, \citenamefont {Ferrari}, \citenamefont {Lepine},
  \citenamefont {Sanz-Vicario}, \citenamefont {Zherebtsov}, \citenamefont
  {Znakovskaya}, \citenamefont {L'Huillier}, \citenamefont {Ivanov},
  \citenamefont {Nisoli}, \citenamefont {Martin},\ and\ \citenamefont
  {Vrakking}}]{Sansone2010}%
  \BibitemOpen
  \bibfield  {author} {\bibinfo {author} {\bibfnamefont {G}~\bibnamefont
  {Sansone}}, \bibinfo {author} {\bibfnamefont {F}~\bibnamefont {Kelkensberg}},
  \bibinfo {author} {\bibfnamefont {J~F}\ \bibnamefont {Perez-Torres}},
  \bibinfo {author} {\bibfnamefont {F}~\bibnamefont {Morales}}, \bibinfo
  {author} {\bibfnamefont {M~F}\ \bibnamefont {Kling}}, \bibinfo {author}
  {\bibfnamefont {W}~\bibnamefont {Siu}}, \bibinfo {author} {\bibfnamefont
  {O}~\bibnamefont {Ghafur}}, \bibinfo {author} {\bibfnamefont {P}~\bibnamefont
  {Johnsson}}, \bibinfo {author} {\bibfnamefont {M}~\bibnamefont {Swoboda}},
  \bibinfo {author} {\bibfnamefont {E}~\bibnamefont {Benedetti}}, \bibinfo
  {author} {\bibfnamefont {F}~\bibnamefont {Ferrari}}, \bibinfo {author}
  {\bibfnamefont {F}~\bibnamefont {Lepine}}, \bibinfo {author} {\bibfnamefont
  {J~L}\ \bibnamefont {Sanz-Vicario}}, \bibinfo {author} {\bibfnamefont
  {S}~\bibnamefont {Zherebtsov}}, \bibinfo {author} {\bibfnamefont
  {I}~\bibnamefont {Znakovskaya}}, \bibinfo {author} {\bibfnamefont
  {A}~\bibnamefont {L'Huillier}}, \bibinfo {author} {\bibfnamefont {M~Yu.}\
  \bibnamefont {Ivanov}}, \bibinfo {author} {\bibfnamefont {M}~\bibnamefont
  {Nisoli}}, \bibinfo {author} {\bibfnamefont {F}~\bibnamefont {Martin}}, \
  and\ \bibinfo {author} {\bibfnamefont {M~J~J}\ \bibnamefont {Vrakking}},\
  }\bibfield  {title} {\enquote {\bibinfo {title} {{Electron localization
  following attosecond molecular photoionization}},}\ }\href
  {http://dx.doi.org/10.1038/nature09084
  http://www.nature.com/nature/journal/v465/n7299/abs/nature09084.html{\#}supplementary-information}
  {\bibfield  {journal} {\bibinfo  {journal} {Nature}\ }\textbf {\bibinfo
  {volume} {465}},\ \bibinfo {pages} {763--766} (\bibinfo {year}
  {2010})}\BibitemShut {NoStop}%
\bibitem [{\citenamefont {Calegari}\ \emph {et~al.}(2014)\citenamefont
  {Calegari}, \citenamefont {Ayuso}, \citenamefont {Trabattoni}, \citenamefont
  {Belshaw}, \citenamefont {{De Camillis}}, \citenamefont {Anumula},
  \citenamefont {Frassetto}, \citenamefont {Poletto}, \citenamefont {Palacios},
  \citenamefont {Decleva}, \citenamefont {Greenwood}, \citenamefont
  {Mart{\'{i}}n},\ and\ \citenamefont {Nisoli}}]{Calegari2014}%
  \BibitemOpen
  \bibfield  {author} {\bibinfo {author} {\bibfnamefont {F}~\bibnamefont
  {Calegari}}, \bibinfo {author} {\bibfnamefont {D}~\bibnamefont {Ayuso}},
  \bibinfo {author} {\bibfnamefont {A}~\bibnamefont {Trabattoni}}, \bibinfo
  {author} {\bibfnamefont {L}~\bibnamefont {Belshaw}}, \bibinfo {author}
  {\bibfnamefont {S}~\bibnamefont {{De Camillis}}}, \bibinfo {author}
  {\bibfnamefont {S}~\bibnamefont {Anumula}}, \bibinfo {author} {\bibfnamefont
  {F}~\bibnamefont {Frassetto}}, \bibinfo {author} {\bibfnamefont
  {L}~\bibnamefont {Poletto}}, \bibinfo {author} {\bibfnamefont
  {A}~\bibnamefont {Palacios}}, \bibinfo {author} {\bibfnamefont
  {P}~\bibnamefont {Decleva}}, \bibinfo {author} {\bibfnamefont {J~B}\
  \bibnamefont {Greenwood}}, \bibinfo {author} {\bibfnamefont {F}~\bibnamefont
  {Mart{\'{i}}n}}, \ and\ \bibinfo {author} {\bibfnamefont {M}~\bibnamefont
  {Nisoli}},\ }\bibfield  {title} {\enquote {\bibinfo {title} {{Ultrafast
  electron dynamics in phenylalanine initiated by attosecond pulses.}}}\ }\href
  {\doibase 10.1126/science.1254061} {\bibfield  {journal} {\bibinfo  {journal}
  {Science}\ }\textbf {\bibinfo {volume} {346}},\ \bibinfo {pages} {336--9}
  (\bibinfo {year} {2014})}\BibitemShut {NoStop}%
\bibitem [{\citenamefont {Schiffrin}\ \emph {et~al.}(2013)\citenamefont
  {Schiffrin}, \citenamefont {Paasch-Colberg}, \citenamefont {Karpowicz},
  \citenamefont {Apalkov}, \citenamefont {Gerster}, \citenamefont {Muhlbrandt},
  \citenamefont {Korbman}, \citenamefont {Reichert}, \citenamefont {Schultze},
  \citenamefont {Holzner}, \citenamefont {Barth}, \citenamefont {Kienberger},
  \citenamefont {Ernstorfer}, \citenamefont {Yakovlev}, \citenamefont
  {Stockman},\ and\ \citenamefont {Krausz}}]{Schiffrin2013}%
  \BibitemOpen
  \bibfield  {author} {\bibinfo {author} {\bibfnamefont {Agustin}\ \bibnamefont
  {Schiffrin}}, \bibinfo {author} {\bibfnamefont {Tim}\ \bibnamefont
  {Paasch-Colberg}}, \bibinfo {author} {\bibfnamefont {Nicholas}\ \bibnamefont
  {Karpowicz}}, \bibinfo {author} {\bibfnamefont {Vadym}\ \bibnamefont
  {Apalkov}}, \bibinfo {author} {\bibfnamefont {Daniel}\ \bibnamefont
  {Gerster}}, \bibinfo {author} {\bibfnamefont {Sascha}\ \bibnamefont
  {Muhlbrandt}}, \bibinfo {author} {\bibfnamefont {Michael}\ \bibnamefont
  {Korbman}}, \bibinfo {author} {\bibfnamefont {Joachim}\ \bibnamefont
  {Reichert}}, \bibinfo {author} {\bibfnamefont {Martin}\ \bibnamefont
  {Schultze}}, \bibinfo {author} {\bibfnamefont {Simon}\ \bibnamefont
  {Holzner}}, \bibinfo {author} {\bibfnamefont {Johannes~V}\ \bibnamefont
  {Barth}}, \bibinfo {author} {\bibfnamefont {Reinhard}\ \bibnamefont
  {Kienberger}}, \bibinfo {author} {\bibfnamefont {Ralph}\ \bibnamefont
  {Ernstorfer}}, \bibinfo {author} {\bibfnamefont {Vladislav~S}\ \bibnamefont
  {Yakovlev}}, \bibinfo {author} {\bibfnamefont {Mark~I}\ \bibnamefont
  {Stockman}}, \ and\ \bibinfo {author} {\bibfnamefont {Ferenc}\ \bibnamefont
  {Krausz}},\ }\bibfield  {title} {\enquote {\bibinfo {title}
  {{Optical-field-induced current in dielectrics}},}\ }\href
  {http://dx.doi.org/10.1038/nature11567
  http://www.nature.com/nature/journal/v493/n7430/abs/nature11567.html{\#}supplementary-information}
  {\bibfield  {journal} {\bibinfo  {journal} {Nature}\ }\textbf {\bibinfo
  {volume} {493}},\ \bibinfo {pages} {70--74} (\bibinfo {year}
  {2013})}\BibitemShut {NoStop}%
\bibitem [{\citenamefont {Schultze}\ \emph {et~al.}(2014)\citenamefont
  {Schultze}, \citenamefont {Ramasesha}, \citenamefont {Pemmaraju},
  \citenamefont {Sato}, \citenamefont {Whitmore}, \citenamefont {Gandman},
  \citenamefont {Prell}, \citenamefont {Borja}, \citenamefont {Prendergast},
  \citenamefont {Yabana}, \citenamefont {Neumark},\ and\ \citenamefont
  {Leone}}]{Schultze2014}%
  \BibitemOpen
  \bibfield  {author} {\bibinfo {author} {\bibfnamefont {Martin}\ \bibnamefont
  {Schultze}}, \bibinfo {author} {\bibfnamefont {Krupa}\ \bibnamefont
  {Ramasesha}}, \bibinfo {author} {\bibfnamefont {C~D}\ \bibnamefont
  {Pemmaraju}}, \bibinfo {author} {\bibfnamefont {S~A}\ \bibnamefont {Sato}},
  \bibinfo {author} {\bibfnamefont {D}~\bibnamefont {Whitmore}}, \bibinfo
  {author} {\bibfnamefont {A}~\bibnamefont {Gandman}}, \bibinfo {author}
  {\bibfnamefont {James~S}\ \bibnamefont {Prell}}, \bibinfo {author}
  {\bibfnamefont {L~J}\ \bibnamefont {Borja}}, \bibinfo {author} {\bibfnamefont
  {D}~\bibnamefont {Prendergast}}, \bibinfo {author} {\bibfnamefont
  {K}~\bibnamefont {Yabana}}, \bibinfo {author} {\bibfnamefont {Daniel~M}\
  \bibnamefont {Neumark}}, \ and\ \bibinfo {author} {\bibfnamefont {Stephen~R}\
  \bibnamefont {Leone}},\ }\bibfield  {title} {\enquote {\bibinfo {title}
  {{Attosecond band-gap dynamics in silicon}},}\ }\href {\doibase
  10.1126/science.1260311} {\bibfield  {journal} {\bibinfo  {journal}
  {Science}\ }\textbf {\bibinfo {volume} {346}},\ \bibinfo {pages} {1348--1352}
  (\bibinfo {year} {2014})}\BibitemShut {NoStop}%
\bibitem [{\citenamefont {Hentschel}\ \emph {et~al.}(2001)\citenamefont
  {Hentschel}, \citenamefont {Kienberger}, \citenamefont {Spielmann},
  \citenamefont {Reider}, \citenamefont {Milosevic}, \citenamefont {Brabec},
  \citenamefont {Corkum}, \citenamefont {Heinzmann}, \citenamefont {Drescher},\
  and\ \citenamefont {Krausz}}]{Hentschel2001}%
  \BibitemOpen
  \bibfield  {author} {\bibinfo {author} {\bibfnamefont {M}~\bibnamefont
  {Hentschel}}, \bibinfo {author} {\bibfnamefont {R}~\bibnamefont
  {Kienberger}}, \bibinfo {author} {\bibfnamefont {C}~\bibnamefont
  {Spielmann}}, \bibinfo {author} {\bibfnamefont {G~A}\ \bibnamefont {Reider}},
  \bibinfo {author} {\bibfnamefont {N}~\bibnamefont {Milosevic}}, \bibinfo
  {author} {\bibfnamefont {T}~\bibnamefont {Brabec}}, \bibinfo {author}
  {\bibfnamefont {P.~B}\ \bibnamefont {Corkum}}, \bibinfo {author}
  {\bibfnamefont {U}~\bibnamefont {Heinzmann}}, \bibinfo {author}
  {\bibfnamefont {M}~\bibnamefont {Drescher}}, \ and\ \bibinfo {author}
  {\bibfnamefont {F}~\bibnamefont {Krausz}},\ }\bibfield  {title} {\enquote
  {\bibinfo {title} {{Attosecond metrology.}}}\ }\href {\doibase
  10.1038/35107000} {\bibfield  {journal} {\bibinfo  {journal} {Nature}\
  }\textbf {\bibinfo {volume} {414}},\ \bibinfo {pages} {509--13} (\bibinfo
  {year} {2001})}\BibitemShut {NoStop}%
\bibitem [{\citenamefont {Sansone}\ \emph {et~al.}(2006)\citenamefont
  {Sansone}, \citenamefont {Benedetti}, \citenamefont {Calegari}, \citenamefont
  {Vozzi}, \citenamefont {Avaldi}, \citenamefont {Flammini}, \citenamefont
  {Poletto}, \citenamefont {Villoresi}, \citenamefont {Altucci}, \citenamefont
  {Velotta}, \citenamefont {Stagira}, \citenamefont {{De Silvestri}},\ and\
  \citenamefont {Nisoli}}]{Sansone2006}%
  \BibitemOpen
  \bibfield  {author} {\bibinfo {author} {\bibfnamefont {Giuseppe}\
  \bibnamefont {Sansone}}, \bibinfo {author} {\bibfnamefont {E}~\bibnamefont
  {Benedetti}}, \bibinfo {author} {\bibfnamefont {F}~\bibnamefont {Calegari}},
  \bibinfo {author} {\bibfnamefont {C}~\bibnamefont {Vozzi}}, \bibinfo {author}
  {\bibfnamefont {L}~\bibnamefont {Avaldi}}, \bibinfo {author} {\bibfnamefont
  {R}~\bibnamefont {Flammini}}, \bibinfo {author} {\bibfnamefont
  {L}~\bibnamefont {Poletto}}, \bibinfo {author} {\bibfnamefont
  {P}~\bibnamefont {Villoresi}}, \bibinfo {author} {\bibfnamefont
  {C}~\bibnamefont {Altucci}}, \bibinfo {author} {\bibfnamefont
  {R}~\bibnamefont {Velotta}}, \bibinfo {author} {\bibfnamefont
  {S}~\bibnamefont {Stagira}}, \bibinfo {author} {\bibfnamefont
  {S}~\bibnamefont {{De Silvestri}}}, \ and\ \bibinfo {author} {\bibfnamefont
  {M}~\bibnamefont {Nisoli}},\ }\bibfield  {title} {\enquote {\bibinfo {title}
  {{Isolated Single-Cycle Attosecond Pulses}},}\ }\href {\doibase
  10.1126/science.1132838} {\bibfield  {journal} {\bibinfo  {journal}
  {Science}\ }\textbf {\bibinfo {volume} {314}},\ \bibinfo {pages} {443--446}
  (\bibinfo {year} {2006})}\BibitemShut {NoStop}%
\bibitem [{\citenamefont {Goulielmakis}\ \emph {et~al.}(2008)\citenamefont
  {Goulielmakis}, \citenamefont {Schultze}, \citenamefont {Hofstetter},
  \citenamefont {Yakovlev}, \citenamefont {Gagnon}, \citenamefont {Uiberacker},
  \citenamefont {Aquila}, \citenamefont {Gullikson}, \citenamefont {Attwood},
  \citenamefont {Kienberger}, \citenamefont {Krausz},\ and\ \citenamefont
  {Kleineberg}}]{Goulielmakis2008}%
  \BibitemOpen
  \bibfield  {author} {\bibinfo {author} {\bibfnamefont {E}~\bibnamefont
  {Goulielmakis}}, \bibinfo {author} {\bibfnamefont {M}~\bibnamefont
  {Schultze}}, \bibinfo {author} {\bibfnamefont {M}~\bibnamefont {Hofstetter}},
  \bibinfo {author} {\bibfnamefont {V~S}\ \bibnamefont {Yakovlev}}, \bibinfo
  {author} {\bibfnamefont {J}~\bibnamefont {Gagnon}}, \bibinfo {author}
  {\bibfnamefont {M}~\bibnamefont {Uiberacker}}, \bibinfo {author}
  {\bibfnamefont {A~L}\ \bibnamefont {Aquila}}, \bibinfo {author}
  {\bibfnamefont {E~M}\ \bibnamefont {Gullikson}}, \bibinfo {author}
  {\bibfnamefont {D~T}\ \bibnamefont {Attwood}}, \bibinfo {author}
  {\bibfnamefont {R}~\bibnamefont {Kienberger}}, \bibinfo {author}
  {\bibfnamefont {F}~\bibnamefont {Krausz}}, \ and\ \bibinfo {author}
  {\bibfnamefont {U}~\bibnamefont {Kleineberg}},\ }\bibfield  {title} {\enquote
  {\bibinfo {title} {{Single-Cycle Nonlinear Optics}},}\ }\href {\doibase
  10.1126/science.1157846} {\bibfield  {journal} {\bibinfo  {journal}
  {Science}\ }\textbf {\bibinfo {volume} {320}},\ \bibinfo {pages} {1614--1617}
  (\bibinfo {year} {2008})}\BibitemShut {NoStop}%
\bibitem [{\citenamefont {Ferrari}\ \emph {et~al.}(2010)\citenamefont
  {Ferrari}, \citenamefont {Calegari}, \citenamefont {Lucchini}, \citenamefont
  {Vozzi}, \citenamefont {Stagira}, \citenamefont {Sansone},\ and\
  \citenamefont {Nisoli}}]{Ferrari2010}%
  \BibitemOpen
  \bibfield  {author} {\bibinfo {author} {\bibfnamefont {F.}~\bibnamefont
  {Ferrari}}, \bibinfo {author} {\bibfnamefont {F.}~\bibnamefont {Calegari}},
  \bibinfo {author} {\bibfnamefont {M.}~\bibnamefont {Lucchini}}, \bibinfo
  {author} {\bibfnamefont {C.}~\bibnamefont {Vozzi}}, \bibinfo {author}
  {\bibfnamefont {S.}~\bibnamefont {Stagira}}, \bibinfo {author} {\bibfnamefont
  {Giuseppe}\ \bibnamefont {Sansone}}, \ and\ \bibinfo {author} {\bibfnamefont
  {M.}~\bibnamefont {Nisoli}},\ }\bibfield  {title} {\enquote {\bibinfo {title}
  {{High-energy isolated attosecond pulses generated by above-saturation
  few-cycle fields}},}\ }\href {\doibase 10.1038/nphoton.2010.250} {\bibfield
  {journal} {\bibinfo  {journal} {Nature Photonics}\ }\textbf {\bibinfo
  {volume} {4}},\ \bibinfo {pages} {875--879} (\bibinfo {year}
  {2010})}\BibitemShut {NoStop}%
\bibitem [{\citenamefont {Zhao}\ \emph {et~al.}(2012)\citenamefont {Zhao},
  \citenamefont {Zhang}, \citenamefont {Chini}, \citenamefont {Wu},
  \citenamefont {Wang},\ and\ \citenamefont {Chang}}]{Zhao2012}%
  \BibitemOpen
  \bibfield  {author} {\bibinfo {author} {\bibfnamefont {Kun}\ \bibnamefont
  {Zhao}}, \bibinfo {author} {\bibfnamefont {Qi}~\bibnamefont {Zhang}},
  \bibinfo {author} {\bibfnamefont {Michael}\ \bibnamefont {Chini}}, \bibinfo
  {author} {\bibfnamefont {Yi}~\bibnamefont {Wu}}, \bibinfo {author}
  {\bibfnamefont {Xiaowei}\ \bibnamefont {Wang}}, \ and\ \bibinfo {author}
  {\bibfnamefont {Zenghu}\ \bibnamefont {Chang}},\ }\bibfield  {title}
  {\enquote {\bibinfo {title} {{Tailoring a 67 attosecond pulse through
  advantageous}},}\ }\href@noop {} {\bibfield  {journal} {\bibinfo  {journal}
  {Optics Letters}\ }\textbf {\bibinfo {volume} {37}},\ \bibinfo {pages}
  {3891--3893} (\bibinfo {year} {2012})}\BibitemShut {NoStop}%
\bibitem [{\citenamefont {Fie{\ss}}\ \emph {et~al.}(2010)\citenamefont
  {Fie{\ss}}, \citenamefont {Schultze}, \citenamefont {Goulielmakis},
  \citenamefont {Dennhardt}, \citenamefont {Gagnon}, \citenamefont
  {Hofstetter}, \citenamefont {Kienberger},\ and\ \citenamefont
  {Krausz}}]{Fieß2010}%
  \BibitemOpen
  \bibfield  {author} {\bibinfo {author} {\bibfnamefont {M.}~\bibnamefont
  {Fie{\ss}}}, \bibinfo {author} {\bibfnamefont {M.}~\bibnamefont {Schultze}},
  \bibinfo {author} {\bibfnamefont {E.}~\bibnamefont {Goulielmakis}}, \bibinfo
  {author} {\bibfnamefont {B.}~\bibnamefont {Dennhardt}}, \bibinfo {author}
  {\bibfnamefont {J.}~\bibnamefont {Gagnon}}, \bibinfo {author} {\bibfnamefont
  {M.}~\bibnamefont {Hofstetter}}, \bibinfo {author} {\bibfnamefont
  {R.}~\bibnamefont {Kienberger}}, \ and\ \bibinfo {author} {\bibfnamefont
  {F.}~\bibnamefont {Krausz}},\ }\bibfield  {title} {\enquote {\bibinfo {title}
  {{Versatile apparatus for attosecond metrology and spectroscopy}},}\ }\href
  {\doibase 10.1063/1.3475689} {\bibfield  {journal} {\bibinfo  {journal}
  {Review of Scientific Instruments}\ }\textbf {\bibinfo {volume} {81}},\
  \bibinfo {pages} {93103} (\bibinfo {year} {2010})}\BibitemShut {NoStop}%
\bibitem [{\citenamefont {Svensson}\ \emph {et~al.}(1988)\citenamefont
  {Svensson}, \citenamefont {Eriksson}, \citenamefont {M{\aa}rtensson},
  \citenamefont {Wendin},\ and\ \citenamefont {Gelius}}]{Svensson1988}%
  \BibitemOpen
  \bibfield  {author} {\bibinfo {author} {\bibfnamefont {Svante}\ \bibnamefont
  {Svensson}}, \bibinfo {author} {\bibfnamefont {Bengt}\ \bibnamefont
  {Eriksson}}, \bibinfo {author} {\bibfnamefont {Nils}\ \bibnamefont
  {M{\aa}rtensson}}, \bibinfo {author} {\bibfnamefont {G{\"{o}}ran}\
  \bibnamefont {Wendin}}, \ and\ \bibinfo {author} {\bibfnamefont {Ulrik}\
  \bibnamefont {Gelius}},\ }\bibfield  {title} {\enquote {\bibinfo {title}
  {{Electron shake-up and correlation satellites and continuum shake-off
  distributions in X-Ray photoelectron spectra of the rare gas atoms}},}\
  }\href {\doibase 10.1016/0368-2048(88)85020-5} {\bibfield  {journal}
  {\bibinfo  {journal} {Journal of Electron Spectroscopy and Related
  Phenomena}\ }\textbf {\bibinfo {volume} {47}},\ \bibinfo {pages} {327--384}
  (\bibinfo {year} {1988})}\BibitemShut {NoStop}%
\bibitem [{\citenamefont {Sisourat}\ \emph {et~al.}(2010)\citenamefont
  {Sisourat}, \citenamefont {Kryzhevoi}, \citenamefont {Kolorenc},
  \citenamefont {Scheit}, \citenamefont {Jahnke},\ and\ \citenamefont
  {Cederbaum}}]{Sisourat2010}%
  \BibitemOpen
  \bibfield  {author} {\bibinfo {author} {\bibfnamefont {Nicolas}\ \bibnamefont
  {Sisourat}}, \bibinfo {author} {\bibfnamefont {Nikolai~V}\ \bibnamefont
  {Kryzhevoi}}, \bibinfo {author} {\bibfnamefont {Premysl}\ \bibnamefont
  {Kolorenc}}, \bibinfo {author} {\bibfnamefont {Simona}\ \bibnamefont
  {Scheit}}, \bibinfo {author} {\bibfnamefont {Till}\ \bibnamefont {Jahnke}}, \
  and\ \bibinfo {author} {\bibfnamefont {Lorenz~S}\ \bibnamefont {Cederbaum}},\
  }\bibfield  {title} {\enquote {\bibinfo {title} {{Ultralong-range energy
  transfer by interatomic Coulombic decay in an extreme quantum system}},}\
  }\href {http://dx.doi.org/10.1038/nphys1685} {\bibfield  {journal} {\bibinfo
  {journal} {Nat Phys}\ }\textbf {\bibinfo {volume} {6}},\ \bibinfo {pages}
  {508--511} (\bibinfo {year} {2010})}\BibitemShut {NoStop}%
\bibitem [{\citenamefont {Tseng}\ \emph {et~al.}(2010)\citenamefont {Tseng},
  \citenamefont {Urban}, \citenamefont {Wang}, \citenamefont {Otero},
  \citenamefont {Tait}, \citenamefont {Alcam{\'{i}}}, \citenamefont
  {{\'{E}}cija}, \citenamefont {Trelka}, \citenamefont {Gallego}, \citenamefont
  {Lin}, \citenamefont {Konuma}, \citenamefont {Starke}, \citenamefont
  {Nefedov}, \citenamefont {Langner}, \citenamefont {W{\"{o}}ll}, \citenamefont
  {Herranz}, \citenamefont {Mart{\'{i}}n}, \citenamefont {Mart{\'{i}}n},
  \citenamefont {Kern},\ and\ \citenamefont {Miranda}}]{Tseng2010}%
  \BibitemOpen
  \bibfield  {author} {\bibinfo {author} {\bibfnamefont {Tzu-Chun}\
  \bibnamefont {Tseng}}, \bibinfo {author} {\bibfnamefont {Christian}\
  \bibnamefont {Urban}}, \bibinfo {author} {\bibfnamefont {Yang}\ \bibnamefont
  {Wang}}, \bibinfo {author} {\bibfnamefont {Roberto}\ \bibnamefont {Otero}},
  \bibinfo {author} {\bibfnamefont {Steven~L}\ \bibnamefont {Tait}}, \bibinfo
  {author} {\bibfnamefont {Manuel}\ \bibnamefont {Alcam{\'{i}}}}, \bibinfo
  {author} {\bibfnamefont {David}\ \bibnamefont {{\'{E}}cija}}, \bibinfo
  {author} {\bibfnamefont {Marta}\ \bibnamefont {Trelka}}, \bibinfo {author}
  {\bibfnamefont {Jos{\'{e}}~Mar{\'{i}}a}\ \bibnamefont {Gallego}}, \bibinfo
  {author} {\bibfnamefont {Nian}\ \bibnamefont {Lin}}, \bibinfo {author}
  {\bibfnamefont {Mitsuharu}\ \bibnamefont {Konuma}}, \bibinfo {author}
  {\bibfnamefont {Ulrich}\ \bibnamefont {Starke}}, \bibinfo {author}
  {\bibfnamefont {Alexei}\ \bibnamefont {Nefedov}}, \bibinfo {author}
  {\bibfnamefont {Alexander}\ \bibnamefont {Langner}}, \bibinfo {author}
  {\bibfnamefont {Christof}\ \bibnamefont {W{\"{o}}ll}}, \bibinfo {author}
  {\bibfnamefont {Mar{\'{i}}a~{\'{A}}ngeles}\ \bibnamefont {Herranz}}, \bibinfo
  {author} {\bibfnamefont {Fernando}\ \bibnamefont {Mart{\'{i}}n}}, \bibinfo
  {author} {\bibfnamefont {Nazario}\ \bibnamefont {Mart{\'{i}}n}}, \bibinfo
  {author} {\bibfnamefont {Klaus}\ \bibnamefont {Kern}}, \ and\ \bibinfo
  {author} {\bibfnamefont {Rodolfo}\ \bibnamefont {Miranda}},\ }\bibfield
  {title} {\enquote {\bibinfo {title} {{Charge-transfer-induced structural
  rearrangements at both sides of organic/metal interfaces}},}\ }\href
  {http://dx.doi.org/10.1038/nchem.591
  http://www.nature.com/nchem/journal/v2/n5/abs/nchem.591.html{\#}supplementary-information}
  {\bibfield  {journal} {\bibinfo  {journal} {Nat Chem}\ }\textbf {\bibinfo
  {volume} {2}},\ \bibinfo {pages} {374--379} (\bibinfo {year}
  {2010})}\BibitemShut {NoStop}%
\bibitem [{\citenamefont {Remacle}\ and\ \citenamefont
  {Levine}(2006)}]{Remacle2006}%
  \BibitemOpen
  \bibfield  {author} {\bibinfo {author} {\bibfnamefont {F}~\bibnamefont
  {Remacle}}\ and\ \bibinfo {author} {\bibfnamefont {R~D}\ \bibnamefont
  {Levine}},\ }\bibfield  {title} {\enquote {\bibinfo {title} {{An electronic
  time scale in chemistry.}}}\ }\href {\doibase 10.1073/pnas.0601855103}
  {\bibfield  {journal} {\bibinfo  {journal} {Proceedings of the National
  Academy of Sciences of the United States of America}\ }\textbf {\bibinfo
  {volume} {103}},\ \bibinfo {pages} {6793--8} (\bibinfo {year}
  {2006})}\BibitemShut {NoStop}%
\bibitem [{\citenamefont {Griffini}\ \emph {et~al.}(2013)\citenamefont
  {Griffini}, \citenamefont {Brambilla}, \citenamefont {Levi}, \citenamefont
  {{Del Zoppo}},\ and\ \citenamefont {Turri}}]{Griffini2013}%
  \BibitemOpen
  \bibfield  {author} {\bibinfo {author} {\bibfnamefont {Gianmarco}\
  \bibnamefont {Griffini}}, \bibinfo {author} {\bibfnamefont {Luigi}\
  \bibnamefont {Brambilla}}, \bibinfo {author} {\bibfnamefont {Marinella}\
  \bibnamefont {Levi}}, \bibinfo {author} {\bibfnamefont {Mirella}\
  \bibnamefont {{Del Zoppo}}}, \ and\ \bibinfo {author} {\bibfnamefont
  {Stefano}\ \bibnamefont {Turri}},\ }\bibfield  {title} {\enquote {\bibinfo
  {title} {{Photo-degradation of a perylene-based organic luminescent solar
  concentrator: Molecular aspects and device implications}},}\ }\href {\doibase
  10.1016/j.solmat.2012.12.021} {\bibfield  {journal} {\bibinfo  {journal}
  {Solar Energy Materials and Solar Cells}\ }\textbf {\bibinfo {volume}
  {111}},\ \bibinfo {pages} {41--48} (\bibinfo {year} {2013})}\BibitemShut
  {NoStop}%
\bibitem [{\citenamefont {Ross}(1985)}]{Ross1985}%
  \BibitemOpen
  \bibfield  {author} {\bibinfo {author} {\bibfnamefont {M}~\bibnamefont
  {Ross}},\ }\bibfield  {title} {\enquote {\bibinfo {title} {{Matter under
  extreme conditions of temperature and pressure}},}\ }\href
  {http://stacks.iop.org/0034-4885/48/i=1/a=001} {\bibfield  {journal}
  {\bibinfo  {journal} {Reports on Progress in Physics}\ }\textbf {\bibinfo
  {volume} {48}},\ \bibinfo {pages} {1} (\bibinfo {year} {1985})}\BibitemShut
  {NoStop}%
\bibitem [{\citenamefont {St{\"{a}}hler}\ \emph {et~al.}(2010)\citenamefont
  {St{\"{a}}hler}, \citenamefont {Bovensiepen},\ and\ \citenamefont
  {Wolf}}]{Stahler2010}%
  \BibitemOpen
  \bibfield  {author} {\bibinfo {author} {\bibfnamefont {Julia}\ \bibnamefont
  {St{\"{a}}hler}}, \bibinfo {author} {\bibfnamefont {Uwe}\ \bibnamefont
  {Bovensiepen}}, \ and\ \bibinfo {author} {\bibfnamefont {Martin}\
  \bibnamefont {Wolf}},\ }\bibfield  {title} {\enquote {\bibinfo {title}
  {{Electron dynamics at polar molecule–metal interfaces: Competition between
  localization, solvation, and transfer}},}\ }\href@noop {} {\bibfield
  {journal} {\bibinfo  {journal} {Dynamics at Solid State Surfaces and
  Interfaces: Current Developments, Volume 1}\ ,\ \bibinfo {pages} {359--379}}
  (\bibinfo {year} {2010})}\BibitemShut {NoStop}%
\bibitem [{\citenamefont {Shanmugam}\ \emph {et~al.}(2013)\citenamefont
  {Shanmugam}, \citenamefont {Jacobs-Gedrim}, \citenamefont {Durcan},\ and\
  \citenamefont {Yu}}]{Shanmugam2013}%
  \BibitemOpen
  \bibfield  {author} {\bibinfo {author} {\bibfnamefont {Mariyappan}\
  \bibnamefont {Shanmugam}}, \bibinfo {author} {\bibfnamefont {Robin}\
  \bibnamefont {Jacobs-Gedrim}}, \bibinfo {author} {\bibfnamefont {Chris}\
  \bibnamefont {Durcan}}, \ and\ \bibinfo {author} {\bibfnamefont {Bin}\
  \bibnamefont {Yu}},\ }\bibfield  {title} {\enquote {\bibinfo {title} {{2D
  layered insulator hexagonal boron nitride enabled surface passivation in dye
  sensitized solar cells}},}\ }\href {\doibase 10.1039/C3NR03767C} {\bibfield
  {journal} {\bibinfo  {journal} {Nanoscale}\ }\textbf {\bibinfo {volume}
  {5}},\ \bibinfo {pages} {11275--11282} (\bibinfo {year} {2013})}\BibitemShut
  {NoStop}%
\bibitem [{\citenamefont {Benjamin}\ \emph {et~al.}(2013)\citenamefont
  {Benjamin}, \citenamefont {Zamponi}, \citenamefont {Juv{\'{e}}},
  \citenamefont {Stingl}, \citenamefont {Woerner}, \citenamefont {Elsaesser},\
  and\ \citenamefont {Chergui}}]{Benjamin2013}%
  \BibitemOpen
  \bibfield  {author} {\bibinfo {author} {\bibfnamefont {Freyer}\ \bibnamefont
  {Benjamin}}, \bibinfo {author} {\bibfnamefont {Flavio}\ \bibnamefont
  {Zamponi}}, \bibinfo {author} {\bibfnamefont {Vincent}\ \bibnamefont
  {Juv{\'{e}}}}, \bibinfo {author} {\bibfnamefont {Johannes}\ \bibnamefont
  {Stingl}}, \bibinfo {author} {\bibfnamefont {Michael}\ \bibnamefont
  {Woerner}}, \bibinfo {author} {\bibfnamefont {Thomas}\ \bibnamefont
  {Elsaesser}}, \ and\ \bibinfo {author} {\bibfnamefont {Majed}\ \bibnamefont
  {Chergui}},\ }\bibfield  {title} {\enquote {\bibinfo {title} {{Ultrafast
  inter-ionic charge transfer of transition-metal complexes mapped by
  femtosecond X-ray powder diffraction}},}\ }\href {\doibase 10.1063/1.4800223}
  {\bibfield  {journal} {\bibinfo  {journal} {The Journal of Chemical Physics}\
  }\textbf {\bibinfo {volume} {138}},\ \bibinfo {pages} {144504} (\bibinfo
  {year} {2013})}\BibitemShut {NoStop}%
\bibitem [{\citenamefont {Fohlisch}\ \emph {et~al.}(2005)\citenamefont
  {Fohlisch}, \citenamefont {Feulner}, \citenamefont {Hennies}, \citenamefont
  {Fink}, \citenamefont {Menzel}, \citenamefont {Sanchez-Portal}, \citenamefont
  {Echenique},\ and\ \citenamefont {Wurth}}]{Fohlisch2005}%
  \BibitemOpen
  \bibfield  {author} {\bibinfo {author} {\bibfnamefont {A}~\bibnamefont
  {Fohlisch}}, \bibinfo {author} {\bibfnamefont {P}~\bibnamefont {Feulner}},
  \bibinfo {author} {\bibfnamefont {F}~\bibnamefont {Hennies}}, \bibinfo
  {author} {\bibfnamefont {A}~\bibnamefont {Fink}}, \bibinfo {author}
  {\bibfnamefont {D}~\bibnamefont {Menzel}}, \bibinfo {author} {\bibfnamefont
  {D}~\bibnamefont {Sanchez-Portal}}, \bibinfo {author} {\bibfnamefont {P~M}\
  \bibnamefont {Echenique}}, \ and\ \bibinfo {author} {\bibfnamefont
  {W}~\bibnamefont {Wurth}},\ }\bibfield  {title} {\enquote {\bibinfo {title}
  {{Direct observation of electron dynamics in the attosecond domain}},}\
  }\href {http://dx.doi.org/10.1038/nature03833} {\bibfield  {journal}
  {\bibinfo  {journal} {Nature}\ }\textbf {\bibinfo {volume} {436}},\ \bibinfo
  {pages} {373--376} (\bibinfo {year} {2005})}\BibitemShut {NoStop}%
\bibitem [{\citenamefont {Boll}\ \emph {et~al.}(2013)\citenamefont {Boll},
  \citenamefont {Anielski}, \citenamefont {Bostedt}, \citenamefont {Bozek},
  \citenamefont {Christensen}, \citenamefont {Coffee}, \citenamefont {De},
  \citenamefont {Decleva}, \citenamefont {Epp}, \citenamefont {Erk},
  \citenamefont {Foucar}, \citenamefont {Krasniqi}, \citenamefont
  {K{\"{u}}pper}, \citenamefont {Rouz{\'{e}}e}, \citenamefont {Rudek},
  \citenamefont {Rudenko}, \citenamefont {Schorb}, \citenamefont {Stapelfeldt},
  \citenamefont {Stener}, \citenamefont {Stern}, \citenamefont {Techert},
  \citenamefont {Trippel}, \citenamefont {Vrakking}, \citenamefont {Ullrich},\
  and\ \citenamefont {Rolles}}]{Boll2013}%
  \BibitemOpen
  \bibfield  {author} {\bibinfo {author} {\bibfnamefont {R}~\bibnamefont
  {Boll}}, \bibinfo {author} {\bibfnamefont {D}~\bibnamefont {Anielski}},
  \bibinfo {author} {\bibfnamefont {C}~\bibnamefont {Bostedt}}, \bibinfo
  {author} {\bibfnamefont {J~D}\ \bibnamefont {Bozek}}, \bibinfo {author}
  {\bibfnamefont {L}~\bibnamefont {Christensen}}, \bibinfo {author}
  {\bibfnamefont {R}~\bibnamefont {Coffee}}, \bibinfo {author} {\bibfnamefont
  {S}~\bibnamefont {De}}, \bibinfo {author} {\bibfnamefont {P}~\bibnamefont
  {Decleva}}, \bibinfo {author} {\bibfnamefont {S~W}\ \bibnamefont {Epp}},
  \bibinfo {author} {\bibfnamefont {B}~\bibnamefont {Erk}}, \bibinfo {author}
  {\bibfnamefont {L}~\bibnamefont {Foucar}}, \bibinfo {author} {\bibfnamefont
  {F}~\bibnamefont {Krasniqi}}, \bibinfo {author} {\bibfnamefont
  {J}~\bibnamefont {K{\"{u}}pper}}, \bibinfo {author} {\bibfnamefont
  {A}~\bibnamefont {Rouz{\'{e}}e}}, \bibinfo {author} {\bibfnamefont
  {B}~\bibnamefont {Rudek}}, \bibinfo {author} {\bibfnamefont {A}~\bibnamefont
  {Rudenko}}, \bibinfo {author} {\bibfnamefont {S}~\bibnamefont {Schorb}},
  \bibinfo {author} {\bibfnamefont {H}~\bibnamefont {Stapelfeldt}}, \bibinfo
  {author} {\bibfnamefont {M}~\bibnamefont {Stener}}, \bibinfo {author}
  {\bibfnamefont {S}~\bibnamefont {Stern}}, \bibinfo {author} {\bibfnamefont
  {S}~\bibnamefont {Techert}}, \bibinfo {author} {\bibfnamefont
  {S}~\bibnamefont {Trippel}}, \bibinfo {author} {\bibfnamefont {M~J~J}\
  \bibnamefont {Vrakking}}, \bibinfo {author} {\bibfnamefont {J}~\bibnamefont
  {Ullrich}}, \ and\ \bibinfo {author} {\bibfnamefont {D}~\bibnamefont
  {Rolles}},\ }\bibfield  {title} {\enquote {\bibinfo {title} {{Femtosecond
  photoelectron diffraction on laser-aligned molecules: Towards time-resolved
  imaging of molecular structure}},}\ }\href
  {https://link.aps.org/doi/10.1103/PhysRevA.88.061402} {\bibfield  {journal}
  {\bibinfo  {journal} {Physical Review A}\ }\textbf {\bibinfo {volume} {88}},\
  \bibinfo {pages} {61402} (\bibinfo {year} {2013})}\BibitemShut {NoStop}%
\bibitem [{\citenamefont {Breidbach}\ and\ \citenamefont
  {Cederbaum}(2005)}]{Breidbach2005}%
  \BibitemOpen
  \bibfield  {author} {\bibinfo {author} {\bibfnamefont {J}~\bibnamefont
  {Breidbach}}\ and\ \bibinfo {author} {\bibfnamefont {L~S}\ \bibnamefont
  {Cederbaum}},\ }\bibfield  {title} {\enquote {\bibinfo {title} {{Universal
  Attosecond Response to the Removal of an Electron}},}\ }\href
  {https://link.aps.org/doi/10.1103/PhysRevLett.94.033901} {\bibfield
  {journal} {\bibinfo  {journal} {Physical Review Letters}\ }\textbf {\bibinfo
  {volume} {94}},\ \bibinfo {pages} {33901} (\bibinfo {year}
  {2005})}\BibitemShut {NoStop}%
\bibitem [{\citenamefont {Reed}\ \emph {et~al.}(2010)\citenamefont {Reed},
  \citenamefont {Uchoa}, \citenamefont {Joe}, \citenamefont {Gan},
  \citenamefont {Casa}, \citenamefont {Fradkin},\ and\ \citenamefont
  {Abbamonte}}]{Reed2010}%
  \BibitemOpen
  \bibfield  {author} {\bibinfo {author} {\bibfnamefont {James~P}\ \bibnamefont
  {Reed}}, \bibinfo {author} {\bibfnamefont {Bruno}\ \bibnamefont {Uchoa}},
  \bibinfo {author} {\bibfnamefont {Young~Il}\ \bibnamefont {Joe}}, \bibinfo
  {author} {\bibfnamefont {Yu}~\bibnamefont {Gan}}, \bibinfo {author}
  {\bibfnamefont {Diego}\ \bibnamefont {Casa}}, \bibinfo {author}
  {\bibfnamefont {Eduardo}\ \bibnamefont {Fradkin}}, \ and\ \bibinfo {author}
  {\bibfnamefont {Peter}\ \bibnamefont {Abbamonte}},\ }\bibfield  {title}
  {\enquote {\bibinfo {title} {{The Effective Fine-Structure Constant of
  Freestanding Graphene Measured in Graphite}},}\ }\href
  {http://science.sciencemag.org/content/330/6005/805.abstract} {\bibfield
  {journal} {\bibinfo  {journal} {Science}\ }\textbf {\bibinfo {volume}
  {330}},\ \bibinfo {pages} {805 -- 808} (\bibinfo {year} {2010})}\BibitemShut
  {NoStop}%
\bibitem [{\citenamefont {Cousin}\ \emph {et~al.}(2014)\citenamefont {Cousin},
  \citenamefont {Silva}, \citenamefont {Teichmann}, \citenamefont {Hemmer},
  \citenamefont {Buades},\ and\ \citenamefont {Biegert}}]{Cousin2014a}%
  \BibitemOpen
  \bibfield  {author} {\bibinfo {author} {\bibfnamefont {S.~L.}\ \bibnamefont
  {Cousin}}, \bibinfo {author} {\bibfnamefont {F.}~\bibnamefont {Silva}},
  \bibinfo {author} {\bibfnamefont {S.}~\bibnamefont {Teichmann}}, \bibinfo
  {author} {\bibfnamefont {M.}~\bibnamefont {Hemmer}}, \bibinfo {author}
  {\bibfnamefont {B.}~\bibnamefont {Buades}}, \ and\ \bibinfo {author}
  {\bibfnamefont {J.}~\bibnamefont {Biegert}},\ }\bibfield  {title} {\enquote
  {\bibinfo {title} {{High-flux table-top soft x-ray source driven by
  sub-2-cycle, CEP stable, 185-$\mu$m 1-kHz pulses for carbon K-edge
  spectroscopy}},}\ }\href
  {http://www.opticsinfobase.org/viewmedia.cfm?uri=ol-39-18-5383{\&}seq=0{\&}html=true}
  {\bibfield  {journal} {\bibinfo  {journal} {Optics Letters}\ }\textbf
  {\bibinfo {volume} {39}},\ \bibinfo {pages} {5383} (\bibinfo {year}
  {2014})}\BibitemShut {NoStop}%
\bibitem [{\citenamefont {Cousin}\ \emph {et~al.}(2015)\citenamefont {Cousin},
  \citenamefont {Silva}, \citenamefont {Teichmann}, \citenamefont {Hemmer},
  \citenamefont {Buades},\ and\ \citenamefont {Biegert}}]{Cousin2015}%
  \BibitemOpen
  \bibfield  {author} {\bibinfo {author} {\bibfnamefont {S.~L.}\ \bibnamefont
  {Cousin}}, \bibinfo {author} {\bibfnamefont {F.}~\bibnamefont {Silva}},
  \bibinfo {author} {\bibfnamefont {S.}~\bibnamefont {Teichmann}}, \bibinfo
  {author} {\bibfnamefont {M.}~\bibnamefont {Hemmer}}, \bibinfo {author}
  {\bibfnamefont {B.}~\bibnamefont {Buades}}, \ and\ \bibinfo {author}
  {\bibfnamefont {J.}~\bibnamefont {Biegert}},\ }\bibfield  {title} {\enquote
  {\bibinfo {title} {{Attosecond pulse generation in the water window}},}\ }in\
  \href@noop {} {\emph {\bibinfo {booktitle} {Ultrafast Optics UFO X}}}\
  (\bibinfo {year} {2015})\BibitemShut {NoStop}%
\bibitem [{\citenamefont {Biegert}(2016)}]{Biegert2016}%
  \BibitemOpen
  \bibfield  {author} {\bibinfo {author} {\bibfnamefont {Jens}\ \bibnamefont
  {Biegert}},\ }\bibfield  {title} {\enquote {\bibinfo {title} {{Isolated
  attosecond soft X-rays and water window XAFS}},}\ }in\ \href@noop {} {\emph
  {\bibinfo {booktitle} {47th Annual Meeting of the APS Division of Atomic,
  Molecular and Optical Physics}}}\ (\bibinfo  {publisher} {Bulletin of the
  American Physical Society},\ \bibinfo {year} {2016})\ p.\ \bibinfo {pages}
  {BAPS.2016.DAMOP.U9.2}\BibitemShut {NoStop}%
\bibitem [{\citenamefont {Kane}\ and\ \citenamefont
  {Trebino}(1993)}]{Kane1993}%
  \BibitemOpen
  \bibfield  {author} {\bibinfo {author} {\bibfnamefont {D~J}\ \bibnamefont
  {Kane}}\ and\ \bibinfo {author} {\bibfnamefont {R}~\bibnamefont {Trebino}},\
  }\bibfield  {title} {\enquote {\bibinfo {title} {{Characterization of
  arbitrary femtosecond pulses using frequency-resolved optical gating}},}\
  }\href@noop {} {\bibfield  {journal} {\bibinfo  {journal} {IEEE Journal of
  Quantum Electronics}\ }\textbf {\bibinfo {volume} {29}},\ \bibinfo {pages}
  {571--579} (\bibinfo {year} {1993})}\BibitemShut {NoStop}%
\bibitem [{\citenamefont {Walmsley}\ and\ \citenamefont
  {Wong}(1996)}]{Walmsley1996}%
  \BibitemOpen
  \bibfield  {author} {\bibinfo {author} {\bibfnamefont {Ian~A.}\ \bibnamefont
  {Walmsley}}\ and\ \bibinfo {author} {\bibfnamefont {Victor}\ \bibnamefont
  {Wong}},\ }\bibfield  {title} {\enquote {\bibinfo {title} {{Characterization
  of the electric field of ultrashort optical pulses}},}\ }\href {\doibase
  10.1364/JOSAB.13.002453} {\bibfield  {journal} {\bibinfo  {journal} {Journal
  of the Optical Society of America B}\ }\textbf {\bibinfo {volume} {13}},\
  \bibinfo {pages} {2453} (\bibinfo {year} {1996})}\BibitemShut {NoStop}%
\bibitem [{\citenamefont {Nisoli}\ \emph {et~al.}(1996)\citenamefont {Nisoli},
  \citenamefont {{De Silvestri}},\ and\ \citenamefont {Svelto}}]{Nisoli1996}%
  \BibitemOpen
  \bibfield  {author} {\bibinfo {author} {\bibfnamefont {M.}~\bibnamefont
  {Nisoli}}, \bibinfo {author} {\bibfnamefont {S.}~\bibnamefont {{De
  Silvestri}}}, \ and\ \bibinfo {author} {\bibfnamefont {O.}~\bibnamefont
  {Svelto}},\ }\bibfield  {title} {\enquote {\bibinfo {title} {{Generation of
  high energy 10 fs pulses by a new pulse compression technique}},}\ }\href
  {\doibase 10.1063/1.116609} {\bibfield  {journal} {\bibinfo  {journal}
  {Applied Physics Letters}\ }\textbf {\bibinfo {volume} {68}},\ \bibinfo
  {pages} {2793} (\bibinfo {year} {1996})}\BibitemShut {NoStop}%
\bibitem [{\citenamefont {Schenkel}\ \emph {et~al.}(2003)\citenamefont
  {Schenkel}, \citenamefont {Biegert}, \citenamefont {Keller}, \citenamefont
  {Vozzi}, \citenamefont {Nisoli}, \citenamefont {Sansone}, \citenamefont
  {Stagira}, \citenamefont {{De Silvestri}},\ and\ \citenamefont
  {Svelto}}]{Schenkel2003}%
  \BibitemOpen
  \bibfield  {author} {\bibinfo {author} {\bibfnamefont {B}~\bibnamefont
  {Schenkel}}, \bibinfo {author} {\bibfnamefont {J}~\bibnamefont {Biegert}},
  \bibinfo {author} {\bibfnamefont {U}~\bibnamefont {Keller}}, \bibinfo
  {author} {\bibfnamefont {C}~\bibnamefont {Vozzi}}, \bibinfo {author}
  {\bibfnamefont {M}~\bibnamefont {Nisoli}}, \bibinfo {author} {\bibfnamefont
  {Giuseppe}\ \bibnamefont {Sansone}}, \bibinfo {author} {\bibfnamefont
  {S}~\bibnamefont {Stagira}}, \bibinfo {author} {\bibfnamefont
  {S}~\bibnamefont {{De Silvestri}}}, \ and\ \bibinfo {author} {\bibfnamefont
  {O}~\bibnamefont {Svelto}},\ }\bibfield  {title} {\enquote {\bibinfo {title}
  {{Generation of 3.8-fs pulses from adaptive compression of a cascaded hollow
  fiber supercontinuum.}}}\ }\href
  {http://www.ncbi.nlm.nih.gov/pubmed/14587798} {\bibfield  {journal} {\bibinfo
   {journal} {Optics Letters}\ }\textbf {\bibinfo {volume} {28}},\ \bibinfo
  {pages} {1987--9} (\bibinfo {year} {2003})}\BibitemShut {NoStop}%
\bibitem [{\citenamefont {Hauri}\ \emph {et~al.}(2004)\citenamefont {Hauri},
  \citenamefont {Kornelis}, \citenamefont {Helbing}, \citenamefont {Heinrich},
  \citenamefont {Couairon}, \citenamefont {Mysyrowicz}, \citenamefont
  {Biegert},\ and\ \citenamefont {Keller}}]{Hauri2004}%
  \BibitemOpen
  \bibfield  {author} {\bibinfo {author} {\bibfnamefont {C.P.}\ \bibnamefont
  {Hauri}}, \bibinfo {author} {\bibfnamefont {W.}~\bibnamefont {Kornelis}},
  \bibinfo {author} {\bibfnamefont {F.W.}\ \bibnamefont {Helbing}}, \bibinfo
  {author} {\bibfnamefont {A}~\bibnamefont {Heinrich}}, \bibinfo {author}
  {\bibfnamefont {A}~\bibnamefont {Couairon}}, \bibinfo {author} {\bibfnamefont
  {A}~\bibnamefont {Mysyrowicz}}, \bibinfo {author} {\bibfnamefont
  {J}~\bibnamefont {Biegert}}, \ and\ \bibinfo {author} {\bibfnamefont
  {U}~\bibnamefont {Keller}},\ }\bibfield  {title} {\enquote {\bibinfo {title}
  {{Generation of intense, carrier-envelope phase-locked few-cycle laser pulses
  through filamentation}},}\ }\href {\doibase 10.1007/s00340-004-1650-z}
  {\bibfield  {journal} {\bibinfo  {journal} {Applied Physics B}\ }\textbf
  {\bibinfo {volume} {79}},\ \bibinfo {pages} {673--677} (\bibinfo {year}
  {2004})}\BibitemShut {NoStop}%
\bibitem [{\citenamefont {Couairon}\ \emph {et~al.}(2005)\citenamefont
  {Couairon}, \citenamefont {Franco}, \citenamefont {Mysyrowicz}, \citenamefont
  {Biegert},\ and\ \citenamefont {Keller}}]{Couairon2005}%
  \BibitemOpen
  \bibfield  {author} {\bibinfo {author} {\bibfnamefont {A}~\bibnamefont
  {Couairon}}, \bibinfo {author} {\bibfnamefont {M}~\bibnamefont {Franco}},
  \bibinfo {author} {\bibfnamefont {A}~\bibnamefont {Mysyrowicz}}, \bibinfo
  {author} {\bibfnamefont {J}~\bibnamefont {Biegert}}, \ and\ \bibinfo {author}
  {\bibfnamefont {U}~\bibnamefont {Keller}},\ }\bibfield  {title} {\enquote
  {\bibinfo {title} {{Pulse self-compression to the single-cycle limit by
  filamentation in a gas with a pressure gradient.}}}\ }\href
  {http://www.ncbi.nlm.nih.gov/pubmed/16208932} {\bibfield  {journal} {\bibinfo
   {journal} {Optics Letters}\ }\textbf {\bibinfo {volume} {30}},\ \bibinfo
  {pages} {2657--9} (\bibinfo {year} {2005})}\BibitemShut {NoStop}%
\bibitem [{\citenamefont {Hauri}\ \emph {et~al.}(2005)\citenamefont {Hauri},
  \citenamefont {Guandalini}, \citenamefont {Eckle}, \citenamefont {Kornelis},
  \citenamefont {Biegert},\ and\ \citenamefont {Keller}}]{Hauri2005a}%
  \BibitemOpen
  \bibfield  {author} {\bibinfo {author} {\bibfnamefont {C}~\bibnamefont
  {Hauri}}, \bibinfo {author} {\bibfnamefont {A}~\bibnamefont {Guandalini}},
  \bibinfo {author} {\bibfnamefont {P}~\bibnamefont {Eckle}}, \bibinfo {author}
  {\bibfnamefont {W}~\bibnamefont {Kornelis}}, \bibinfo {author} {\bibfnamefont
  {J}~\bibnamefont {Biegert}}, \ and\ \bibinfo {author} {\bibfnamefont
  {U}~\bibnamefont {Keller}},\ }\bibfield  {title} {\enquote {\bibinfo {title}
  {{Generation of intense few-cycle laser pulses through filamentation -
  parameter dependence.}}}\ }\href
  {http://www.ncbi.nlm.nih.gov/pubmed/19498780} {\bibfield  {journal} {\bibinfo
   {journal} {Optics Express}\ }\textbf {\bibinfo {volume} {13}},\ \bibinfo
  {pages} {7541--7} (\bibinfo {year} {2005})}\BibitemShut {NoStop}%
\bibitem [{\citenamefont {Guandalini}\ \emph {et~al.}(2006)\citenamefont
  {Guandalini}, \citenamefont {Eckle}, \citenamefont {Anscombe}, \citenamefont
  {Schlup}, \citenamefont {Biegert},\ and\ \citenamefont
  {Keller}}]{Guandalini2006}%
  \BibitemOpen
  \bibfield  {author} {\bibinfo {author} {\bibfnamefont {A}~\bibnamefont
  {Guandalini}}, \bibinfo {author} {\bibfnamefont {P}~\bibnamefont {Eckle}},
  \bibinfo {author} {\bibfnamefont {M}~\bibnamefont {Anscombe}}, \bibinfo
  {author} {\bibfnamefont {P}~\bibnamefont {Schlup}}, \bibinfo {author}
  {\bibfnamefont {J}~\bibnamefont {Biegert}}, \ and\ \bibinfo {author}
  {\bibfnamefont {U}~\bibnamefont {Keller}},\ }\bibfield  {title} {\enquote
  {\bibinfo {title} {{5.1 Fs Pulses Generated By Filamentation and Carrier
  Envelope Phase Stability Analysis}},}\ }\href {\doibase
  10.1088/0953-4075/39/13/S01} {\bibfield  {journal} {\bibinfo  {journal}
  {Journal of Physics B: Atomic, Molecular and Optical Physics}\ }\textbf
  {\bibinfo {volume} {39}},\ \bibinfo {pages} {S257--S264} (\bibinfo {year}
  {2006})}\BibitemShut {NoStop}%
\bibitem [{\citenamefont {Wirth}\ \emph {et~al.}(2011)\citenamefont {Wirth},
  \citenamefont {Hassan}, \citenamefont {Grgura{\v{s}}}, \citenamefont
  {Gagnon}, \citenamefont {Moulet}, \citenamefont {Luu}, \citenamefont {Pabst},
  \citenamefont {Santra}, \citenamefont {Alahmed}, \citenamefont {Azzeer},
  \citenamefont {Yakovlev}, \citenamefont {Pervak}, \citenamefont {Krausz},\
  and\ \citenamefont {Goulielmakis}}]{Wirth2011}%
  \BibitemOpen
  \bibfield  {author} {\bibinfo {author} {\bibfnamefont {A}~\bibnamefont
  {Wirth}}, \bibinfo {author} {\bibfnamefont {M~Th.}\ \bibnamefont {Hassan}},
  \bibinfo {author} {\bibfnamefont {I}~\bibnamefont {Grgura{\v{s}}}}, \bibinfo
  {author} {\bibfnamefont {J}~\bibnamefont {Gagnon}}, \bibinfo {author}
  {\bibfnamefont {A}~\bibnamefont {Moulet}}, \bibinfo {author} {\bibfnamefont
  {T~T}\ \bibnamefont {Luu}}, \bibinfo {author} {\bibfnamefont {S}~\bibnamefont
  {Pabst}}, \bibinfo {author} {\bibfnamefont {R}~\bibnamefont {Santra}},
  \bibinfo {author} {\bibfnamefont {Z~A}\ \bibnamefont {Alahmed}}, \bibinfo
  {author} {\bibfnamefont {A~M}\ \bibnamefont {Azzeer}}, \bibinfo {author}
  {\bibfnamefont {V~S}\ \bibnamefont {Yakovlev}}, \bibinfo {author}
  {\bibfnamefont {V}~\bibnamefont {Pervak}}, \bibinfo {author} {\bibfnamefont
  {F}~\bibnamefont {Krausz}}, \ and\ \bibinfo {author} {\bibfnamefont
  {E}~\bibnamefont {Goulielmakis}},\ }\bibfield  {title} {\enquote {\bibinfo
  {title} {{Synthesized Light Transients}},}\ }\href {\doibase
  10.1126/science.1210268} {\bibfield  {journal} {\bibinfo  {journal}
  {Science}\ }\textbf {\bibinfo {volume} {334}},\ \bibinfo {pages} {195--200}
  (\bibinfo {year} {2011})}\BibitemShut {NoStop}%
\bibitem [{\citenamefont {Chen}\ \emph {et~al.}(2014)\citenamefont {Chen},
  \citenamefont {Mancuso}, \citenamefont {Hern{\'{a}}ndez-Garc{\'{i}}a},
  \citenamefont {Dollar}, \citenamefont {Galloway}, \citenamefont
  {Popmintchev}, \citenamefont {Huang}, \citenamefont {Walker}, \citenamefont
  {Plaja}, \citenamefont {Jaro{\'{n}}-Becker}, \citenamefont {Becker},
  \citenamefont {Murnane}, \citenamefont {Kapteyn},\ and\ \citenamefont
  {Popmintchev}}]{Chen2014}%
  \BibitemOpen
  \bibfield  {author} {\bibinfo {author} {\bibfnamefont {Ming~Chang}\
  \bibnamefont {Chen}}, \bibinfo {author} {\bibfnamefont {Christopher}\
  \bibnamefont {Mancuso}}, \bibinfo {author} {\bibfnamefont {Carlos}\
  \bibnamefont {Hern{\'{a}}ndez-Garc{\'{i}}a}}, \bibinfo {author}
  {\bibfnamefont {Franklin}\ \bibnamefont {Dollar}}, \bibinfo {author}
  {\bibfnamefont {Ben}\ \bibnamefont {Galloway}}, \bibinfo {author}
  {\bibfnamefont {Dimitar}\ \bibnamefont {Popmintchev}}, \bibinfo {author}
  {\bibfnamefont {Pei-Chi}\ \bibnamefont {Huang}}, \bibinfo {author}
  {\bibfnamefont {Barry}\ \bibnamefont {Walker}}, \bibinfo {author}
  {\bibfnamefont {Luis}\ \bibnamefont {Plaja}}, \bibinfo {author}
  {\bibfnamefont {Agnieszka}\ \bibnamefont {Jaro{\'{n}}-Becker}}, \bibinfo
  {author} {\bibfnamefont {Andreas}\ \bibnamefont {Becker}}, \bibinfo {author}
  {\bibfnamefont {Margaret~M}\ \bibnamefont {Murnane}}, \bibinfo {author}
  {\bibfnamefont {Henry~C}\ \bibnamefont {Kapteyn}}, \ and\ \bibinfo {author}
  {\bibfnamefont {Tenio}\ \bibnamefont {Popmintchev}},\ }\bibfield  {title}
  {\enquote {\bibinfo {title} {{Generation of bright isolated attosecond soft
  X-ray pulses driven by multicycle midinfrared lasers.}}}\ }\href {\doibase
  10.1073/pnas.1407421111} {\bibfield  {journal} {\bibinfo  {journal}
  {Proceedings of the National Academy of Sciences of the United States of
  America}\ }\textbf {\bibinfo {volume} {111}},\ \bibinfo {pages} {1--7}
  (\bibinfo {year} {2014})}\BibitemShut {NoStop}%
\bibitem [{\citenamefont {Corkum}(1993)}]{Corkum1993}%
  \BibitemOpen
  \bibfield  {author} {\bibinfo {author} {\bibfnamefont {PB}~\bibnamefont
  {Corkum}},\ }\bibfield  {title} {\enquote {\bibinfo {title} {{Plasma
  Perspective on Strong-Field Multiphoton Ionization}},}\ }\href
  {http://link.aps.org/doi/10.1103/PhysRevLett.71.1994} {\bibfield  {journal}
  {\bibinfo  {journal} {Physical Review Letters}\ }\textbf {\bibinfo {volume}
  {71}},\ \bibinfo {pages} {1994--1997} (\bibinfo {year} {1993})}\BibitemShut
  {NoStop}%
\bibitem [{\citenamefont {Kim}\ \emph {et~al.}(2013)\citenamefont {Kim},
  \citenamefont {Zhang}, \citenamefont {Ruchon},\ and\ \citenamefont
  {Hergott}}]{Kim2013}%
  \BibitemOpen
  \bibfield  {author} {\bibinfo {author} {\bibfnamefont {KT}~\bibnamefont
  {Kim}}, \bibinfo {author} {\bibfnamefont {Chunmei}\ \bibnamefont {Zhang}},
  \bibinfo {author} {\bibfnamefont {Thierry}\ \bibnamefont {Ruchon}}, \ and\
  \bibinfo {author} {\bibfnamefont {JF}~\bibnamefont {Hergott}},\ }\bibfield
  {title} {\enquote {\bibinfo {title} {{Photonic streaking of attosecond pulse
  trains}},}\ }\href {\doibase 10.1038/nphoton.2013.170} {\bibfield  {journal}
  {\bibinfo  {journal} {Nature Photonics}\ }\textbf {\bibinfo {volume} {7}},\
  \bibinfo {pages} {6--11} (\bibinfo {year} {2013})}\BibitemShut {NoStop}%
\bibitem [{\citenamefont {Itatani}\ \emph {et~al.}(2002)\citenamefont
  {Itatani}, \citenamefont {Qu{\'{e}}r{\'{e}}}, \citenamefont {Yudin},
  \citenamefont {Ivanov}, \citenamefont {Krausz},\ and\ \citenamefont
  {Corkum}}]{Itatani2002}%
  \BibitemOpen
  \bibfield  {author} {\bibinfo {author} {\bibfnamefont {J}~\bibnamefont
  {Itatani}}, \bibinfo {author} {\bibfnamefont {F}~\bibnamefont
  {Qu{\'{e}}r{\'{e}}}}, \bibinfo {author} {\bibfnamefont {G~L}\ \bibnamefont
  {Yudin}}, \bibinfo {author} {\bibfnamefont {M~Yu}\ \bibnamefont {Ivanov}},
  \bibinfo {author} {\bibfnamefont {F}~\bibnamefont {Krausz}}, \ and\ \bibinfo
  {author} {\bibfnamefont {P~B}\ \bibnamefont {Corkum}},\ }\bibfield  {title}
  {\enquote {\bibinfo {title} {{Attosecond streak camera.}}}\ }\href {\doibase
  10.1103/PhysRevLett.88.173903} {\bibfield  {journal} {\bibinfo  {journal}
  {Physical Review Letters}\ }\textbf {\bibinfo {volume} {88}},\ \bibinfo
  {pages} {173903} (\bibinfo {year} {2002})}\BibitemShut {NoStop}%
\bibitem [{\citenamefont {Paul}\ \emph {et~al.}(2001)\citenamefont {Paul},
  \citenamefont {Toma}, \citenamefont {Breger}, \citenamefont {Mullot},
  \citenamefont {Aug{\'{e}}}, \citenamefont {Balcou}, \citenamefont {Muller},\
  and\ \citenamefont {Agostini}}]{Paul2001}%
  \BibitemOpen
  \bibfield  {author} {\bibinfo {author} {\bibfnamefont {P~M}\ \bibnamefont
  {Paul}}, \bibinfo {author} {\bibfnamefont {E~S}\ \bibnamefont {Toma}},
  \bibinfo {author} {\bibfnamefont {P}~\bibnamefont {Breger}}, \bibinfo
  {author} {\bibfnamefont {G}~\bibnamefont {Mullot}}, \bibinfo {author}
  {\bibfnamefont {F}~\bibnamefont {Aug{\'{e}}}}, \bibinfo {author}
  {\bibfnamefont {Ph.}\ \bibnamefont {Balcou}}, \bibinfo {author}
  {\bibfnamefont {H~G}\ \bibnamefont {Muller}}, \ and\ \bibinfo {author}
  {\bibfnamefont {P}~\bibnamefont {Agostini}},\ }\bibfield  {title} {\enquote
  {\bibinfo {title} {{Observation of a Train of Attosecond Pulses from High
  Harmonic Generation}},}\ }\href {\doibase 10.1126/science.1059413} {\bibfield
   {journal} {\bibinfo  {journal} {Science}\ }\textbf {\bibinfo {volume}
  {292}},\ \bibinfo {pages} {1689--1692} (\bibinfo {year} {2001})}\BibitemShut
  {NoStop}%
\bibitem [{\citenamefont {Mairesse}\ and\ \citenamefont
  {Qu{\'{e}}r{\'{e}}}(2005)}]{Mairesse2005}%
  \BibitemOpen
  \bibfield  {author} {\bibinfo {author} {\bibfnamefont {Y.}~\bibnamefont
  {Mairesse}}\ and\ \bibinfo {author} {\bibfnamefont {F.}~\bibnamefont
  {Qu{\'{e}}r{\'{e}}}},\ }\bibfield  {title} {\enquote {\bibinfo {title}
  {{Frequency-resolved optical gating for complete reconstruction of attosecond
  bursts}},}\ }\href {\doibase 10.1103/PhysRevA.71.011401} {\bibfield
  {journal} {\bibinfo  {journal} {Physical Review A - Atomic, Molecular, and
  Optical Physics}\ }\textbf {\bibinfo {volume} {71}},\ \bibinfo {pages} {1--4}
  (\bibinfo {year} {2005})}\BibitemShut {NoStop}%
\bibitem [{\citenamefont {Kienberger}\ \emph {et~al.}(2004)\citenamefont
  {Kienberger}, \citenamefont {Goulielmakis}, \citenamefont {Uiberacker},
  \citenamefont {Baltuska}, \citenamefont {Yakovlev}, \citenamefont {Bammer},
  \citenamefont {Scrinzi}, \citenamefont {Westerwalbesloh}, \citenamefont
  {Kleineberg}, \citenamefont {Heinzmann}, \citenamefont {Drescher},\ and\
  \citenamefont {Krausz}}]{Kienberger2004}%
  \BibitemOpen
  \bibfield  {author} {\bibinfo {author} {\bibfnamefont {R}~\bibnamefont
  {Kienberger}}, \bibinfo {author} {\bibfnamefont {E}~\bibnamefont
  {Goulielmakis}}, \bibinfo {author} {\bibfnamefont {M}~\bibnamefont
  {Uiberacker}}, \bibinfo {author} {\bibfnamefont {A}~\bibnamefont {Baltuska}},
  \bibinfo {author} {\bibfnamefont {V}~\bibnamefont {Yakovlev}}, \bibinfo
  {author} {\bibfnamefont {F}~\bibnamefont {Bammer}}, \bibinfo {author}
  {\bibfnamefont {A}~\bibnamefont {Scrinzi}}, \bibinfo {author} {\bibfnamefont
  {Th.}\ \bibnamefont {Westerwalbesloh}}, \bibinfo {author} {\bibfnamefont
  {U}~\bibnamefont {Kleineberg}}, \bibinfo {author} {\bibfnamefont
  {U}~\bibnamefont {Heinzmann}}, \bibinfo {author} {\bibfnamefont
  {M}~\bibnamefont {Drescher}}, \ and\ \bibinfo {author} {\bibfnamefont
  {F}~\bibnamefont {Krausz}},\ }\bibfield  {title} {\enquote {\bibinfo {title}
  {{Atomic transient recorder}},}\ }\href
  {http://dx.doi.org/10.1038/nature02277} {\bibfield  {journal} {\bibinfo
  {journal} {Nature}\ }\textbf {\bibinfo {volume} {427}},\ \bibinfo {pages}
  {817--821} (\bibinfo {year} {2004})}\BibitemShut {NoStop}%
\bibitem [{\citenamefont {Goulielmakis}\ \emph {et~al.}(2007)\citenamefont
  {Goulielmakis}, \citenamefont {Yakovlev}, \citenamefont {Cavalieri},
  \citenamefont {Uiberacker}, \citenamefont {Pervak}, \citenamefont
  {Apolonski}, \citenamefont {Kienberger}, \citenamefont {Kleineberg},\ and\
  \citenamefont {Krausz}}]{Goulielmakis2007}%
  \BibitemOpen
  \bibfield  {author} {\bibinfo {author} {\bibfnamefont {E}~\bibnamefont
  {Goulielmakis}}, \bibinfo {author} {\bibfnamefont {V~S}\ \bibnamefont
  {Yakovlev}}, \bibinfo {author} {\bibfnamefont {A~L}\ \bibnamefont
  {Cavalieri}}, \bibinfo {author} {\bibfnamefont {M}~\bibnamefont
  {Uiberacker}}, \bibinfo {author} {\bibfnamefont {V}~\bibnamefont {Pervak}},
  \bibinfo {author} {\bibfnamefont {A}~\bibnamefont {Apolonski}}, \bibinfo
  {author} {\bibfnamefont {R}~\bibnamefont {Kienberger}}, \bibinfo {author}
  {\bibfnamefont {U}~\bibnamefont {Kleineberg}}, \ and\ \bibinfo {author}
  {\bibfnamefont {F}~\bibnamefont {Krausz}},\ }\bibfield  {title} {\enquote
  {\bibinfo {title} {{Attosecond control and measurement: lightwave
  electronics.}}}\ }\href {\doibase 10.1126/science.1142855} {\bibfield
  {journal} {\bibinfo  {journal} {Science}\ }\textbf {\bibinfo {volume}
  {317}},\ \bibinfo {pages} {769--775} (\bibinfo {year} {2007})}\BibitemShut
  {NoStop}%
\bibitem [{\citenamefont {Cavalieri}\ \emph {et~al.}(2007)\citenamefont
  {Cavalieri}, \citenamefont {M{\"{u}}ller}, \citenamefont {Uphues},
  \citenamefont {Yakovlev}, \citenamefont {Baltuska}, \citenamefont {Horvath},
  \citenamefont {Schmidt}, \citenamefont {Bl{\"{u}}mel}, \citenamefont
  {Holzwarth}, \citenamefont {Hendel}, \citenamefont {Drescher}, \citenamefont
  {Kleineberg}, \citenamefont {Echenique}, \citenamefont {Kienberger},
  \citenamefont {Krausz},\ and\ \citenamefont {Heinzmann}}]{Cavalieri2007}%
  \BibitemOpen
  \bibfield  {author} {\bibinfo {author} {\bibfnamefont {A~L}\ \bibnamefont
  {Cavalieri}}, \bibinfo {author} {\bibfnamefont {N}~\bibnamefont
  {M{\"{u}}ller}}, \bibinfo {author} {\bibfnamefont {Th}~\bibnamefont
  {Uphues}}, \bibinfo {author} {\bibfnamefont {V~S}\ \bibnamefont {Yakovlev}},
  \bibinfo {author} {\bibfnamefont {A}~\bibnamefont {Baltuska}}, \bibinfo
  {author} {\bibfnamefont {B}~\bibnamefont {Horvath}}, \bibinfo {author}
  {\bibfnamefont {B}~\bibnamefont {Schmidt}}, \bibinfo {author} {\bibfnamefont
  {L}~\bibnamefont {Bl{\"{u}}mel}}, \bibinfo {author} {\bibfnamefont
  {R}~\bibnamefont {Holzwarth}}, \bibinfo {author} {\bibfnamefont
  {S}~\bibnamefont {Hendel}}, \bibinfo {author} {\bibfnamefont {M}~\bibnamefont
  {Drescher}}, \bibinfo {author} {\bibfnamefont {U}~\bibnamefont {Kleineberg}},
  \bibinfo {author} {\bibfnamefont {P~M}\ \bibnamefont {Echenique}}, \bibinfo
  {author} {\bibfnamefont {R}~\bibnamefont {Kienberger}}, \bibinfo {author}
  {\bibfnamefont {F}~\bibnamefont {Krausz}}, \ and\ \bibinfo {author}
  {\bibfnamefont {U}~\bibnamefont {Heinzmann}},\ }\bibfield  {title} {\enquote
  {\bibinfo {title} {{Attosecond spectroscopy in condensed matter.}}}\ }\href
  {\doibase 10.1038/nature06229} {\bibfield  {journal} {\bibinfo  {journal}
  {Nature}\ }\textbf {\bibinfo {volume} {449}},\ \bibinfo {pages} {1029--1032}
  (\bibinfo {year} {2007})}\BibitemShut {NoStop}%
\bibitem [{\citenamefont {Zhao}\ \emph {et~al.}(2017)\citenamefont {Zhao},
  \citenamefont {Wei}, \citenamefont {Wu},\ and\ \citenamefont
  {Lin}}]{Zhao2017}%
  \BibitemOpen
  \bibfield  {author} {\bibinfo {author} {\bibfnamefont {Xi}~\bibnamefont
  {Zhao}}, \bibinfo {author} {\bibfnamefont {Hui}\ \bibnamefont {Wei}},
  \bibinfo {author} {\bibfnamefont {Yan}\ \bibnamefont {Wu}}, \ and\ \bibinfo
  {author} {\bibfnamefont {C.~D.}\ \bibnamefont {Lin}},\ }\bibfield  {title}
  {\enquote {\bibinfo {title} {{Phase-retrieval algorithm for the
  characterization of broadband single attosecond pulses}},}\ }\href {\doibase
  10.1103/PhysRevA.95.043407} {\bibfield  {journal} {\bibinfo  {journal}
  {Physical Review A}\ }\textbf {\bibinfo {volume} {95}},\ \bibinfo {pages}
  {043407} (\bibinfo {year} {2017})}\BibitemShut {NoStop}%
\bibitem [{\citenamefont {Pabst}\ and\ \citenamefont
  {Dahlstr{\"{o}}m}(2016)}]{Pabst2016}%
  \BibitemOpen
  \bibfield  {author} {\bibinfo {author} {\bibfnamefont {Stefan}\ \bibnamefont
  {Pabst}}\ and\ \bibinfo {author} {\bibfnamefont {Jan~Marcus}\ \bibnamefont
  {Dahlstr{\"{o}}m}},\ }\bibfield  {title} {\enquote {\bibinfo {title}
  {{Eliminating the dipole phase in attosecond pulse characterization using
  Rydberg wave packets}},}\ }\href
  {https://link.aps.org/doi/10.1103/PhysRevA.94.013411} {\bibfield  {journal}
  {\bibinfo  {journal} {Physical Review A}\ }\textbf {\bibinfo {volume} {94}},\
  \bibinfo {pages} {13411} (\bibinfo {year} {2016})}\BibitemShut {NoStop}%
\bibitem [{\citenamefont {Pabst}\ and\ \citenamefont
  {Dahlstr{\"{o}}m}(2017)}]{Pabst2017}%
  \BibitemOpen
  \bibfield  {author} {\bibinfo {author} {\bibfnamefont {Stefan}\ \bibnamefont
  {Pabst}}\ and\ \bibinfo {author} {\bibfnamefont {J~M}\ \bibnamefont
  {Dahlstr{\"{o}}m}},\ }\bibfield  {title} {\enquote {\bibinfo {title}
  {{Characterizing attosecond pulses in the soft x-ray regime}},}\ }\href
  {http://stacks.iop.org/0953-4075/50/i=10/a=104002} {\bibfield  {journal}
  {\bibinfo  {journal} {Journal of Physics B: Atomic, Molecular and Optical
  Physics}\ }\textbf {\bibinfo {volume} {50}},\ \bibinfo {pages} {104002}
  (\bibinfo {year} {2017})}\BibitemShut {NoStop}%
\bibitem [{\citenamefont {Krause}\ \emph {et~al.}(1992)\citenamefont {Krause},
  \citenamefont {Schafer},\ and\ \citenamefont {Kulander}}]{Krause1992}%
  \BibitemOpen
  \bibfield  {author} {\bibinfo {author} {\bibfnamefont {JL}~\bibnamefont
  {Krause}}, \bibinfo {author} {\bibfnamefont {KJ}~\bibnamefont {Schafer}}, \
  and\ \bibinfo {author} {\bibfnamefont {KC}~\bibnamefont {Kulander}},\
  }\bibfield  {title} {\enquote {\bibinfo {title} {{High-Order Harmonic
  Generation from Atoms and Ions in the High Intensity Regime}},}\ }\href
  {http://link.aps.org/doi/10.1103/PhysRevLett.68.3535} {\bibfield  {journal}
  {\bibinfo  {journal} {Physical Review Letters}\ }\textbf {\bibinfo {volume}
  {68}},\ \bibinfo {pages} {3535--3538} (\bibinfo {year} {1992})}\BibitemShut
  {NoStop}%
\bibitem [{\citenamefont {Frolov}\ \emph {et~al.}(2008)\citenamefont {Frolov},
  \citenamefont {Manakov},\ and\ \citenamefont {Starace}}]{Frolov2008}%
  \BibitemOpen
  \bibfield  {author} {\bibinfo {author} {\bibfnamefont {M.}~\bibnamefont
  {Frolov}}, \bibinfo {author} {\bibfnamefont {N.}~\bibnamefont {Manakov}}, \
  and\ \bibinfo {author} {\bibfnamefont {Anthony}\ \bibnamefont {Starace}},\
  }\bibfield  {title} {\enquote {\bibinfo {title} {{Wavelength Scaling of
  High-Harmonic Yield: Threshold Phenomena and Bound State Symmetry
  Dependence}},}\ }\href {\doibase 10.1103/PhysRevLett.100.173001} {\bibfield
  {journal} {\bibinfo  {journal} {Physical Review Letters}\ }\textbf {\bibinfo
  {volume} {100}},\ \bibinfo {pages} {173001} (\bibinfo {year}
  {2008})}\BibitemShut {NoStop}%
\bibitem [{\citenamefont {Austin}\ and\ \citenamefont
  {Biegert}(2012)}]{Austin2012}%
  \BibitemOpen
  \bibfield  {author} {\bibinfo {author} {\bibfnamefont {Dane}\ \bibnamefont
  {Austin}}\ and\ \bibinfo {author} {\bibfnamefont {Jens}\ \bibnamefont
  {Biegert}},\ }\bibfield  {title} {\enquote {\bibinfo {title} {{Strong-field
  approximation for the wavelength scaling of high-harmonic generation}},}\
  }\href@noop {} {\bibfield  {journal} {\bibinfo  {journal} {Physical Review
  A}\ }\textbf {\bibinfo {volume} {86}} (\bibinfo {year} {2012})}\BibitemShut
  {NoStop}%
\bibitem [{\citenamefont {Popmintchev}\ \emph {et~al.}(2009)\citenamefont
  {Popmintchev}, \citenamefont {Chen}, \citenamefont {Bahabad}, \citenamefont
  {Gerrity}, \citenamefont {Sidorenko}, \citenamefont {Cohen}, \citenamefont
  {Christov}, \citenamefont {Murnane},\ and\ \citenamefont
  {Kapteyn}}]{Popmintchev2009}%
  \BibitemOpen
  \bibfield  {author} {\bibinfo {author} {\bibfnamefont {Tenio}\ \bibnamefont
  {Popmintchev}}, \bibinfo {author} {\bibfnamefont {Ming-Chang}\ \bibnamefont
  {Chen}}, \bibinfo {author} {\bibfnamefont {Alon}\ \bibnamefont {Bahabad}},
  \bibinfo {author} {\bibfnamefont {Michael}\ \bibnamefont {Gerrity}}, \bibinfo
  {author} {\bibfnamefont {Pavel}\ \bibnamefont {Sidorenko}}, \bibinfo {author}
  {\bibfnamefont {Oren}\ \bibnamefont {Cohen}}, \bibinfo {author}
  {\bibfnamefont {Ivan~P}\ \bibnamefont {Christov}}, \bibinfo {author}
  {\bibfnamefont {Margaret~M}\ \bibnamefont {Murnane}}, \ and\ \bibinfo
  {author} {\bibfnamefont {Henry~C}\ \bibnamefont {Kapteyn}},\ }\bibfield
  {title} {\enquote {\bibinfo {title} {{Phase matching of high harmonic
  generation in the soft and hard X-ray regions of the spectrum.}}}\ }\href
  {\doibase 10.1073/pnas.0903748106} {\bibfield  {journal} {\bibinfo  {journal}
  {Proceedings of the National Academy of Sciences of the United States of
  America}\ }\textbf {\bibinfo {volume} {106}},\ \bibinfo {pages} {10516--21}
  (\bibinfo {year} {2009})}\BibitemShut {NoStop}%
\bibitem [{\citenamefont {Yeh}\ and\ \citenamefont {Lindau}(1985)}]{Yeh1985}%
  \BibitemOpen
  \bibfield  {author} {\bibinfo {author} {\bibfnamefont {J~J}\ \bibnamefont
  {Yeh}}\ and\ \bibinfo {author} {\bibfnamefont {I}~\bibnamefont {Lindau}},\
  }\bibfield  {title} {\enquote {\bibinfo {title} {{Atomic subshell
  photoionization cross sections and asymmetry parameters: 1 ⩽ Z ⩽ 103}},}\
  }\href {\doibase http://dx.doi.org/10.1016/0092-640X(85)90016-6} {\bibfield
  {journal} {\bibinfo  {journal} {Atomic Data and Nuclear Data Tables}\
  }\textbf {\bibinfo {volume} {32}},\ \bibinfo {pages} {1--155} (\bibinfo
  {year} {1985})}\BibitemShut {NoStop}%
\bibitem [{\citenamefont {Biegert}\ and\ \citenamefont
  {Jens}(2014)}]{Biegert2014}%
  \BibitemOpen
  \bibfield  {author} {\bibinfo {author} {\bibfnamefont {Dane R~Austin}\
  \bibnamefont {Biegert}}\ and\ \bibinfo {author} {\bibnamefont {Jens}},\
  }\bibfield  {title} {\enquote {\bibinfo {title} {{Attosecond pulse shaping
  using partial phase matching}},}\ }\href
  {http://stacks.iop.org/1367-2630/16/i=11/a=113011} {\bibfield  {journal}
  {\bibinfo  {journal} {New Journal of Physics}\ }\textbf {\bibinfo {volume}
  {16}},\ \bibinfo {pages} {113011} (\bibinfo {year} {2014})}\BibitemShut
  {NoStop}%
\bibitem [{\citenamefont {Kim}\ \emph {et~al.}(2004)\citenamefont {Kim},
  \citenamefont {Kim}, \citenamefont {Baik}, \citenamefont {Umesh},\ and\
  \citenamefont {Nam}}]{Kim2004}%
  \BibitemOpen
  \bibfield  {author} {\bibinfo {author} {\bibfnamefont {Kyung~Taec}\
  \bibnamefont {Kim}}, \bibinfo {author} {\bibfnamefont {Chul~Min}\
  \bibnamefont {Kim}}, \bibinfo {author} {\bibfnamefont {Moon~Gu}\ \bibnamefont
  {Baik}}, \bibinfo {author} {\bibfnamefont {G.}~\bibnamefont {Umesh}}, \ and\
  \bibinfo {author} {\bibfnamefont {Chang~Hee}\ \bibnamefont {Nam}},\
  }\bibfield  {title} {\enquote {\bibinfo {title} {{Single sub-50-attosecond
  pulse generation from chirp-compensated harmonic radiation using material
  dispersion}},}\ }\href {\doibase 10.1103/PhysRevA.69.051805} {\bibfield
  {journal} {\bibinfo  {journal} {Physical Review A - Atomic, Molecular, and
  Optical Physics}\ }\textbf {\bibinfo {volume} {69}},\ \bibinfo {pages}
  {051805--1} (\bibinfo {year} {2004})}\BibitemShut {NoStop}%
\bibitem [{\citenamefont {Doumy}\ \emph {et~al.}(2009)\citenamefont {Doumy},
  \citenamefont {Wheeler}, \citenamefont {Roedig}, \citenamefont {Chirla},
  \citenamefont {Agostini},\ and\ \citenamefont {Dimauro}}]{Doumy2009}%
  \BibitemOpen
  \bibfield  {author} {\bibinfo {author} {\bibfnamefont {G.}~\bibnamefont
  {Doumy}}, \bibinfo {author} {\bibfnamefont {J.}~\bibnamefont {Wheeler}},
  \bibinfo {author} {\bibfnamefont {C.}~\bibnamefont {Roedig}}, \bibinfo
  {author} {\bibfnamefont {R.}~\bibnamefont {Chirla}}, \bibinfo {author}
  {\bibfnamefont {P.}~\bibnamefont {Agostini}}, \ and\ \bibinfo {author}
  {\bibfnamefont {L.~F.}\ \bibnamefont {Dimauro}},\ }\bibfield  {title}
  {\enquote {\bibinfo {title} {{Attosecond synchronization of high-order
  harmonics from midinfrared drivers}},}\ }\href {\doibase
  10.1103/PhysRevLett.102.093002} {\bibfield  {journal} {\bibinfo  {journal}
  {Physical Review Letters}\ }\textbf {\bibinfo {volume} {102}},\ \bibinfo
  {pages} {2--5} (\bibinfo {year} {2009})}\BibitemShut {NoStop}%
\bibitem [{\citenamefont {Mairesse}\ \emph {et~al.}(2003)\citenamefont
  {Mairesse}, \citenamefont {de~Bohan}, \citenamefont {Frasinski},
  \citenamefont {Merdji}, \citenamefont {Dinu}, \citenamefont {Monchicourt},
  \citenamefont {Breger}, \citenamefont {Kova{\v{c}}ev}, \citenamefont
  {Ta{\"{i}}eb}, \citenamefont {Carr{\'{e}}}, \citenamefont {Muller},
  \citenamefont {Agostini},\ and\ \citenamefont
  {Sali{\`{e}}res}}]{Mairesse2003a}%
  \BibitemOpen
  \bibfield  {author} {\bibinfo {author} {\bibfnamefont {Y}~\bibnamefont
  {Mairesse}}, \bibinfo {author} {\bibfnamefont {A}~\bibnamefont {de~Bohan}},
  \bibinfo {author} {\bibfnamefont {L~J}\ \bibnamefont {Frasinski}}, \bibinfo
  {author} {\bibfnamefont {H}~\bibnamefont {Merdji}}, \bibinfo {author}
  {\bibfnamefont {L~C}\ \bibnamefont {Dinu}}, \bibinfo {author} {\bibfnamefont
  {P}~\bibnamefont {Monchicourt}}, \bibinfo {author} {\bibfnamefont
  {P}~\bibnamefont {Breger}}, \bibinfo {author} {\bibfnamefont {M}~\bibnamefont
  {Kova{\v{c}}ev}}, \bibinfo {author} {\bibfnamefont {R}~\bibnamefont
  {Ta{\"{i}}eb}}, \bibinfo {author} {\bibfnamefont {B}~\bibnamefont
  {Carr{\'{e}}}}, \bibinfo {author} {\bibfnamefont {H~G}\ \bibnamefont
  {Muller}}, \bibinfo {author} {\bibfnamefont {P.}~\bibnamefont {Agostini}}, \
  and\ \bibinfo {author} {\bibfnamefont {P}~\bibnamefont {Sali{\`{e}}res}},\
  }\bibfield  {title} {\enquote {\bibinfo {title} {{Attosecond Synchronization
  of High-Harmonic Soft X-rays}},}\ }\href {\doibase 10.1126/science.1090277}
  {\bibfield  {journal} {\bibinfo  {journal} {Science}\ }\textbf {\bibinfo
  {volume} {302}},\ \bibinfo {pages} {1540--1543} (\bibinfo {year}
  {2003})}\BibitemShut {NoStop}%
\bibitem [{\citenamefont {Ko}\ \emph {et~al.}(2012)\citenamefont {Ko},
  \citenamefont {Kim},\ and\ \citenamefont {Nam}}]{Ko2012}%
  \BibitemOpen
  \bibfield  {author} {\bibinfo {author} {\bibfnamefont {Dong~Hyuk}\
  \bibnamefont {Ko}}, \bibinfo {author} {\bibfnamefont {Kyung~Taec}\
  \bibnamefont {Kim}}, \ and\ \bibinfo {author} {\bibfnamefont {Chang~Hee}\
  \bibnamefont {Nam}},\ }\bibfield  {title} {\enquote {\bibinfo {title}
  {{Attosecond-chirp compensation with material dispersion to produce near
  transform-limited attosecond pulses}},}\ }\href {\doibase
  10.1088/0953-4075/45/7/074015} {\bibfield  {journal} {\bibinfo  {journal}
  {Journal of Physics B: Atomic, Molecular and Optical Physics}\ }\textbf
  {\bibinfo {volume} {45}},\ \bibinfo {pages} {074015} (\bibinfo {year}
  {2012})}\BibitemShut {NoStop}%
\bibitem [{\citenamefont {Gagnon}\ and\ \citenamefont
  {Yakovlev}(2009)}]{Gagnon2009}%
  \BibitemOpen
  \bibfield  {author} {\bibinfo {author} {\bibfnamefont {Justin}\ \bibnamefont
  {Gagnon}}\ and\ \bibinfo {author} {\bibfnamefont {Vladislav~S}\ \bibnamefont
  {Yakovlev}},\ }\bibfield  {title} {\enquote {\bibinfo {title} {{The
  robustness of attosecond streaking measurements.}}}\ }\href {\doibase
  10.1364/OE.17.017678} {\bibfield  {journal} {\bibinfo  {journal} {Optics
  Express}\ }\textbf {\bibinfo {volume} {17}},\ \bibinfo {pages} {17678--17693}
  (\bibinfo {year} {2009})}\BibitemShut {NoStop}%
\bibitem [{\citenamefont {Gagnon}\ \emph {et~al.}(2008)\citenamefont {Gagnon},
  \citenamefont {Goulielmakis},\ and\ \citenamefont {Yakovlev}}]{Gagnon2008}%
  \BibitemOpen
  \bibfield  {author} {\bibinfo {author} {\bibfnamefont {J.}~\bibnamefont
  {Gagnon}}, \bibinfo {author} {\bibfnamefont {E.}~\bibnamefont
  {Goulielmakis}}, \ and\ \bibinfo {author} {\bibfnamefont {V.~S.}\
  \bibnamefont {Yakovlev}},\ }\bibfield  {title} {\enquote {\bibinfo {title}
  {{The accurate FROG characterization of attosecond pulses from streaking
  measurements}},}\ }\href {\doibase 10.1007/s00340-008-3063-x} {\bibfield
  {journal} {\bibinfo  {journal} {Applied Physics B: Lasers and Optics}\
  }\textbf {\bibinfo {volume} {92}},\ \bibinfo {pages} {25--32} (\bibinfo
  {year} {2008})}\BibitemShut {NoStop}%
\bibitem [{\citenamefont {Becker}\ and\ \citenamefont
  {Shirley}(1996)}]{Becker1996}%
  \BibitemOpen
  \bibfield  {author} {\bibinfo {author} {\bibfnamefont {U.}~\bibnamefont
  {Becker}}\ and\ \bibinfo {author} {\bibfnamefont {D.~A.}\ \bibnamefont
  {Shirley}},\ }\bibfield  {title} {\enquote {\bibinfo {title} {{Partial Cross
  Sections and Angular Distributions}},}\ }in\ \href {\doibase
  10.1007/978-1-4613-0315-2_5} {\emph {\bibinfo {booktitle} {VUV and Soft X-Ray
  Photoionization}}}\ (\bibinfo  {publisher} {Springer},\ \bibinfo {address}
  {Boston, MA},\ \bibinfo {year} {1996})\ pp.\ \bibinfo {pages}
  {135--180}\BibitemShut {NoStop}%
\bibitem [{\citenamefont {Lindle}\ \emph {et~al.}(1986)\citenamefont {Lindle},
  \citenamefont {Heimann}, \citenamefont {Ferrett}, \citenamefont {Kobrin},
  \citenamefont {Truesdale}, \citenamefont {Becker}, \citenamefont {Kerkhoff},\
  and\ \citenamefont {Shirley}}]{Lindle1986}%
  \BibitemOpen
  \bibfield  {author} {\bibinfo {author} {\bibfnamefont {D~W}\ \bibnamefont
  {Lindle}}, \bibinfo {author} {\bibfnamefont {P~A}\ \bibnamefont {Heimann}},
  \bibinfo {author} {\bibfnamefont {T~A}\ \bibnamefont {Ferrett}}, \bibinfo
  {author} {\bibfnamefont {P~H}\ \bibnamefont {Kobrin}}, \bibinfo {author}
  {\bibfnamefont {C~M}\ \bibnamefont {Truesdale}}, \bibinfo {author}
  {\bibfnamefont {U}~\bibnamefont {Becker}}, \bibinfo {author} {\bibfnamefont
  {H~G}\ \bibnamefont {Kerkhoff}}, \ and\ \bibinfo {author} {\bibfnamefont
  {D~A}\ \bibnamefont {Shirley}},\ }\bibfield  {title} {\enquote {\bibinfo
  {title} {{Photoemission from the 3d and 3p subshells of Kr}},}\ }\href
  {https://link.aps.org/doi/10.1103/PhysRevA.33.319} {\bibfield  {journal}
  {\bibinfo  {journal} {Physical Review A}\ }\textbf {\bibinfo {volume} {33}},\
  \bibinfo {pages} {319--323} (\bibinfo {year} {1986})}\BibitemShut {NoStop}%
\bibitem [{\citenamefont {Ricz}\ \emph {et~al.}(2003)\citenamefont {Ricz},
  \citenamefont {Sankari}, \citenamefont {K{\"{o}}v{\'{e}}r}, \citenamefont
  {Jurvansuu}, \citenamefont {Varga}, \citenamefont {Nikkinen}, \citenamefont
  {Ricsoka}, \citenamefont {Aksela},\ and\ \citenamefont {Aksela}}]{Ricz2003}%
  \BibitemOpen
  \bibfield  {author} {\bibinfo {author} {\bibfnamefont {S}~\bibnamefont
  {Ricz}}, \bibinfo {author} {\bibfnamefont {R}~\bibnamefont {Sankari}},
  \bibinfo {author} {\bibfnamefont {{\'{A}}}~\bibnamefont {K{\"{o}}v{\'{e}}r}},
  \bibinfo {author} {\bibfnamefont {M}~\bibnamefont {Jurvansuu}}, \bibinfo
  {author} {\bibfnamefont {D}~\bibnamefont {Varga}}, \bibinfo {author}
  {\bibfnamefont {J}~\bibnamefont {Nikkinen}}, \bibinfo {author} {\bibfnamefont
  {T}~\bibnamefont {Ricsoka}}, \bibinfo {author} {\bibfnamefont
  {H}~\bibnamefont {Aksela}}, \ and\ \bibinfo {author} {\bibfnamefont
  {S}~\bibnamefont {Aksela}},\ }\bibfield  {title} {\enquote {\bibinfo {title}
  {{Strong nondipole effect created by multielectron correlation in 5s
  photoionization of xenon}},}\ }\href
  {https://link.aps.org/doi/10.1103/PhysRevA.67.012712} {\bibfield  {journal}
  {\bibinfo  {journal} {Physical Review A}\ }\textbf {\bibinfo {volume} {67}},\
  \bibinfo {pages} {12712} (\bibinfo {year} {2003})}\BibitemShut {NoStop}%
\bibitem [{\citenamefont {Verhoef}\ \emph {et~al.}(2011)\citenamefont
  {Verhoef}, \citenamefont {Mitrofanov}, \citenamefont {Nguyen}, \citenamefont
  {Krikunova}, \citenamefont {Fritzsche}, \citenamefont {Kabachnik},
  \citenamefont {A},\ and\ \citenamefont {Baltu{\v{s}}ka}}]{Verhoef2011}%
  \BibitemOpen
  \bibfield  {author} {\bibinfo {author} {\bibfnamefont {A~J}\ \bibnamefont
  {Verhoef}}, \bibinfo {author} {\bibfnamefont {A~V}\ \bibnamefont
  {Mitrofanov}}, \bibinfo {author} {\bibfnamefont {X~T}\ \bibnamefont
  {Nguyen}}, \bibinfo {author} {\bibfnamefont {M}~\bibnamefont {Krikunova}},
  \bibinfo {author} {\bibfnamefont {S}~\bibnamefont {Fritzsche}}, \bibinfo
  {author} {\bibfnamefont {N~M}\ \bibnamefont {Kabachnik}}, \bibinfo {author}
  {\bibfnamefont {M~Drescher}\ \bibnamefont {A}}, \ and\ \bibinfo {author}
  {\bibfnamefont {A.}~\bibnamefont {Baltu{\v{s}}ka}},\ }\bibfield  {title}
  {\enquote {\bibinfo {title} {{Time-and-energy-resolved measurement of Auger
  cascades following Kr 3d excitation by attosecond pulses}},}\ }\href
  {http://stacks.iop.org/1367-2630/13/i=11/a=113003} {\bibfield  {journal}
  {\bibinfo  {journal} {New Journal of Physics}\ }\textbf {\bibinfo {volume}
  {13}},\ \bibinfo {pages} {113003} (\bibinfo {year} {2011})}\BibitemShut
  {NoStop}%
\bibitem [{\citenamefont {Jauhiainen}\ \emph {et~al.}(1995)\citenamefont
  {Jauhiainen}, \citenamefont {Kivimaki}, \citenamefont {S}, \citenamefont
  {Sairanen},\ and\ \citenamefont {Aksela}}]{Jauhiainen1995}%
  \BibitemOpen
  \bibfield  {author} {\bibinfo {author} {\bibfnamefont {J.}~\bibnamefont
  {Jauhiainen}}, \bibinfo {author} {\bibfnamefont {A}~\bibnamefont {Kivimaki}},
  \bibinfo {author} {\bibfnamefont {Aksela}\ \bibnamefont {S}}, \bibinfo
  {author} {\bibfnamefont {O~P}\ \bibnamefont {Sairanen}}, \ and\ \bibinfo
  {author} {\bibfnamefont {H}~\bibnamefont {Aksela}},\ }\bibfield  {title}
  {\enquote {\bibinfo {title} {{Auger and Coster-Kronig decay of the 3p hole
  states in krypton}},}\ }\href {http://stacks.iop.org/0953-4075/28/i=18/a=012}
  {\bibfield  {journal} {\bibinfo  {journal} {Journal of Physics B: Atomic,
  Molecular and Optical Physics}\ }\textbf {\bibinfo {volume} {28}},\ \bibinfo
  {pages} {4091} (\bibinfo {year} {1995})}\BibitemShut {NoStop}%
\bibitem [{\citenamefont {Jonauskas}\ \emph {et~al.}(2011)\citenamefont
  {Jonauskas}, \citenamefont {Ku{\v{c}}as},\ and\ \citenamefont
  {Karazija}}]{Jonauskas2011}%
  \BibitemOpen
  \bibfield  {author} {\bibinfo {author} {\bibfnamefont {V}~\bibnamefont
  {Jonauskas}}, \bibinfo {author} {\bibfnamefont {S}~\bibnamefont
  {Ku{\v{c}}as}}, \ and\ \bibinfo {author} {\bibfnamefont {R}~\bibnamefont
  {Karazija}},\ }\bibfield  {title} {\enquote {\bibinfo {title} {{Auger decay
  of 3{\$}p{\$}-ionized krypton}},}\ }\href
  {https://link.aps.org/doi/10.1103/PhysRevA.84.053415} {\bibfield  {journal}
  {\bibinfo  {journal} {Physical Review A}\ }\textbf {\bibinfo {volume} {84}},\
  \bibinfo {pages} {53415} (\bibinfo {year} {2011})}\BibitemShut {NoStop}%
\bibitem [{\citenamefont {Kheifets}(2013)}]{Kheifets2013}%
  \BibitemOpen
  \bibfield  {author} {\bibinfo {author} {\bibfnamefont {A~S}\ \bibnamefont
  {Kheifets}},\ }\bibfield  {title} {\enquote {\bibinfo {title} {{Time delay in
  valence-shell photoionization of noble-gas atoms}},}\ }\href
  {https://link.aps.org/doi/10.1103/PhysRevA.87.063404} {\bibfield  {journal}
  {\bibinfo  {journal} {Physical Review A}\ }\textbf {\bibinfo {volume} {87}},\
  \bibinfo {pages} {63404} (\bibinfo {year} {2013})}\BibitemShut {NoStop}%
\bibitem [{\citenamefont {Calegari}\ \emph {et~al.}(2013)\citenamefont
  {Calegari}, \citenamefont {Lucchini}, \citenamefont {Sansone}, \citenamefont
  {Stagira}, \citenamefont {Vozzi},\ and\ \citenamefont
  {Nisoli}}]{Calegari2013}%
  \BibitemOpen
  \bibfield  {author} {\bibinfo {author} {\bibfnamefont {F}~\bibnamefont
  {Calegari}}, \bibinfo {author} {\bibfnamefont {M}~\bibnamefont {Lucchini}},
  \bibinfo {author} {\bibfnamefont {G}~\bibnamefont {Sansone}}, \bibinfo
  {author} {\bibfnamefont {S}~\bibnamefont {Stagira}}, \bibinfo {author}
  {\bibfnamefont {C}~\bibnamefont {Vozzi}}, \ and\ \bibinfo {author}
  {\bibfnamefont {M}~\bibnamefont {Nisoli}},\ }\bibfield  {title} {\enquote
  {\bibinfo {title} {{Attosecond Pulse Characterization}},}\ }in\ \href
  {\doibase 10.1007/978-3-642-37623-8_5} {\emph {\bibinfo {booktitle}
  {Attosecond Physics: Attosecond Measurements and Control of Physical
  Systems}}},\ \bibinfo {editor} {edited by\ \bibinfo {editor} {\bibfnamefont
  {Luis}\ \bibnamefont {Plaja}}, \bibinfo {editor} {\bibfnamefont {Ricardo}\
  \bibnamefont {Torres}}, \ and\ \bibinfo {editor} {\bibfnamefont {Amelle}\
  \bibnamefont {Za{\"{i}}r}}}\ (\bibinfo  {publisher} {Springer Berlin
  Heidelberg},\ \bibinfo {address} {Berlin, Heidelberg},\ \bibinfo {year}
  {2013})\ pp.\ \bibinfo {pages} {69--80}\BibitemShut {NoStop}%
\bibitem [{\citenamefont {Teichmann}\ \emph {et~al.}(2016)\citenamefont
  {Teichmann}, \citenamefont {Silva}, \citenamefont {Cousin}, \citenamefont
  {Hemmer},\ and\ \citenamefont {Biegert}}]{Teichmann2016}%
  \BibitemOpen
  \bibfield  {author} {\bibinfo {author} {\bibfnamefont {S.~M.}\ \bibnamefont
  {Teichmann}}, \bibinfo {author} {\bibfnamefont {F.}~\bibnamefont {Silva}},
  \bibinfo {author} {\bibfnamefont {S.~L.}\ \bibnamefont {Cousin}}, \bibinfo
  {author} {\bibfnamefont {M.}~\bibnamefont {Hemmer}}, \ and\ \bibinfo {author}
  {\bibfnamefont {J.}~\bibnamefont {Biegert}},\ }\bibfield  {title} {\enquote
  {\bibinfo {title} {{0.5-keV Soft X-ray attosecond continua}},}\ }\href
  {\doibase 10.1038/ncomms11493} {\bibfield  {journal} {\bibinfo  {journal}
  {Nature Communications}\ }\textbf {\bibinfo {volume} {7}},\ \bibinfo {pages}
  {11493} (\bibinfo {year} {2016})}\BibitemShut {NoStop}%
\end{thebibliography}%
\end{document}